\documentclass[12pt]{article}
\usepackage{rotating,latexsym,amssymb}
\usepackage{epstopdf}
\usepackage{graphicx}
\usepackage{mathptmx} % fr das Times-\mu (�tere Variante w�e mathptm)
\usepackage{times}
\usepackage{lineno}
\usepackage{natbib}
\newcommand{\mc}[3]{\multicolumn{#1}{#2}{#3}}

\newcommand{\kms}{$\mathrm {km\,s^{-1}}$}

\newcommand{\QI}{\,Q_{\mathrm{I}}}
\newcommand{\QE}{\,Q_{\mathrm{E}}}
\newcommand{\QC}{\,Q_{\mathrm{C}}}

\linenumbers

\begin{document}
\bibliographystyle{apalike}

%{\small \em Submitted to Planetary and Space Science \hfill \today}

\medskip

\begin{center}
{\Large \bf Galileo dust data from the jovian system: 2000 to 2003}
\end{center}

\bigskip

{\bf
        H.~Kr\"uger$^{a,b,}$\footnote{{\em Correspondence to:} 
Harald Kr\"uger, krueger@mps.mpg.de}, 
        D.~Bindschadler$^c$,
        S.~F.~Dermott$^d$,
%        H.~Fech\-tig$^b$,
        A.~L.~Graps$^{e}$,
        E.~Gr\"un$^{b,f}$,
        B.~A.~Gustaf\-son$^d$,
        D.~P.~Hamilton$^g$, 
        M.~S.~Hanner$^h$,
        M.~Hor\'anyi$^f$,
        J.~Kissel$^a$,
%        B.~A.~Lind\-blad$^h$,
        D.~Linkert$^b$,
        G.~Linkert$^b$,
        I.~Mann$^{i}$,
        J.~A.~M.~McDonnell$^{j}$,
        R.~Moissl$^{a}$,
        G.~E.~Mor\-fill$^{k}$, 
        C.~Polanskey$^c$,
        M.~Roy$^{c}$, 
        G.~Schwehm$^{l}$ and
        R.~Srama$^{b,m}$ 
        }

\bigskip

\small
\begin{tabular}{ll}
a)& Max-Planck-Institut f\"ur Sonnensystemforschung, 37191 Katlenburg-Lindau, \\ 
  &  Germany\\
b)& Max-Planck-Institut f\"ur Kernphysik, 69029 Heidelberg, Germany\\
c)& Jet Propulsion Laboratory, Pasadena, California 91109, USA\\
d)& University of Florida, 211 SSRB, Campus, Gainesville, FL\,32609, USA \\
%e)& Istituto di Fisica dello Spazio Interplanetario, INAF - ARTOV,
% 00133 Roma, Italy \\
e)& Department of Space Studies, Southwest Research Institute, 1050 Walnut \\ 
  & Street, Suite 300, Boulder, Colorado, 80302, USA \\
f)& Laboratory for Atmospheric and Space Physics, Univ. of Colorado, Boulder, \\ 
  & CO\,80309, USA\\
g)& University of Maryland, College Park, MD\,20742-2421, USA\\
h)& Astronomy Dept. 619 LGRT, University of Massachusetts, Amherst MA 01003, USA \\
%h)& Lund Observatory, 221 Lund, Sweden\\
%i)& Department of Earth and Planetary Sciences, Faculty of Science, \\ 
%  & Kobe University,  Nada, Kobe 657-8501, Japan \\
i)& School of Science and Engineering, Kindai University, \\
  & Kowakae 3-4-1, Higashi-Osaka, Osaka, 577-8502, Japan \\
j)& Planetary and Space Science Research Institute, The Open University, \\
  & Milton Keynes, MK7 6AA, UK\\
k)& Max-Planck-Institut f\"ur Extraterrestrische Physik, 85748 Garching, 
                                                                   Germany\\ 
l)& ESAC, PO Box 78, 28691 Villanueva de la Ca{\~n}ada, Spain \\
m)& Universit{\"a}t Stuttgart, Institut f{\"u}r Raumfahrtsysteme, Pfaffenwaldring 31, \\ 
  & 70569 Stuttgart, Germany \\
\end{tabular}

\normalsize

\bigskip
\clearpage

\begin{abstract}

The Galileo spacecraft was the first man-made 
satellite of Jupiter, orbiting the planet between December 1995 and September 2003. 
The spacecraft was equipped with a highly sensitive dust detector that
monitored the jovian dust environment between approximately 2 and 370
$\mathrm{R_J}$ (jovian radius $\mathrm{R_J = 71492\,km}$). The Galileo
dust detector was a twin of the one flying on board the Ulysses spacecraft.
This is the tenth in a series of papers dedicated to presenting Galileo and 
Ulysses dust data. Here we present data from the Galileo dust instrument for the 
period January 2000 to September 2003 until Galileo was destroyed in a planned
impact with Jupiter. 
%In this time interval the spacecraft completed nine
%revolutions about Jupiter. Data were obtained as high 
%resolution realtime science data or recorded data during 570 days (representing
%42\% of the entire period), or via memory readouts during the remaining times.
%Because the data transmission rate of the spacecraft was very low, the 
%complete data sets  
%(i.~e. all parameters measured by the instrument during impact of a dust 
%particle) of only  2\% (5389) of all particles detected could be transmitted 
%to Earth; the other particles were only counted. Together with the data of 
%2883 particles detected 
The previous Galileo dust data set contains  data  of 2883 particles detected 
during Galileo's interplanetary cruise and 12978 
particles detected in the jovian system between 1996 and 1999.
In this paper we report on the data of additional 5389 particles 
measured between 2000 and the end of the mission in 2003.
The majority of the 21250 particles for which the full set of measured 
impact parameters (impact time, impact direction, charge rise times, charge amplitudes, etc.) 
was transmitted to Earth were tiny grains (about 10~nm in radius),
most of them originating from Jupiter's innermost Galilean moon Io. 
They were detected throughout the jovian system and the 
impact rates frequently exceeded $\rm 10\,min^{-1}$. Surprisingly large 
impact rates up to $\rm 100 \,min^{-1}$ occurred in August/September 2000 when 
Galileo was far away ($\mathrm{\approx 280\,R_J}$) from Jupiter, implying dust
ejection rates in excess of $\mathrm{100\,kg\,s^{-1}}$. This peak in dust emission
appears to coincide with strong changes in the release of neutral gas from the Io
torus. Strong variability in the Io dust flux was measured on timescales of
days to weeks, indicating
large variations in the dust release from Io or the Io torus or both on such 
short timescales.
Galileo has detected a large number of bigger micron-sized particles 
mostly in the region between the Galilean moons. A surprisingly large number of such 
bigger grains was measured in March 2003 within a 4-day interval when Galileo
was outside Jupiter's magnetosphere at approximately $\mathrm{350\,R_J}$ jovicentric 
distance. 
Two passages of Jupiter's gossamer rings in 2002 and 2003 provided the first 
actual comparison of in-situ dust data from a planetary ring with the results 
inferred from inverting optical images. 
Strong electronics degradation of the dust instrument due to the harsh radiation 
environment of Jupiter led to increased calibration uncertainties of the dust data.
%lead to significant changes in the measured charge
%amplitudes and rise times used to calibrate dust impact speed and mass. 
%This made the mass and speed calibration of the dust impacts after 2000 invalid.
\end{abstract}

\clearpage

\section{Introduction}

\label{introduction}

The Galileo spacecraft was the first artifical satellite  
orbiting Jupiter. Galileo had a
highly sensitive impact ionization dust detector on board
which was identical with the dust detector of the Ulysses 
spacecraft \citep{gruen1992a,gruen1992b,gruen1995a}. 
Dust data from both spacecraft were used for the analysis of 
e.\,g. the interplanetary dust complex, 
dust related to asteroids and comets, 
interstellar dust grains sweeping through the solar system, 
and various dust phenomena in the environment of Jupiter.
References can be found in \citet{krueger1999a,krueger1999b}.

In Section~\ref{sec_results} we summarize results that
are related to dust in the Jupiter system. 
A comprehensive overview of the investigation of dust in the 
jovian system was given by \citet{krueger2003c} and \citet{krueger2004a}.

\subsection{Summary of results from the Galileo dust investigations at Jupiter}

\label{sec_results}

The Jupiter system was found to be a strong source of dust when in 1992 Ulysses flew by the 
planet and discovered streams 
of dust particles emanating from 
the giant planet's magnetosphere  \citep{gruen1993a}. These were 
later confirmed by Galileo \citep{gruen1996b,gruen1996c} and measured
again by Ulysses in 2003-05 during its second flyby at the planet
\citep{krueger2006c,flandes2007,flandes2009}. 
At least four dust populations were identified
in the Jupiter system with Galileo \citep{gruen1997b,gruen1998}: 

i) Streams of dust particles with high and variable impact rates 
throughout Jupiter's magnetosphere. They are the extension of streams 
discovered with Ulysses outside Jupiter's magnetosphere. The particles are 
about 10\,nm in radius \citep{zook1996}
and they mostly originate from the 
innermost Galilean moon Io \citep{graps2000a}. Because
of their small sizes the charged grains strongly interact with Jupiter's 
magnetosphere \citep{horanyi1997,gruen1998,heck1998}, and they
are a natural laboratory to study dust-plasma interactions. The dust streams 
mostly show a dust-in-plasma behavior while 
only some portions of those Galileo orbits displaying the highest dust stream 
fluxes (Galileo orbits E4, G7, G8, C21) satisfy the minimum requirements for 
a dusty plasma  \citep{graps2006}.
The dust streams served as a monitor 
of Io's volcanic plume activity \citep{krueger2003d} and as 
probes of the Io plasma torus \citep{krueger2003a}. Dust 
charging mechanisms in Io's plumes and in the jovian magnetosphere
were investigated by \citet{graps2001a} and \citet{flandes2004}. 
Dust measurements of the Cassini spacecraft at its Jupiter flyby
in 2000 showed that the grains are mostly composed of sodium chloride
(NaCl) formed by condensation in Io's volcanic plumes \citep{postberg2006}.

ii) Dust clouds surrounding the Galilean moons which consist of mostly 
sub-micron grains \citep{krueger1999d,krueger2000a,krueger2003b}. These 
grains were ejected from the moons' surfaces by hypervelocity 
impacts of interplanetary dust particles
\citep{krivov2003,sremcevic2003,sremcevic2005}.

iii) Bigger micron-sized grains  forming a tenuous dust ring 
between the Galilean moons. This group is composed of two sub-populations, 
one orbiting Jupiter on prograde orbits and a second one on 
retrograde orbits \citep{colwell1998a,thiessenhusen2000}. Most of the 
prograde population is maintained by grains escaping
from the clouds that surround the Galilean moons 
\citep{krivov2002a,krivov2002b}.

iv) On 5 November 2002 and 21 September 2003 -- before Galileo  was 
destroyed in a planned impact with Jupiter -- the spacecraft traversed Jupiter's
gossamer ring twice and provided the first in-situ measurements of a
dusty planetary ring \citep{krueger2003c,moissl2005,hamilton2008,krueger2009b} 
which is also accessible with astronomical imaging techniques. 
These fly-throughs revealed previously unknown structures in the gossamer rings: 
a drop in the dust density between the moons Amalthea and Thebe, grains orbiting
Jupiter on highly inclined orbits and an increase in the number of small 
grains in the inner regions of the rings as compared to the regions further
away from the planet. All these features can nicely be explained by electromagnetic
forces on the grains that shape the gossamer rings \citep{hamilton2008}.

\subsection{The Galileo and Ulysses dust data papers}

This is the tenth paper in a series dedicated to presenting both 
raw and reduced data from the Galileo and Ulysses dust instruments. 
\citet[][hereafter Paper~I]{gruen1995a} described the reduction 
process of Galileo and Ulysses dust data.
In the even-numbered Papers~II, IV, VI and VIII 
\citep{gruen1995b,krueger1999a,krueger2001a,krueger2006a} 
we presented the Galileo data set spanning the 
ten year time period from October 1989 to December 1999. 
The present paper extends the 
Galileo data set from January 2000 to September 2003, which covers
the Galileo Millenium mission and two traverses of Jupiter's gossamer ring
until the spacecraft impacted Jupiter on 21 September 2003.
Companion odd-numbered 
Papers~III, V, VII, IX and XI
\citep{gruen1995c,krueger1999b,krueger2001b,krueger2006b,krueger2010b} provide
the entire dust data set measured with 
Ulysses  between 1990 and  2007.
An overview of our 
Galileo dust data papers and mission highlights is given in Table~\ref{papers}.

\begin{center} \fbox{\bf Insert Table~\ref{papers}} \end{center}

The main data products are a table of the number of all impacts 
determined from the particle accumulators and a table of both raw and
reduced data of all ``big'' impacts received on the ground. The 
information presented in these papers is similar to data which we 
are submitting to the various data archiving centres (Planetary 
Data System, NSSDC, etc.). The only difference is that the paper version 
does not contain the full data set of the large number of ``small'' 
particles, and the numbers of impacts deduced from the accumulators
are typically averaged over several days. Electronic access to the 
complete data set including the numbers of impacts deduced from the 
accumulators in full time resolution is also possible via 
the world wide web: http://www.mpi-hd.mpg.de/dustgroup/.

This paper is organised similarly to our previous papers. Section~\ref{mission} 
gives a brief overview of the Galileo mission with particular emphasis on
the time period 2000-2003,
the dust instrument operation and lists important mission events in the time
interval 2000-2003 considered in this paper.
A description of the new Galileo dust data set for 2000-2003
together with a discussion of the detected noise and dust impact rates is given 
in Section~\ref{events}. Section~\ref{analysis} analyses and discusses 
various characteristics of the new data set. 
Finally, in Section~\ref{discussion} we discuss
results on jovian dust achieved with this new data set, and in Section~\ref{sec_summary}
we summarise our results.

\section{Mission and instrument operations} \label{mission}

\subsection{Galileo mission}

Galileo was launched on 18 October 1989. Two flybys at 
Earth and one at Venus between 1990 and 1992 gave the spacecraft 
enough energy to leave the inner solar system. During its 
interplanetary voyage Galileo had close encounters with the 
asteroids Gaspra and Ida. On 7 December 1995 the spacecraft 
arrived at Jupiter and was injected into a highly elliptical orbit 
about the planet, becoming the first spacecraft orbiting a planet 
of the outer solar system. Galileo performed 34 revolutions about 
Jupiter until 21 September 2003 when the spacecraft was destroyed in 
a planned impact with Jupiter.

Galileo's trajectory during its orbital tour about 
Jupiter from January 2000 to  September 2003 is shown in 
Figure~\ref{trajectory}. Galileo had regular close flybys at 
Jupiter's Galilean moons. Eight such encounters occurred in the 
2000-2003 interval (1 at Europa, 4 at Io,
2 at Ganymede, 1 at Callisto) plus one at Amalthea (Table~\ref{event_table_1}).
Galileo orbits 
are labelled with the first letter of the Galilean 
moon which was the encounter target during that orbit,
followed by the orbit number. For example, 
``G29'' refers to a Ganymede flyby in orbit 29. Satellite flybys always occurred 
within two days of 
Jupiter closest approach (pericentre passage). Detailed descriptions of 
the Galileo mission and the spacecraft were given by 
\citet{johnson1992} and  \citet{damario1992}. 

\begin{center} \fbox{\bf Insert Table~\ref{event_table_1}} \end{center}
\begin{center} \fbox{\bf Insert Figure~\ref{trajectory}} \end{center}

Galileo was a dual spinning spacecraft with an antenna that pointed 
antiparallel to the positive spin axis. During most of the initial 3 years 
of the mission the antenna pointed towards the Sun 
(Paper~II). Since 1993 the antenna was usually pointed towards Earth.
Deviations from the Earth pointing direction 
in 2000-2003, the time period considered 
in this paper, are shown in Figure~\ref{pointing}.	
Sharp spikes in the pointing deviation occurred when the spacecraft was turned
away from the nominal Earth direction for dedicated imaging observations with 
Galileo's cameras or for orbit trim maneuvers with the spacecraft thrusters. 
These spikes lasted typically several hours. From January to September 2003, 
the Galileo pointing deviated significantly from the Earth pointing direction 
for a long time interval. Table~\ref{event_table_1} 
lists significant mission and dust instrument events for 2000-2003. 
Comprehensive lists of earlier events can be found in Papers~II, IV, VI and VIII. 

\begin{center} \fbox{\bf Insert Figure~\ref{pointing}} \end{center}

\subsection{Dust detection geometry}  \label{det_geom}

The Dust Detector System (DDS) was mounted on the spinning section of 
Galileo and the sensor axis was offset by $60^{\circ}$
from the positive spin axis (an angle of $55^{\circ}$
was erroneously stated in publications before). A schematic view 
of the Galileo spacecraft and the geometry of dust detection
is shown in the inset in  Figure~\ref{trajectory}.

The rotation angle measured the viewing 
direction of the dust sensor at the time of a dust impact. 
During one spin revolution of the spacecraft the rotation angle scanned 
through a complete circle of $360^{\circ}$. At rotation angles of 
$90^{\circ}$ and $270^{\circ}$ the sensor axis lay nearly 
in the ecliptic plane, and at $0^{\circ}$ it was close to the ecliptic 
north direction. 
DDS rotation angles are taken positive around the negative spin axis of 
the spacecraft which pointed towards Earth. This is done to facilitate 
comparison of the Galileo spin 
angle data with those taken by Ulysses, which, unlike
Galileo, had its positive spin axis pointed towards Earth \citep{gruen1995a}.

The nominal field-of-view (FOV) of the DDS sensor target 
was $140^{\circ}$. A smaller FOV applies to a 
subset of jovian dust stream particle impacts -- the 
so-called class~3 impacts in amplitude range AR1 \citep[][{\em cf.}~Paper~I and 
Section~\ref{events} for a definition of
these parameters]{krueger1999c} while the nominal target 
size should be applied to class~2 jovian dust stream 
impacts. For all impacts
which are not due to jovian dust stream particles
a larger FOV of $180^{\circ}$ should be 
applied because the inner sensor side wall turned out to be 
almost as sensitive to larger dust impacts as the target itself 
\citep{altobelli2004a,willis2004,willis2005}. These different
sensor fields-of-view and the corresponding target sizes are 
summarised in Table~\ref{tab_fov}.

\begin{center} \fbox{\bf Insert Table~\ref{tab_fov}} \end{center}
 
During one spin revolution
of the spacecraft the sensor axis scanned a cone with $120^{\circ}$
opening angle towards the anti-Earth direction. Dust particles that arrived
from within $10^{\circ}$ of the positive spin axis (anti-Earth direction) 
could be detected at all rotation angles, whereas those that arrived at angles 
from $10^{\circ}$ to $130^{\circ}$ from the positive spin axis could 
be detected over only a limited range of rotation angles. Note that
these angles refer to the nominal sensor field-of-view of $140^{\circ}$.

\subsection{Data transmission}

\label{sec_transmission}

In June 1990 the dust instrument was reprogrammed for 
the first time after launch and since then the instrument memory could store 46 
instrument data frames (with each frame comprising the
complete data set of an impact or noise event, consisting of 128 bits,
plus ancillary and engineering data; {\em cf.}~Papers~I and II). The dust instrument
time-tagged each impact event with an 8 bit word allowing for the identification of 
256 unique steps. In 1990 the step size of this time 
word was set to 4.3~hours. Hence, the total accumulation time after 
which the time word was reset and the time labels of older impact events 
became ambiguous was $\rm 256 \times 4.3\,h \simeq 46\,$days.  

During a large fraction of Galileo's orbital mission about Jupiter dust 
detector data were transmitted to Earth in the so-called realtime 
science mode (RTS). 
In RTS mode, dust data were read out either every 7.1 or every 21.2 minutes -- 
depending on the spacecraft data transmission rate -- and directly transmitted 
to Earth with a rate of 3.4 or 1.1 bits per second, respectively. Additionally,
Galileo had the so-called record mode. In this mode  data were read out from 
the dust instrument memory
with  24 bits per second, recorded on Galileo's tape recorder 
and transmitted to Earth up to several weeks later. Recorded data were 
received during three satellite flybys in 2000 during short periods 
of  $\mathrm{ \sim \pm 1/2\,\,hour}$ around closest 
approach to the satellite, and 
for $\rm \sim 3.8$ hours during Galileo's gossamer ring passage on 5 November 2002 
(Table~\ref{event_table_1}).
Details of the
various data transmission modes of Galileo are also given in Table~\ref{tab_data_modes}.

\begin{center} \fbox{\bf Insert Table~\ref{tab_data_modes}} \end{center}

%Only during short time intervals RTS data 
%were stored on Galileo's tape recorder and transmitted to Earth later. 
%Sometimes RTS data for short time intervals were  stored on the tape recorder
%and transmitted later but this did not change the labelling -- they are 
%also called RTS.  

In RTS and record mode the time between two readouts of the 
instrument memory determined the number of events in a given time period 
for which their complete information could be transmitted.
Thus, the complete information on each impact was
transmitted to Earth when the impact rate was below one impact per
either 7.1 or 21.2 minutes in RTS mode or one impact per minute in
record mode, respectively (Table~\ref{tab_data_modes}). If the impact rate exceeded these 
values, the detailed information of older events was lost because 
the full data set of only the latest event was stored in the dust
instrument memory. 

Furthermore, in RTS and record mode the time between two readouts
also defined the accuracy with which the impact time is known.
Hence, the uncertainty in the impact time is 7.1 or 21.2 minutes in RTS mode 
and about one minute in record mode, respectively. 
%During MROs 
%the complete instrument memory is
%readout and  the information of all 40 impacts is transmitted to Earth.
%If too many impacts occur between two MROs the detailed information of the
%older particles is lost.

In RTS and record mode only seven instrument data frames were read out at a 
time and transmitted to Earth rather than the complete instrument memory.  Six 
of the frames contained the information of the six most recent events in 
each amplitude range.
The seventh frame belonged
to an older event read out from the instrument memory (FN=7) and was transmitted in
addition to the six new events. The position in the instrument memory
from which this seventh frame was read changed for each readout so that after
40 readouts the complete instrument memory was transmitted (note that the 
contents of the memory may have changed significantly during the time period of 40 
readouts if high event rates occurred). 

RTS data were usually obtained when Galileo was in the inner jovian system
where relatively high dust impact rates occurred. During time intervals when
Galileo was in the outer jovian magnetosphere dust data were usually received 
as instrument memory-readouts (MROs).  MROs returned event data which had accumulated in the 
instrument memory over time. The contents of all 46 instrument data 
frames of the dust instrument was read out during an MRO and transmitted to Earth. 
%The complete data set of only 40 different events (dust impacts
%or noise events) is contained in an MRO because the information of the 
%latest event in each 
%of the six amplitude ranges (Paper~I) is stored twice (the so-called
%A range). 
If too many events occurred between two MROs, the data sets of the oldest 
events became overwritten in the memory and were lost. Although
the entire memory was read out during an MRO,
the number of data sets of new events that could be 
transmitted to Earth in a given time period was much smaller than with RTS
data because MROs occurred much less frequently (Table~\ref{tab_data_modes}).
During times when only MROs occurred, the accuracy of the impact time was defined
by the increment of the instrument's internal clock, i.e. 4.3~hours. 

In 2000-2003, RTS and record data were obtained during a period of 570 days 
(Figure~\ref{trajectory}) which amounts to about 40\% of the total almost 4-year
period.
% (note that during the entire Galileo Jupiter mission from 1996 to 2003
%this number is 40\%). 
During the remaining times when the dust instrument was  operated in neither RTS nor
record mode, a total of 59 MROs occurred at approximately 2 to 3 week intervals. 
Until the end of 2002, MROs were frequent enough so that usually no ambiguities 
in the time-tagging occurred (i.e. MROs occurred at intervals smaller than
46 days). 

The last MRO for the entire Galileo mission occurred at the end of 2002 on day 02-363. 
In 2003 we received dust data neither as MROs nor as record data. Only RTS data were
received during rather short time intervals: 
about one week from 03-063 to 03-070 and a total of about two days between 03-255 and 
03-264 before the spacecraft hit
Jupiter (Table~\ref{event_table_1}). No dust data were obtained outside these 
intervals in 2003.

Several resets of the dust instrument's internal clock occurred during the long periods
without data transmission in 2003, leading to ambiguities in the impact
time of some dust impacts. One clock reset occurred during the first data gap between 
02-363 and 03-063 and four resets in the second gap between 03-070 and 03-255. Furthermore,
due to data transmission problems, the time tagging was lost for the events
transmitted in the interval 03-063 to 03-070. Consequently, the impact time of two events 
which occurred between 02-363 and 03-063 is completely unknown. We have therefore 
set their impact time to 03-030 (these grains are indicated by horizontal 
bars in Figure~\ref{rot_angle}). For seven data sets transmitted between 03-063 and 03-070
the impact time could be determined with an accuracy of approximately one day 
from the time tagging of test pulses that were routinely performed by the dust instrument
(see also Section~\ref{sec_j35}).

\subsection{Dust instrument operation}

During Galileo's earlier orbital mission about Jupiter strong 
channeltron noise was usually recorded while Galileo was 
within about 20$\rm R_J$ distance from Jupiter 
(Jupiter radius, $\rm R_J = 71,492\,km$). The details are 
described in Papers~VI and VIII and not repeated here. Furthermore,
due to degradation of the channeltron, the high
voltage setting (HV) had to be raised two times in 1999 (Paper~VIII).
At the beginning of the year 2000, i.e. at the beginning of the time 
period considered in this paper, the dust instrument was 
operated in the following nominal configuration:
the channeltron high voltage was set to 1250~V (HV~=~4), 
the event definition status was set such that only 
the ion-collector channel could initiate a measurement cycle 
(EVD~=~I) and the detection thresholds for the charges on the 
ion-collector, channeltron, 
electron-channel and entrance grid were set (SSEN~=~0,~1,~1,~1). 
This configuration effectively prevented dead time of the instrument
due to channeltron noise (serious channeltron noise rates with $\mathrm{CN\,>\,10}$
occurred only during 
seven short time intervals in orbit A34 on day 02-309 when Galileo was 
inside Io's orbit and lasted only between several seconds and less than a minute. 
The resulting dead time is negligible because of its random occurrence and 
short duration). Due to degradation of the
channeltron (Section~\ref{el_deg}) the channeltron high
voltage was raised two additional times on days 00-309 and 01-352 in order 
to maintain a rather constant instrument sensitivity for dust 
impacts (Table~\ref{event_table_1}).

During the Jupiter orbital tour of Galileo, orbit trim maneuvers 
(OTMs) were 
executed around perijove and apojove passages to target
the spacecraft to close encounters with the Galilean 
moons. Many of these maneuvers required changes in the 
spacecraft attitude off the nominal Earth pointing direction
(Figure~\ref{pointing}).  Additionally, dedicated spacecraft turns occurred 
typically in the inner jovian system within a few days around  
perijove passage to allow for imaging observations with Galileo's cameras or
to maintain the nominal Earth pointing direction. 
%Specifically large turns 
%happened on 4 September 1997 (97-247), 10 September 1997 (97-253), and
%7 November 1997 (97-311). During
%these turns the spacecraft spin axis was oriented $\rm 64^{\circ}$,
%$\rm 51^{\circ}$ and $\rm 40^{\circ}$ away 
%from the Earth direction, respectively. All three attitude changes were 
%large enough that DDS could record impacts of dust stream particles at times
%when these grains would have been undetectable with the 
%nominal spacecraft orientation (Figures~\ref{rate} and \ref{rate_highres}). 

In the time interval considered in this paper a total of five spacecraft 
anomalies (safings) occurred 
on days 00-055, 02-017, 02-047, 02-274, and 02-309 (Table~\ref{event_table_1}). 
Three of these anomalies 
occurred in the inner jovian system in the region where the 
highest radiation levels were collected by the spacecraft, and 
recovery usually took several days. Although the dust instrument continued to measure 
dust impacts, the collected data could not be transmitted to Earth 
during the recovery and most of them were lost. 

No reprogramming of the instrument's onboard computer was necessary in the
2000-2003 time interval. In fact, the last reprogramming for the entire
Galileo mission took place on 4 December 1996 when two overflow counters 
were added for the so-called AR1 impacts in classes~2 and 3 (Paper~VI). With 
these overflow counters, all
accumulator overflows could be recognized in these two channels in the 2000-2003 
interval. It is very unlikely that unrecognized overflows occurred in the
higher amplitude ranges. The only exception is day 02-309 when Galileo was 
in the gossamer ring region and the instrument continued to collect data 
after the spacecraft anomaly (see also Section~\ref{sec_gossamer}). Here unrecognized 
overflows have likely occurred 
in amplitude range AR1, class~1 (channel AC11) and amplitude range AR2 (except
channel AC32), while the higher
amplitude ranges AR3 and AR4 were most likely free of overflows. See Section~\ref{class_and_noise}
for a description of the amplitude ranges and quality classes of dust impacts.

\subsection{Dust instrument electronics degradation}   \label{el_deg}

Analysis of the impact charges and rise times measured by the dust instrument
revealed strong degradation of the instrument electronics 
which was most likely caused by the harsh 
radiation environment in the inner jovian magnetosphere.
A detailed analysis was published by 
\citet{krueger2005a}. Here we recall the most 
significant results: 
a) the sensitivity of the instrument for dust impacts and 
noise had dropped. 
b) the amplification of the charge amplifiers had 
degraded, leading to reduced impact charge values $\QI$ and $\QE$. 
c) drifts in the target and ion collector rise time signals lead
to prolonged rise times $\rm t_I$ and $\rm t_E$. 
d) degradation of the channeltron required increases in the channeltron
high voltage (Table~\ref{event_table_1}).
In particular, 
a) requires a time-dependent correction when comparing 
dust fluxes early in the Galileo Jupiter mission with later 
measurements; b) and c) affect the mass and speed calibration of
the dust instrument. After 2000, masses and speeds derived from the instrument 
calibration have to be taken with caution because the electronics
degradation was very severe. Only in cases where impact 
speeds are known from other arguments can corrected masses 
of particles be derived (e.g. the dust cloud measurements in the 
vicinity of the Galilean moons or Galileo's gossamer ring passages).
On the other hand, given the uncertainty in the impact calibration 
of a factor of two in the
speed and that of a factor of ten in the mass, the increased 
uncertainty due to the electronics degradation was comparatively small
before 2000 (it should be noted that the dust data until
end 1999 published earlier -- Papers~II, IV, VI and VIII -- remain unchanged).
In particular, no corrections for dust fluxes, 
grain speeds and masses are necessary until end 1999 and results 
obtained with this data set in earlier publications remain valid.
Beginning in 2000, however, the degradation became so severe that the calibrated 
speeds and masses have to be considered as lower and upper limits,
respectively (see also Section~\ref{sec_tables}). 

\section{Impact events} \label{events}     

\subsection{Event classification and noise}

\label{class_and_noise}

The dust instrument classified all events -- real dust impacts and noise 
events -- into one of 24 different categories (6 amplitude ranges for
the charge measured on the ion collector grid 
and 4 event classes) and counted them in 24 corresponding 8 bit accumulators 
(Paper~I).  In interplanetary space most of the 24 categories were 
relatively free from noise and only sensitive to real dust impacts.
The details of the noise behaviour in interplanetary space can be
found in Papers~II and IV.
%except for extreme situations like the crossings of the radiation 
%belts of Earth, Venus (Paper~II) and Jupiter (Paper~IV). During most 
%of Galileo's initial three years of interplanetary cruise only the lowest 
%amplitude and class categories -- AC01 (event class 0, amplitude 
%range 1, AR1), AC11, and AC02 -- were contaminated by noise events 
%(Paper~II). In July 1994 the onboard classification scheme 
%of DDS was changed to identify smaller dust impacts in the data. 
%With the modified scheme noise events 
%were usually restricted to class 0 but may have occurred in all amplitude ranges. 
%All events in higher quality 
%classes detected in the low radiation environment of interplanetary 
%space were true dust impacts 
%(class~0 may still contain unrecognized dust impacts). 

In the extreme radiation environment of the jovian system, a 
different noise response of the instrument was recognized: 
especially within about 20 $\rm R_J$ from Jupiter 
class~1 and class~2 were contaminated with noise while class~3 was almost
always noise-free \citep{krueger1999c}. 
%However, this noise was different from 
%the channeltron noise recorded in the G1 orbit (Paper~VI).
Analysis of the dust data set from Galileo's entire Jupiter mission
showed that noise events could reliably be eliminated from class~2 
\citep{krueger2005a}
while most class~1 events detected in the jovian environment showed signatures of
being noise events. For most of 
Galileo's Jupiter mission we therefore consider
the class~3 and the noise-removed class~2 
impacts as the complete set of dust data.
Apart from a missing third charge signal -- class~3 has three charge 
signals and class~2 only two -- there is no physical difference between
dust impacts categorized into class~2 or class~3. In particular, 
we usually classify all class 1 and class 0 events detected in the jovian 
environment as noise.

The only exceptions are the passages through Jupiter's gossamer rings in 2002 and 2003
where a somewhat different noise response of the instrument was recognized \citep{moissl2005}. 
Here, good dust impacts could also be identified in class~1. In Table~\ref{tab_gossamer_noise} 
we show the noise identification scheme applied to the data from the gossamer ring passages 
obtained while Galileo was within Io's orbit. 

\begin{center} \fbox{\bf Insert Table~\ref{tab_gossamer_noise}} \end{center}

To summarise, noise was removed from the data set we present here with 
two different criteria: data obtained outside Io's orbit were processed 
according to the criteria derived by \citet{krueger2005a}, while data 
obtained inside Io's orbit were noise-removed with the criteria 
of  
\citet{moissl2005} (Table~\ref{tab_gossamer_noise}). 
Degradation of the instrument electronics was taken into account 
beginning in 1997 (Paper~VIII). 
%In particular, the data set of 1996 published in Paper~VI
%is not affected by electronics degradation and 
%remains unchanged. 
The derivation of the noise contamination factor 
$f_{\rm noi}$ for class~2 was described in Paper~VI
and is not repeated here. 

In this paper the terms ''small`` and ''big`` have the
same meaning as in Papers~IV, VI and VIII (which is different from the terminology 
of Paper~II). Here, we call all particles  in the 
amplitude ranges 2 and higher (AR2-6) ''big``. Particles in
the lowest amplitude range (AR1) are called ''small``. This distinction
separates the small jovian dust stream particles from bigger grains 
which are mostly detected between the Galilean moons (see also 
Section~\ref{sec_rate}).

Table~\ref{rate_table} lists the number of all dust impacts and noise events identified 
with the dust instrument in the 2000-2003 interval as deduced from the 
accumulators of classes 2 and 3. Depending on the event rate the numbers
are given in intervals from half a day to a few weeks (the numbers with
the highest time resolution are available in electronic form
only and are provided to the data archiving centres). For impacts in these 
two classes in the lowest amplitude range AR1 the complete data sets for 
only 2\% of all detected events were transmitted, the 
remaining 98\% of events were only counted. About 32\% of all data sets for 
events in the higher amplitude ranges were transmitted. We give only the number 
of events in classes 2 and 3 because they have been shown to contain
real dust impacts during the entire Jupiter mission: class~3 is almost always 
noise free (although 
 \citet{krueger1999c} found indications for a very small number of noise 
events in class~3, AR1, in 
the inner jovian system). Class~2 is strongly contaminated by noise events 
in the inner jovian system (within about $\rm 15\,R_J$ from Jupiter). 

\begin{center} \fbox{\bf Insert Table~\ref{rate_table}} \end{center}

In the 2000-2003 interval Galileo had a total of eight targeted flybys at 
the Galilean moons plus one at Amalthea (Table~\ref{event_table_1}).
During the flybys at the Galilean moons no ejecta particles from 
the moons could be
detected because of unfavourable detection geometry. During the Amalthea
flyby in A34, however, the dust instrument had the right detection geometry. 
Taking the recently determined mass of Amalthea \citep{anderson2005},
its Hill radius is $\mathrm{r_{Hill} \sim 130\,km}$, only slightly
larger than the moon itself. Galileo's closest approach distance was 244\,km 
from the moon's centre so that the spacecraft did not cross the Hill 
sphere  where an increased dust density was expected.
In fact, no increase in the
dust impact rate could be identified, consistent with our expectations
\citep{krueger2009b}.

\subsection{Dust impact rates}

\label{sec_rate}

Figure~\ref{rate} shows the 
dust impact rate recorded by the dust instrument in 2000-2003 as deduced from 
the class~2 and 3 accumulators. The 
impact rate measured in the lowest amplitude range (AR1) and the 
one measured in the higher amplitude ranges (AR2-6) are 
shown separately because they reflect two distinct populations 
of dust. Until early 2002 AR1 contains mostly stream particles which were 
measured throughout the jovian system. 
Bigger particles (AR2-6) 
were mostly detected in the region between the Galilean
moons. 

Between the perijove passages I33 and A34 in 2002 a low background 
rate of a few times $\mathrm{10^{-4}\,min^{-1}}$ was 
measured in AR1 which is at least an order of magnitude higher than dust
impact rates measured with Galileo and Ulysses in interplanetary space \citep{gruen1997a}.
These impacts show a broad distribution over all rotation angles 
(Figure~\ref{rot_angle}) while stream particles were expected to
approach from rotation angles around $90^{\circ}$ most of the time in 2002,
similar to the earlier Galileo orbits in 2000 and 2001. These 
grains could be stream particles approaching from a much broader range of 
directions as was reported from the dust measurements with Cassini
during Jupiter flyby (Sascha Kempf, personal communication). 

During the gossamer ring passages impacts were measured in all 
amplitude ranges AR1-4 (Section~\ref{sec_gossamer}). 
Note that the impact rate in AR1 was usually at least one to two orders of
magnitude higher than that for the big particles.
Diagrams showing the AR1 impact rate 
with a much higher time resolution in the inner jovian system are given in 
Figure~\ref{rate_highres}, and Galileo's gossamer ring passages are 
discussed in detail by \citet{krueger2009b}.

\begin{center} \fbox{\bf Insert Figure~\ref{rate}} \end{center}
\begin{center} \fbox{\bf Insert Figure~\ref{rate_highres}} \end{center}

In the inner jovian system the impact rates of AR1 particles 
frequently exceeded $\rm 10\,min^{-1}$. An exceptionally large
dust impact rate was recorded during
the orbit G28 in the outer jovian system when Galileo was approximately 
$\mathrm{280\,R_J}$ away
from Jupiter (Section~\ref{sec_g28} and Figure~\ref{rate_g28}).
This represents one of the highest dust ejection
rates of Io recorded during the entire Galileo Jupiter mission 
and is likely connected with a single strong volcanic eruption on Io 
\citep{krueger2003d,geissler2004}.

\subsection{Event tables}

\label{sec_tables}

Table~\ref{dust_impacts} lists the data sets for all 224 big particles detected 
%in classes~2 and 3 
between 
1 January 2000 and 21 September 2003 for which the complete information exists.
Class~1 and class~2 particles were separated from noise by applying the criteria developed 
by \citet{krueger1999c,krueger2005a} and \citet{moissl2005} 
(Section~\ref{class_and_noise}).
We do not list the small stream particles (AR1) in Table~\ref{dust_impacts} because their 
masses and velocities are outside the calibrated range of the dust instrument and they are 
by far too numerous to be listed here. The complete information of a total 
of 5165 small (AR1) 
dust particles was transmitted in 2000-2003. These are mostly stream particles which are 
believed to be about 10~nm in size and their velocities exceed 
200\,\kms\ \citep{zook1996}. Any masses and velocities derived 
for these particles with existing calibration algorithms would 
be unreliable. The full data set for all 5389 particles is submitted 
to the data archiving centres and is available in electronic form. 
A total number of 7566 events (dust plus noise in all amplitude ranges and classes) 
were transmitted in 2000-2003, each with a complete data set.

\begin{center} \fbox{\bf Insert Table~\ref{dust_impacts}} \end{center}

In Table~\ref{dust_impacts} dust particles are identified by their sequence number 
and their impact time. Gaps in the sequence number are due to the 
omission of the small particles. The time error value (TEV) which 
was introduced for the data set from the Jupiter mission because of the large 
differences in the timing accuracy of the dust instrumnet in the various data readout modes is
listed next (see Table~\ref{tab_data_modes} and Paper~VI for details). 
Then the event category -- class (CLN) 
and amplitude range (AR) -- are given. Raw data as transmitted to Earth 
are displayed in the next columns: sector value (SEC) which is the
spacecraft spin orientation at the time of impact, 
impact charge numbers (IA, EA, CA) and rise times (IT, ET), time
difference and coincidence of electron and ion signals (EIT, EIC),
coincidence of ion and channeltron signal (IIC), charge reading at
the entrance grid (PA) and time (PET) between this signal and
the impact. Then the instrument configuration is given: event
definition (EVD), charge sensing thresholds (ICP, ECP, CCP, PCP) and
channeltron high voltage step (HV). See Paper~I for further
explanation of the instrument parameters, except TEV which was introduced
in Paper~VI.

The next four columns in Table~\ref{dust_impacts} give information about Galileo's orbit: 
ecliptic longitude and latitude (LON, LAT) and distance from Jupiter ($\rm D_{Jup}$, 
in $\rm R_J$). The next column gives the 
rotation angle (ROT) as described in Section~\ref{mission}. 
Whenever this value is unknown, ROT is arbitrarily set to
999. This occurs 71 times in the full data set that includes 
the small particles. Then follows the pointing direction of the instrument at 
the time of particle impact in ecliptic longitude and latitude 
($\rm S_{LON}$, $\rm S_{LAT}$).
When ROT is not valid, $\rm S_{LON}$ and $\rm S_{LAT}$ are also useless
and set to 999. 
Mean
impact velocity ($v$) and velocity error factor (VEF, 
i.e. multiply or divide stated velocity by VEF to obtain upper or lower 
limits) as well as mean particle mass ($m$) and mass error factor (MEF) are 
given in the last columns. 
For VEF $> 6$, both velocity and mass estimates are invalid and should be discarded. 

Beginning in 2000 the degradation of the dust instrument electronics 
became very severe, leading to artificially too long rise times and reduced charge 
amplitudes.
The calibrated mass and speed values for VEF $< 6$ listed in Table~\ref{dust_impacts} should thus
be considered as lower limits for the impact velocity and upper limits for the 
particle mass throughout the 2000-2003 interval.

%This occurs for 924 impacts. 

No intrinsic dust charge values are given \citep{svestka1996}. 
Even though the charge carried by the dust grains is expected to be 
larger in the jovian magnetosphere than in interplanetary space the
charge measured on the entrance grid of the dust instrument did not give
any convincing results yet. 
Reliable charge measurements for interplanetary 
dust grains and for dust in Saturn's E ring were recently reported for the 
Cassini dust detector \citep{kempf2004,kempf2006a}. 
These measurements may lead to an improved unterstanding of the charge 
measurements of Ulysses and Galileo in the future. 

Entries for the parameter PA in Table~\ref{dust_impacts} sometimes have values between 49 and 63 
although the highest possible value allowed by the instrument electronics 
is 48 (Paper~I). This is also 
inherent in all Galileo and Ulysses data sets published earlier (Papers~II to IX) 
and it is due to a bit flip. According to our present understanding the correct PA 
values are obtained by subtracting 32 from all entries which have values between 
49 and 63. Values of 48 and lower should remain unchanged. 

\section{Analysis} \label{analysis}

The positive charge measured on the ion collector, $\QI$, is 
the most important impact parameter determined by the dust instrument because it is 
rather insensitive to noise. Figure~\ref{nqi} shows the distribution of 
$\QI$ for the full 2000-2003 data set (small and big particles together).
Ion impact charges were only detected over four orders of magnitude instead 
of the entire range of six 
orders of magnitude the instrument could measure. Note that the saturation limit  
of the instrument was at about $\sim 10^{-8}\,\mathrm{C}$ but the maximum measured 
charge was $ \QI = 9.7 \times 10^{-11}\,\mathrm{C}$, 
well below the saturation limit. This is most likely due to
instrument degradation \citep[Section~\ref{el_deg} and][]{krueger2005a}.

The impact charge 
distribution of the big particles ($\QI > 10^{-13}\,\mathrm{C}$) follows a 
power law with index $-0.15$ and is shown as a dashed line in Figure~\ref{nqi}
(if we exclude the particles from the region inside Io's orbit the slope is reduced 
somewhat to $-0.04$). This slope
is flatter than the values of approximately $-1/3$ derived for the jovian system 
from the 1996-1999 Galileo data set (Papers~VI and VIII). Whether this 
flattening is due to changes in the particle properties or due to electronics 
degradation remains unclear. 
%and also close to the 
%Galileo value of $-1/3$ given in Paper~II for the inner solar system. 
%Values 
%derived for the outer solar system are somewhat steeper: $-1/2$ 
%(Ulysses, Paper~III) and $-0.43$ (Galileo, Paper~IV), respectively. 
%Since the impact charge depends on the impact speed and the mass
%of the grains ($\QI \propto m\,v^{\alpha}$, $\alpha \simeq 3.5$), 
%the slopes indicate that, on average, bigger and/or faster particles were
%detected in the inner solar system and in the jovian environment 
%than in the interplanetary space of the outer solar system. It is
%consistent with a smaller relative contribution of interstellar grains 
%in the inner solar system and in the jovian system which 
%preferentially occur in the intermediate impact charge range 
%$ 10^{-13}\,{\rm C} < \QI < 10^{-11}\,{\rm C}$ (AR2-3). 
Note that the jovian 
stream particles (AR1) were excluded from the power law fit. 

\begin{center} \fbox{\bf Insert Figure~\ref{nqi}} \end{center}

In Figure~\ref{nqi} the small stream particles ($\QI < 10^{-13}\,\rm C$) 
are squeezed into the two leftmost histogram bins. In order to investigate 
their behaviour in more
detail we show their number per individual digital step separately in 
Figure~\ref{nqi2}. The distribution flattens 
for impact charges below $\rm 2\times 10^{-14}\,C$. Such a flattening
was also evident in the earlier data sets (Papers~II, IV, VI and VIII),
indicating the sensitivity threshold of the dust instrument may not be sharp.
% and it is consistent with the number of impacts with the lowest impact charges 
%$\QI$ not being complete due to the low data transmission capability of Galileo. 
The impact charge distribution for small particles 
with $\QI > 2\times 10^{-14}\,\rm C$ follows a power law with index 
$-4.7$. It is very close to the slope found from the
1996 Galileo data set ($-4.5$, Paper~VI) and somewhat steeper than the value 
measured in 1997-1999 ($-3.6$, Paper~VIII). The charge distibution 
strongly increases towards smaller impact charges.
%It nevertheless indicates that 
%the size distribution of the stream particles rises strongly towards 
%smaller particles. 
Note that the distribution of the stream particles is much 
steeper than that of the big particles shown in Figure~\ref{nqi}. 
Interestingly, if we restrict the time interval to the period between 
00-220 and 00-250 
when Galileo was outside the jovian magnetosphere in orbit G28 the 
stream particles show a somewhat steeper slope of $-5.9$ (not shown here).

\begin{center} \fbox{\bf Insert Figure~\ref{nqi2}} \end{center}

The ratio of the channeltron charge $\QC$ and the ion collector
charge $\QI$ is a measure of the channeltron amplification $A$ which
is an important parameter for dust impact identification (Paper~I).
The in-flight channeltron amplification was monitored 
in Papers~II, IV, VI and VIII for the initial ten years of the 
Galileo mission to identify possible degrading of the 
channeltron. In the earlier mission the amplification $ A = \QC/\QI$ 
for a channeltron high voltage setting of 1020~V (HV~=~2) 
determined from impacts with $
10^{-12}{\rm\, C} \le \QI \le 10^{-10}{\rm\, C}$ was in 
the range $1.4 \lesssim A \lesssim 1.8$. No
significant channeltron degradation was evident until the end of 1996. In
the 1997-1999 interval (Paper~VIII) a value of $ A \simeq 0.7$ was found which
indicated serious channeltron degradation. 
As a consequence, the channeltron high voltage 
was raised two times (on days 99-305 and 99-345) to 
return to the original amplification factor. 

Here we repeat the same analysis for the 2000-2003 interval.
Figure~\ref{qiqc}
shows the charge ratio $ \QC/\QI$ as a function of 
$\QI$ for a constant high voltage, HV,  as in the previous 
papers. Here we show data for HV~=~6. The charge 
ratio $\QC/\QI$ determined for $10^{-12}{\rm\,C} \le \QI 
\le 10^{-10}{\rm\,C}$ is $ A \simeq 1.6$ and is obtained from 65 impacts. The 
data for HV~=~4 and HV~=~5 (time intervals 00-001 to 00-209 and 00-209 to 01-352)
give $ A \simeq 1.3$ and $ A \simeq 0.5$, respectively. These values, however, are 
derived from only 9 and 15 impacts, respectively, and therefore have 
a much lower statistical significance.
The amplification for HV~=~6 is close to the value 
from the interplanetary cruise and the early Jupiter mission,
showing that the original channeltron amplification 
could be roughly reestablished. Details of the
dust instrument degradation due to the harsh radiation environment
in the jovian magnetosphere are described by 
\citet[][see also Section~\ref{el_deg}]{krueger2005a}.
It should be noted that the ratio $\QC/\QI$ is entirely determined by the
instrument performance. It does 
not depend upon the properties of the detected particles.

\begin{center} \fbox{\bf Insert Figure~\ref{qiqc}} \end{center}

Figure~\ref{mass_speed} displays the calibrated masses and 
velocities of all 5389 dust grains detected in the 2000-2003
interval. Although the range of impact velocities calibrated 
in the laboratory extended
from 2 to 70\,\kms, the measured impact speeds ranged only
up to about 20\,\kms. This is caused by the degradation of the
dust instrument electronics which lead to extended rise time 
measurements and, hence, impact velocities which are 
artificially too low, and calibrated grain masses 
artificially too large. This becomes apparent when comparing
Figure~\ref{mass_speed} with the corresponding figures in the
earlier Papers II, IV, VI and VIII where the measured range
of impact speeds extends up to 70\,\kms. {\em Therefore, due to the strong 
electronics degradation, all calibrated impact speeds and
masses in the time interval considered in this paper should
be considered as lower and upper limits, respectively.}
Any clustering of the velocity values is due to discrete steps in 
the rise time measurement but this quantization is much smaller than the
velocity uncertainty. For further details of the mass and 
velocity calibration the reader is referred to the description
of the mass-velocity diagrams in our earlier papers.

\begin{center} \fbox{\bf Insert Figure~\ref{mass_speed}} \end{center}

%the masses vary over 8 orders of magnitude from $\rm 10^{-7}\,g$ to 
%$\rm 10^{-15}\,g$. The mean errors are a factor of 2 for the 
%velocity and a factor of 10 for the mass. Impact 
%velocities below about  3\,\kms\ should be treated with caution.
%Anomalous impacts onto the sensor grids or structures 
%other than the target generally lead to prolonged rise times of the 
%charge signals. This in turn results in artificially low impact 
%velocities and high dust particles masses. 

The impact direction of the dust particles detected in the 2000-2003 interval
is shown in Figures~\ref{rot_angle} and \ref{rot_angle_highres}. On the
inbound trajectory, when Galileo approached Jupiter, the dust stream
particles (AR1) were mainly detected from rotation angles 
$\mathrm{270\pm 70^{\circ}}$ while on the outbound trajectory the 
streams were detectable from $\mathrm{90\pm 70^{\circ}}$. Before 2000
the detection geometry of the streams was such that the grains could only
be detected during a very limited period of time around perijove passage
(Paper~VIII, Table~4 therein). This changed in 2000 when the streams became detectable from 
rotation angles $\mathrm{90\pm 70^{\circ}}$ during almost the entire orbit
of Galileo. This is best seen in orbits G28 to C30 in 2000 and 2001. 
Big particles were, as in the earlier periods, mostly detected in the
inner jovian system when Galileo was close to Jupiter with the
exception of several impacts recorded in March 2003 at about $\mathrm{350\,R_J}$
from Jupiter (Section~\ref{sec_j35}). Note that an error occurred in
our earlier rotation angle plots in Paper~VIII (Figure~9 in that paper). 
The corrected figure is shown in the Appendix.

\begin{center} \fbox{\bf Insert Figure~\ref{rot_angle}} \end{center}

\begin{center} \fbox{\bf Insert Figure~\ref{rot_angle_highres}} \end{center}

%The mass range populated by the particles is very similar to
%that reported for the 1996 measurements from the jovian system (Paper~VI).
%However the largest and smallest masses are at the edges of the 
%calibrated velocity range of DDS and, hence, they 
%are the most uncertain. 
%Any clustering of the velocity values is due to discrete steps in 
%the rise time measurement but this quantization is much smaller than the
%velocity uncertainty. In addition, masses and velocities in the lowest 
%amplitude range (AR1, particles indicated by plus signs) should be 
%treated with caution. These are mostly jovian stream 
%particles (Section~\ref{jov_streams}) for which we have clear indications 
%that their masses and velocities are outside the calibrated range of DDS 
%\citep[][J. C. Liou, personal communication]{zook1996}.
%The grains are probably much faster and smaller 
%than implied by Figure~\ref{mass_speed}. On the other hand, the 
%analysis of the dust clouds surrounding the Galilean moons has shown
%that the mass and velocity calibration is valid for the bigger particles
%\citep{krueger2000a,krueger2003b}. 
%For many particles in the lowest two amplitude ranges (AR1 and 
%AR2) the velocity had to be computed solely from the ion charge signal 
%which leads to the striping in the lower mass range in 
%Figure~\ref{mass_speed} (most 
%prominent above 10\,\kms). In the 
%higher amplitude ranges the velocity could normally be calculated 
%as the geometric mean from both the target and the ion charge 
%signals which leads to  
%a more continuous distribution in the mass-velocity plane. 

\section{Discussion} \label{discussion}

The dust data set from Galileo's entire Jupiter mission is 
a unique set of dust measurements from the jovian system for many years to 
come. Various jovian dust populations were investigated 
during the last 15 years which  we have
summarised in Section~\ref{introduction}. The present paper finalises our
series of Galileo dust data papers and we discuss 
some particular aspects of the 2000-2003 data set. 

%In this Section
%we analyse the variability of the impact rate of jovian dust streams particles 
%on time scales of days to weeks. Here 
%the 2000-2003 data set gives particularly interesting results. 
%We discuss an unexpectedly large impact rate of dust stream particles detected 
%when Galileo was far away from Jupiter in August/September 2000 (G28 orbit) which
%was likely due to strong dust ejection from Io.
%We briefly describe joint Galileo-Cassini dust measurements from Cassini's
%flyby at Jupiter in 2000/2001 and an increased number of large dust impacts
%in March 2003 again far away from Jupiter. 
%Finally, we present Galileo's unique in-situ dust measurements obtained 
%during two passages through the planet's gossamer rings (a much more 
%detailed analysis of these measurements was recently 
%published by \citet{krueger2009b}). 

\subsection{Variability of Io's dust emission}

\label{sec_plume_monitoring}

%Aus Jupiter-Buch-Kapitel:
Imaging observations of Io with 
Voyager, Galileo, Cassini and New Horizons detected at least
17 volcanic centres with related plumes 
\citep{porco2003,mcewen2004,spencer2007,geissler2008}.
Most of the plumes were sensed through the scattering of sunlight by
dust particles entrained within the plumes, and
ring-shaped surface deposits on Io suggest that other 
plumes have been recently active as well.
%Aus Vulkan-Monitor-Paper:
The dust data from the entire Galileo Jupiter mission are a unique record of the dust 
ejected from Io. In particular, as the plumes are the most plausible sources 
of the grains \citep{graps2000a}, the dust measurements monitor plume activity \citep{krueger2003d}. 

The Galileo dust data show a large orbit-to-orbit
variation due to both systematic and stochastic changes. Systematic effects
include Io's orbital motion, changes in the geometry of Galileo's orbit and in the
magnetic field configuration due to the rotation of Jupiter. Stochastic variations 
include fluctuations of Io's volcanic activity, changes of the particle charging in the
Io torus, variations in grain release  
from the torus, and the deformation of the outer magnetosphere 
in response to the variable solar wind conditions. It should be emphasized that the
mechanisms acting on the grains in the Io torus and in particular the connected temporal
variability are presently 
not well understood. By combining the entire 
Galileo dust data set, the variability due to stochastic processes could be
removed and a strong flux variation with jovian local time showed up 
\citep{krueger2003a}, confirming earlier predictions \citep{horanyi1997}.

Dust emission rates of Io were derived 
by \citet{krueger2003d}. After removal of the systematic variations, the
total dust emission rate of Io turned out to be between $10^{-3}$ and
$\mathrm{10\,kg\,s^{-1}}$, with  typical values in the range 0.1 to $\mathrm{1\,kg\,s^{-1}}$.
Exceptionally high dust emission rates occurred during
orbits E4 (1996), C21 (1999), G28, and, to a lesser extent, also during
G29 and C30. Some of these peaks in the dust emission
could be related to specific plume sightings or other markers of
volcanic activity on Io: The Pele plume is one of the most powerful
plumes and the most steady high-temperature volcanic centre on Io. Surface 
changes at the Pele site were detected frequently, whereas detections of the
Pele plume are relatively rare. Two detections 
of the Pele plume are coincident
with our measurements of high dust fluxes in E4 and G29, while a low dust 
flux in E6 may be explained by the absence of the Pele plume \citep{mcewen1998,porco2003}. 
In August/September 2000 (orbit G28; Section~\ref{sec_g28}) when Galileo was far away 
from Jupiter, a large dust 
flux was observed which is likely connected with surface changes observed at the site of 
the Tvashtar plume \citep{krueger2003d}. 

Here we investigate the orbit-to-orbit variability of the dust emission pattern
on much shorter timescales of  days to weeks. As in earlier works \citep{krueger2003d}
we assume a particle radius $s\,=\,10\,\mathrm{nm}$, grain density 
$\rho\,=\,1.5\,\mathrm{g\,cm^{-3}}$, dust grain charging to +5V in the 
Io torus, and calculate the effective dust sensor area from the particle
dynamics based on the model of \citet{horanyi1997}. We divide
the measured dust impact rate by the effective sensor area to obtain the dust flux
$f$ ($\mathrm{m^{-2}\,s^{-1}}$) as a function of distance $d$ from Jupiter. 
If we assume that Io's dust emission, the dust charging, ejection 
conditions from the plasma torus and the grain speed remain constant over 
the time interval considered, 
we expect a ``dilution'' of the dust with $d^{-2}$. Dynamical modelling implies 
that -- after the grains are released from the Io torus -- 
the major acceleration occurs within approximately $\mathrm{10\,R_J}$ from Jupiter 
so that their  speed remains basically unchanged 
further away from the planet. Finally, the variation of the dust flux with jovian 
local time is usually below a factor of five \citep{krueger2003a} and thus of minor significance here. 
With all these assumptions, we expect a variation of the
dust flux with $d^{-2}$. It should be emphasized that here we use exactly the 
same assumptions for calculating dust emission rates as \citet{krueger2003d}.

\begin{center} \fbox{\bf Insert Figure~\ref{rate_g29}} \end{center}

In Table~\ref{tab_slopes} we list the slopes of power law fits $f \propto d^{\alpha}$ to the 
derived dust flux profiles. We only considered Galileo orbits 
where sufficiently long data sets for at least two days are available so that 
meaningful flux profiles could
be obtained. Large variations in the flux profiles are obvious from Table~\ref{tab_slopes}. 
Given the overall uncertainties we believe that slopes in the range 
$-3 \lesssim \alpha \lesssim -1$ are still compatible with a rather constant dust ejection
rate from Io and the Io torus ($\alpha = -2$). In Figure~\ref{rate_g29} we show the dust flux 
during the G29 orbit
as an example. Here the power law fit to the data gives a slope $\alpha \approx -2$, indicating 
that the dust release from the Io torus stayed remarkably constant for a 
rather long period of more than two months.

Large deviations from this simple and ideal case with constant dust ejection 
are also obvious in the table. For example, orbits E4, E19, I32 and A34 show very flat profiles 
in the range $-1 \lesssim \alpha \lesssim 0$, implying that during these orbits stronger dust emissions 
occurred when Galileo was far away from Jupiter than when the spacecraft was closer to the planet. 
On the other
hand, during orbits G2, G8, E14, E16, E18 and E26 Galileo experienced a stronger
dust ejection  when the spacecraft was in the inner jovian system 
(power law slopes 
$-4 \lesssim \alpha \lesssim -7$). Note that the time coverage of these
data sets usually ranges from  days to a few weeks, indicating
that Io's plume activity or the dust charging and release from the Io torus, or both
frequently changed on such rather short timescales. 

\begin{center} \fbox{\bf Insert Table~\ref{tab_slopes}} \end{center}

Dust production rates of Io calculated with the method described above are also listed in
Table~\ref{tab_slopes}. It should be emphasized that within less than a week the dust 
release frequently changed by approximately a factor of 10, and
the absolute levels of the dust emission  may have been vastly different 
from one Galileo orbit to the next. For a detailed discussion of the total 
dust ejection rates from Io and correlations with individual plume sightings
the reader is referred to \citet{krueger2003d} who showed that all  intervals 
with elevated dust emission exceeding $\mathrm{\sim 1\,kg\,s^{-1}}$ (six intervals in total) 
can be connected with giant plume eruptions or large area surface changes on Io or both. 
See also Section~\ref{sec_g28}.

%In many cases, the eruptions of giant plumes are in agreement 
%with the time periods when our in-situ measurements showed episodes 
%of elevated dust emissions \citep{krueger2003d}.
%These coincidences
%indicate that the surface changes observed by \citet{geissler2003b} were 
%indeed due to very powerful plume eruptions, and
%the extent of the changes implies that these eruptions 
%must have been among the most powerful during the 
%Galileo mission. 
%The lack of plume sightings and/or recognition of large-area
%surface changes in some time intervals when strong dust emission was 
%observed 
%may point to volcanic eruptions on Io which were only 
%recognised because of increased  dust emission. Here, the dust
%measurements may eventually serve as a monitor of Io's volcanic
%activity.

\subsection{Io's dust emission in August/September 2000}

\label{sec_g28}

In summer 2000 (orbit G28) Galileo left the jovian magnetosphere for the
first time since it was injected into the jovian system in 1995 and reached 
a jovicentric distance of
$\mathrm{\sim 280\,R_J}$ (0.13~AU). In August/September 2000,
around Galileo's apojove, the dust instrument measured 
a surprisingly large dust impact rate exceeding $\mathrm{10\,min^{-1}}$ 
% and measured a very high dust flux of up to $\rm \sim 10\,m^{-2}\, s^{-1}$
for about two months (Figure~\ref{rate_g28}).
Similarly high fluxes
were also recorded with the Cassini dust instrument at 
$\mathrm{\sim 0.3\,AU}$ from Jupiter when the spacecraft was
approaching the planet in September 2000 (Sascha Kempf, personal communication). 
The dust emission from Io derived from the Galileo measurements 
by \citet{krueger2003d} in this time period 
exceeds $\sim 100\,\rm kg\, s^{-1}$. Later, when Galileo approached 
Jupiter again, the dust flux profile showed a surprisingly steep drop 
(slope $\alpha \approx 10$), implying a huge decrease in Io's dust 
emission.  

\begin{center} \fbox{\bf Insert Figure~\ref{rate_g28}} \end{center}

%Text z. T. aus Amara's paper:
Frequency analysis of the Galileo dust data from the first three years  
of the Galileo Jupiter mission (1996-1998) revealed strong 5 and 10 hour periodicities 
which were due to Jupiter's 
rotation \citep{graps2000a}. A weak ''Io footprint'' with approximately 42 hour 
frequency caused by this moon's orbital motion about Jupiter and harmonics with 
Jupiter's rotation frequencies were also revealed. These data were collected mostly 
in the inner jovian magnetosphere between 10 and $\mathrm{60\,R_J}$. In the data 
obtained during the 
later Galileo orbits in 1999 and 2000 the Io footprint became more prominent 
and was evident during most Galileo orbits from E19 to G29  \citep{graps2001a}. 

In the data from a total of 26 Galileo 
orbits measured between 1996 and 2000, a total 
of 11 orbits showed a clear modulation with Io's frequency, 3 showed a weak
Io modulation, while the remaining 12 orbits showed no Io signature at all \citep{graps2001a}.
In many, but not all, cases the missing Io signature coincided with time periods when 
a rather weak dust flux was measured. 
%Due to lower dust production in Io's plumes
%and/or less efficient dust charging and release from the Io torus, the torus may 
%more closely resemble a smeared-out ring source 
%for Io dust rather than a point source during these intervals. 

In the data set  
from August/September 2000, collected between days 00-220 and 00-250 at much 
larger jovicentric distances, Io's signature dominated all other
frequency signatures including the 5 and 10 hour periods caused by Jupiter's rotation  
\citep{graps2001b}. These data provide direct evidence for Io being the
source for the majority of the jovian dust stream particles during this time period. 
The presence of Io's orbital frequency implies that Io is a localised source of 
charged dust particles because charged dust from diffuse sources would couple to 
Jupiter's magnetic field and appear in frequency space with Jupiter's 
rotation frequency and its harmonics.

The period of strong dust emission seen in August/September 2000 coincided with 
enhanced neutral gas production from the Io torus, suggesting a coupling
mechanism between gas and dust ejection, although the relation between the
dust emissions and the production of neutral gas is not known \citep{delamere2004}. 
Furthermore, there was a significant reduction in the neutral source beginning in 
October 2000, again coinciding with the strong drop in the dust emission as
derived from our Galileo dust data.

\subsection{Galileo-Cassini joint dust stream measurements}

%Text aus Jupiter-Buch-Kapitel 2004:
On 30 December 2000 the Cassini spacecraft flew by Jupiter, 
providing a unique opportunity for a two-spacecraft time-of-flight measurement 
(Cassini-Galileo) of particles from one collimated stream from the jovian dust 
streams. The goal was to detect particles 
in a stream first with Galileo when the spacecraft 
was inside the jovian magnetosphere close to the orbit of 
Europa (about $\mathrm{12\,R_J}$), and particles in potentially the 
same stream later 
by Cassini outside the magnetosphere (at $\mathrm{140\,R_J}$)
\citep[see ][for a preliminary analysis]{graps2001b}.

The Cassini data from the Jupiter flyby imply that particles of different sizes have 
different phases with respect to Jupiter's rotation (Sascha Kempf, personal
communication), a result which is also seen in earlier 
Galileo data \citep{gruen1998}. Comparison of 
the measurements from both dust instruments, however, is hampered 
by the higher detection sensitivity of the 
Cassini detector with respect to the Galileo sensor. Both 
instruments have detected stream particles with different 
sizes and, hence likely different phases. The analysis is ongoing 
(Hsiang-Wen Hsu, personal communication),
and more detailed modelling to describe the phase relation 
of different-sized particles taking into account the 3-dimensional
structure of the dust emission pattern from the jovian
system is necessary. Our present preliminary analysis 
indicates particle speeds of about $\mathrm{400\,km\,s^{-1}}$. This value 
is in agreement with speeds for 10 nm particles as derived 
from dynamical modelling \citep{hamilton1993a,horanyi1993a}, 
and earlier studies of the jovian dust stream dynamics \citep{zook1996}. 

\subsection{Large dust grains far from Jupiter}

\label{sec_j35}

On 29 December 2002 (day 02-363) the last MRO of the dust instrument memory occurred 
for the remainder
of the Galileo mission. The next time we received dust data was during the time 
interval 4 to 11 March 2003 (days 03-063 to 03-070). These data were obtained as RTS
data. We identified a total number of nine large dust impacts in amplitude 
ranges AR2-4 which occurred  between 29 December 2002 and 11 March 
2003. Due to corruption of the readings from
the instrument's internal clock and one clock reset in this time interval, two of these 
impacts have an exceptionally large uncertainty in the impact time of  
66 days. We could reconstruct the impact time of the remaining seven
impacts with a higher accuracy from accumulator readings obtained with test pulses 
which were routinely performed by the dust instrument \citep[see][for more details]{krueger2005a}. 
This gave  impact times for five impacts with 
about one day uncertainty and for two impacts with 4.3 hour uncertainty (Table~\ref{dust_impacts}). 

The reconstruction of these partially corrupted data implies that at least 
seven impacts occurred during a period of only four days when Galileo was 
outside Jupiter's magnetosphere in interplanetary space at approximately 
$\mathrm{350\,R_J}$ from Jupiter. This is a surprisingly large number of 
impacts at such a large distance 
from Jupiter given the Galileo measurements from the earlier Jupiter mission 
(Papers~VI and VIII) and from Galileo's interplanetary cruise. Potential sources 
for these grains are, for example, collisional ejecta from an (unknown) small jovian satellite 
or a cometary trail crossed by the spacecraft. Judging from the
impact charge distribution of the measured grains, jovian stream particles (Figure~\ref{nqi2})  can
be most likely ruled out because a much larger number of impacts should
have occurred in the lower amplitude range AR1. In fact, only  few impacts were 
recognized in AR1  during this time. A more detailed analysis of these impacts 
has to be postponed to a future investigation. 

\subsection{Galileo's gossamer ring passages}

\label{sec_gossamer}

%Aus Habil:
On 5 November 2002 (orbit A34, day 02-309) Galileo traversed Jupiter's gossamer
rings for the first time and approached the planet to $\mathrm{2\,R_J}$. 
During this ring passage the spacecraft had a close flyby at Amalthea at 
244\,km distance from the moon's centre, well outside  Amaltheas's Hill 
sphere. During approach to Jupiter dust
data were collected with the highest possible rate (record mode;
Section~\ref{sec_transmission}) while
Galileo was within Io's orbit (i.e. within $\mathrm{\sim 5.9\,R_J}$).
Shortly 
%16\,min 
after Amalthea flyby a spacecraft anomaly at $\mathrm{2.33\,R_J}$ jovicentric 
distance prevented the collection of further Galileo dust data. 
Although the dust instrument continued to measure dust
impacts after the anomaly, the data were not written to
the tape recorder on board and, hence, the majority of them were lost. 
Only the data sets of a few dust impacts were received from an MRO
on day 02-322. These events could be located to have happened during the 
gossamer ring passage but their impact time is uncertain by a few hours 
(Table~\ref{dust_impacts}).
The traverse of the optically visible ring from its 
outer edge at 
$\mathrm{\sim 3.75\,R_J}$ until the spacecraft anomaly occurred 
lasted about 100\,min, and the total gossamer ring traverse from
$\mathrm{\sim 3.75\,R_J}$ inbound to $\mathrm{\sim 3.75\,R_J}$ outbound
took approximately six hours.

During the A34 ring passage the lowest amplitude range 
in class~2 (AC21) was strongly contaminated with noise, while 
the higher amplitude ranges showed little or 
no noise contamination. In addition,
many class~1 events recognised within Io's orbit 
showed signatures of being true dust impacts as well. The noise 
identification scheme applied to the dust data from both 
Galileo gossamer ring passages is described in 
Section~\ref{class_and_noise} and given in Table ~\ref{tab_gossamer_noise}.

With the new noise identification scheme, 
complete data sets of 90 dust impacts were identified in the Galileo
recorded data from the gossamer ring region.
Several hundred more events were counted only and their data sets were lost,
in particular in AR1.
The completeness of the transmitted ring data varied between 100\% in the highest
amplitude ranges (AR2-4) in the faint ring extension beyond Thebe's orbit
down to only 4\% for the lowest amplitude range (AR1) in the more populated
Amalthea ring. 

In record mode, the dust instrument memory was read out 
once per minute, and this readout frequency determined the spatial 
resolution of the measurements: within one minute Galileo moved 
about 1,800\,km through the ring which corresponds to
about 1,100\,km (or $\mathrm{0.015\,R_J}$) in radial direction. This is the highest 
spatial resolution achievable in the ring with the Galileo in-situ measurements.

Dust measurements in the gossamer rings were also obtained during Galileo's
second ring traverse on 21 September 2003 (orbit J35)
a few hours before Galileo impacted Jupiter.
The data sets of about 20 dust impacts were successfully transmitted to 
Earth as RTS data. This time the spatial resolution was only about 14,000\,km
(or $\mathrm{0.2\,R_J}$).

The data from both gossamer ring traverses allowed for the first actual comparison of 
in-situ measurements with the properties inferred from 
inverting optical 
images. A detailed analysis of this data was published by 
\citet{krueger2009b}. Below we summarise the most important results.

Images of the rings imply inclinations of the grain orbits of $i\approx 1^{\circ}$ 
for the visible 5 to $\mathrm{10\mu m}$ grains \citep{showalter2008}.
%(Figure~\ref{fig_orbitplot_2002}). 
The expected rotation angle for ring particles on
circular prograde uninclined jovicentric orbits was 
$ \simeq 90^{\circ}$.
%, and that for retrograde orbits was $ \simeq~270^{\circ}$.
The rotation angles measured within Io's orbit and in particular 
during the ring passages were 
%are shown in Figure~\ref{rot_angle_gossamer}.
-- to a first approximation -- consistent with
these expectations. However, the width of the rotation angle distribution
was much wider than the expected width for the geometry conditions during
both gossamer ring passages. 

%\begin{center} \fbox{\bf Insert Figure~\ref{rot_angle_gossamer}} \end{center}

%The measured impact directions also show a lot of additional structure.
%The approach direction of the dust particles with respect to the spacecraft spin 
%axis was such that shading by the magnetometer boom was expected \citep{krueger2009b}.
%Shading by the boom is indeed evident by a void in dust impacts with 
%rotation angles ranging from $80^{\circ}$ to $100^{\circ}$.
%Similar shading was also seen in 
%measurements of the jovian dust streams \citep{krueger1999c}.

%For the geometry conditions during the first gossamer ring passage the 
%expected width of the rotation angle 
%distribution was approximately $100^{\circ}$.
%Hence, the distribution of measured rotation angles should 
%cover the range from $ 40^{\circ}$ to $ 140^{\circ}$.
%It is also obvious from Figure~\ref{rot_angle_gossamer} 
%that the rotation angle distribution is significantly 
%broader than that:
%about half of the impacts were detected with rotation angles further away
%from Jupiter's equatorial plane (rotation angles 
%$ \gtrsim 140^{\circ}$ or $ \lesssim 40^{\circ}$). 
%In fact, the entire distribution is even broader than 
%$180^{\circ}$. 
%This means that particles seem to have approached from directions 
%more or less perpendicular to the jovian equatorial plane. 
%On the other hand, there were
%no impacts from the retrograde direction 
%($\Theta \simeq 270^{\circ}$).

What was the reason for such a broad distribution in impact directions?
One possibility was the sensor side wall which was very 
sensitive to dust impacts \citep{altobelli2004a,willis2005}. 
Taking the sensor side wall into account (Table~\ref{tab_fov}),
the expected width in rotation angle was still significantly smaller than the 
observed width. Another potential explanation was impacts onto 
nearby spacecraft structures
like the magnetometer boom, the EPD and PLS instruments which 
masquerade as particles 
with high inclinations. We are convinced that such an explanation 
can be ruled out  for two reasons \citep{moissl2005}: First, the
impact parameters (charge rise times, charge signal coincidences, etc.) of grains 
measured with rotation angles outside the nominal field-of-view for 
low-inclination particles do not show significant differences  compared to
gains inside the nominal field-of-view. Second, the data from both Galileo
ring traverses show similarly broad rotation angle patterns although they
had different detection geometries. During the first flyby the magnetometer 
boom obscured the field-of-view while during the
second flyby this was not the case \citep{krueger2009b}. 

The most likely explanation for the
observed structure in the rotation angle pattern is the particle dynamics: 
The wide range in impact directions as well as a drop measured in the 
impact rate profile immediately interior to Thebe's orbit and a gradual 
increase in the relative abundance of small particles closer to Jupiter
can best be explained by a shadow resonance caused by 
varying particle charge on the day and night side of Jupiter, driving particles onto high 
inclination orbits \citep{hamilton2008}. In fact, inclinations up to $20^{\circ}$ nicely 
explain the measured impact directions for most grains. 

Comparison of our in-situ measurements with 
imaging observations showed that the in-situ measurements preferentially probe
the large population of small sub-micron particles while the images are 
sensitive to larger grains with radii of at least several microns.
  The grains form a halo of material faint enough to be invisible to imaging, 
but populated enough to be detectable with the Galileo sensor. The faint gossamer ring 
extension previously imaged to about $\mathrm{3.75\,R_J}$ was detected out to at least 
$\mathrm{5\,R_J}$, indicating 
that ejecta from Thebe spread much further and particle orbits get higher eccentricities than 
previously known. Both the gap in the ring and the faint ring extension indicate that the 
grain dynamics is strongly influenced by electromagnetic forces. 
For a more detailed discussion of the ring particle dynamics the reader is referred 
to \citet{hamilton2008}.

%The distribution of impact directions shows even more
%structure:  The rotation angle pattern for each 
%individual class seems to differ (Figure~\ref{rot_angle_gossamer}). 
%Whether these are instrumental 
%effects or due to the relatively small number of impacts is presently not known. 

\section{Conclusions}

\label{sec_summary}

In this paper, which is the tenth  in a series of Galileo and Ulysses
dust data papers, we present data from the Galileo dust instrument for the 
period January 2000 to September 2003. In this time interval the
spacecraft completed nine revolutions about Jupiter in 
the jovicentric distance range between 2 and $\rm 370\,R_J$
(Jupiter radius, $\rm R_J = 71,492\,km$). On 21 September 2003 
Galileo was destroyed in a planned impact with Jupiter.

The data sets of a total of 5389 (or 2\% of the total) recorded 
dust impacts were transmitted to Earth in this period. Many more 
impacts (98\%) were 
counted with the accumulators of the instrument but their
complete information was lost because of the low data transmission
capability of the Galileo spacecraft. Together with 15861
impacts recorded in interplanetary space and in the Jupiter system
between Galileo's launch in October 1989 and December 1999 published 
earlier
\citep{gruen1995b,krueger1999a,krueger2001a,krueger2006a}, the complete data set 
of dust impacts measured with the dust detector during Galileo's 
entire mission contains 21250 impacts. 

Galileo has been an extremely successful dust detector, measuring 
dust streams flowing away from Jupiter, a tenuous dust ring throughout 
the jovian magnetosphere and Jupiter's gossamer rings 
over the almost four year timespan of data considered
in this paper.

Most of the time the jovian dust streams dominated the overall 
impact rate, reaching maxima of more than $\mathrm{10\,min^{-1}}$ 
in the inner jovian system.
A surprisingly large impact rate up to $\mathrm{100\,min^{-1}}$   was 
measured in August/September 2000 (G28 orbit) 
when the spacecraft was at about $\mathrm{280\,R_J}$ distance from Jupiter.
This strong
dust emission was most likely connected with a heavy volcanic
eruption on Io \citep{krueger2003d,geissler2003,geissler2004}. A strong 
variation in the release of neutral gas from the Io torus in this time 
interval was also reported by \citet{delamere2004}.

Io's dust emission as derived from the measured dust fluxes varied by many 
orders of magnitude, with typical values ranging between 0.01 to  
$\mathrm{1\,kg\,s^{-1}}$ of dust ejected. In August/September 2000
the derived dust emission exceeded 
$\mathrm{100\,kg\,s^{-1}}$. 
The investigation of the dust impact rate profiles measured for the 
jovian stream particles as a function of 
radial distance from Jupiter revealed
large orbit-to-orbit variations and variability by a factor of 10 or more on 
timescales of days to a few weeks. This implies strong variability of the dust
release from Io or the Io torus 
or variability of the jovian magnetosphere
on such short timescales. 

A surprisingly large number of impacts of bigger micron-sized dust grains 
was detected within a 4-day time interval far away from 
Jupiter in March 2003 when Galileo was in interplanetary space. The source
of these grains remains unclear.

Finally, in November 2002 and September 2003 Galileo traversed Jupiter's gossamer 
rings twice, providing the first actual opportunity to compare
in-situ dust measurements with the results obtained from remote imaging.
These flybys revealed previously unknown structures in the gossamer rings 
\citep{krueger2009b}: 
a drop in the dust density between the moons Amalthea and Thebe, grains orbiting
Jupiter on highly inclined orbits and an increase in the number of small 
grains in the inner regions of the rings as compared to the regions further
away from the planet. All these features can nicely be explained by electromagnetic
forces on the grains that shape the gossamer rings \citep{hamilton2008}.

Strong degradation of the dust instrument electronics was recognised in 
the Galileo dust data \citep{krueger2005a}.
It was most likely caused by the harsh
radiation environment in the jovian magnetosphere and lead to a degradation
of the  
instrument sensitivity for noise and dust detection during the Galileo 
mission. 
The Galileo data set 
obtained until the end of 1999 (Papers~VI and VIII) 
was not seriously affected by this degradation. 
%In particular, no correction for dust fluxes, grain speeds and masses 
%are necessary and results obtained with this data set in earlier
%publications remain valid. 
%Also, the Galileo dust data published in earlier papers in this
%series (Papers~II, IV and VI) remain unchanged.
%It should be noted, however, that dust fluxes
%calculated with a sensitive area taking into account only
%the sensor target overestimate the true dust fluxes and number densities 
%by about 20\% because the inner sensor side walls turned out to
%be as sensitive to dust impacts as the sensor target itself
%\citep{altobelli2004a,willis2005}.
In the time interval 2000 to 2003 which is the subject of this paper, however, the 
electronics degradation became so severe that the instrument
calibration does not give reliable impact speeds and masses of the dust particles
anymore. Instead, only lower limits for the impact speed and upper limits for the 
grain mass, respectively, can be given. The only exception are dust impacts for which
their impact speeds can be derived from other means \citep[e.g. impacts in the
gossamer rings;][]{krueger2009b}. 
On the other hand, a reduction of the channeltron amplification was
counterbalanced by four increases of the channeltron high voltage  
during the entire Jupiter mission (two in 1999, one each in 2000 and 2001)
to maintain stable instrument operation. 

Even though this is the final paper in our serious of Galileo dust data
papers published during the last 15 years, the evaluation of this unique data 
set is continuing. A list of specific open questions raised in this and earlier 
data papers includes: 
\begin{itemize}
\item {\em Electromagnetic interaction and phase relation of different sized stream 
particles:} Dust grains with different sizes  have a
different susceptibility to electromagnetic interaction with the jovian magnetosphere.
Different-sized grains released from a source in the inner jovian system at the same 
time are expected to arrive at Galileo at a different phase of Jupiter's rotation 
\citep{gruen1998}. This rather simple picture is further complicated by the grains' 
charging history. Studies of the phase relation may lead to better constraints
of the grain size distribution and may give new insights into the grains' electromagnetic
interaction. The phase relation may turn out to be essential to understand
the Galileo-Cassini joint dust streams measurements.
\item {\em Galileo-Cassini joint dust streams measurements:} Being originally designed
as a two-spacecraft time-of-flight measurement of one collimated stream 
from the jovian dust streams, the analysis of this data set 
turned out to be more complicated than anticipated. More detailed modelling of 
the 3-dimensional structure of the dust stream emission pattern from the jovian
system is necessary to describe the phase relation of different-sized particles
and to understand these unique measurements.  
\item {\em ''Big`` micron-sized particles:} Impacts of micron-sized dust grains
were preferentially detected in the inner jovian system between the Galilean
moons. Two sub-populations -- one orbiting Jupiter on prograde and one on 
retrograde orbits -- were identified in earlier analyses \citep{thiessenhusen2000}. 
The derived ratio in number density was approximately 
4:1 with the majority of grains being on prograde orbits. At the time, however, only
about half of the entire Galileo dust data set from Jupiter was available. Given 
that the detection geometry of the dust instrument changed with time during the mission, 
re-evaluation of the full data set from the entire Galileo Jupiter mission 
would be worthwhile to verify the abundance of grains on retrograde orbits. 
\item {\em Dust-plasma interaction:} Very preliminary comparison of the Galileo dust
measurements from the gossamer ring passages with energetic particle data from
the same period has revealed some interesting correlations between both data sets
(Norbert Krupp, personal communication). New insights into the dust-plasma 
interaction and particle dynamics can be expected from combined studies of the dust 
data and other Galileo particles and fields data.
\end{itemize}

\clearpage

\hspace{1cm}

{\bf Acknowledgements.}
We dedicate this work to the memory of Dietmar Linkert who passed away in 
spring 2009. He was Principal Engineer for space instruments at MPI f\"ur 
Kernphysik  including the
dust instruments flown on the HEOS-2, Helios, Galileo, Ulysses and Cassini missions. 
His friends and colleagues around the world appreciated his experience and 
sought his professional advice.
The authors wish to thank the Galileo project at NASA/JPL for effective 
and successful mission operations. This research was supported by 
the German Bundesministerium f\"ur Bildung und Forschung through Deutsches
Zentrum f\"ur Luft- und Raumfahrt e.V. (DLR, grant 50\,QJ\,9503\,3). 
Support by MPI f\"ur Kernphysik and MPI f\"ur Sonnensystemforschung 
is also gratefully acknowledged.

\section*{ERRATUM}

Due to an error in Paper~VIII, all panels of Figure~9 in that paper 
have  wrong labels on the vertical axis. Furthermore, the third panel (data of 1999) erroneously
shows the dataset of 1997. We apologize for this error and show the corrected plots in Figure~\ref{rot_old}.

\begin{center} \fbox{\bf Insert Figure~\ref{rot_old}} \end{center}

\clearpage

%\bibliography{/Users/krueger/hWork/latex/bib/pape,/Users/krueger/hWork/latex/bib/references}

\begin{thebibliography}{}

\bibitem[{Altobelli} et~al., 2004]{altobelli2004a}
{Altobelli}, N., {Moissl}, R., {Kr{\"u}ger}, H., {Landgraf}, M., and
  {Gr{\"u}n}, E. (2004).
\newblock {Influence of wall impacts on the Ulysses dust detector in modelling
  the interstellar dust flux}.
\newblock {\em Planetary and Space Science}, 52:1287--1295.

\bibitem[{Anderson} et~al., 2005]{anderson2005}
{Anderson}, J.~D., {Johnson}, T.~V., {Schubert}, G., {Asmar}, S., {Jacobson},
  R.~A., {Johnston}, D., {Lau}, E.~L., {Lewis}, G., {Moore}, W.~B., {Taylor},
  A., {Thomas}, P.~C., and {Weinwurm}, G. (2005).
\newblock {Amalthea's Density Is Less Than That of Water}.
\newblock {\em Science}, 308:1291--1293.

\bibitem[{Colwell} et~al., 1998]{colwell1998a}
{Colwell}, J.~E., {Hor\'anyi}, M., and {Gr{\"u}n}, E. (1998).
\newblock {Capture of interplanetary and interstellar dust by the Jovian
  magnetosphere}.
\newblock {\em Science}, 280:88--91.

\bibitem[{D'Amario} et~al., 1992]{damario1992}
{D'Amario}, L.~A., {Bright}, L.~E., and {Wolf}, A.~A. (1992).
\newblock {Galileo trajectory design}.
\newblock {\em Space Science Reviews}, 60:23--78.

\bibitem[{Delamere} et~al., 2004]{delamere2004}
{Delamere}, P.~A., {Steffl}, A., and {Bagenal}, F. (2004).
\newblock {Modeling temporal variability of plasma conditions in the Io torus
  during the Cassini era}.
\newblock {\em Journal of Geophysical Research (Space Physics)},
  109:10216--10224.

\bibitem[{Flandes}, 2005]{flandes2004}
{Flandes}, A. (2005).
\newblock {\em Dust dynamics in the jovian system}.
\newblock PhD thesis, Universidad Nacional Autonoma de Mexico.

\bibitem[{Flandes} and {Kr\"uger}, 2007]{flandes2007}
{Flandes}, A. and {Kr\"uger}, H. (2007).
\newblock {Solar wind modulation of Jupiter dust stream detection}.
\newblock In {H. Kr{\"u}ger and A. L. Graps}, editor, {\em {Dust in planetary
  systems}}, pages 87--90. ESA SP-643.

\bibitem[{Flandes} et~al., 2009]{flandes2009}
{Flandes}, A., {Kr\"uger}, H., Hamilton, D.~P., and Vald{\'e}s-Galicia, J.~F.
  (2009).
\newblock {Magnetic field modulated dust streams from Jupiter in interplanetary
  space}.
\newblock {\em Icarus}.
\newblock in preparation.

\bibitem[{Geissler}, 2003]{geissler2003}
{Geissler}, P.~E. (2003).
\newblock {Volcanic Activity on Io During the Galileo Era}.
\newblock {\em {Annual Reviews of Earth and Planetary Sciences}}, 31:175--211.

\bibitem[{Geissler} et~al., 2004]{geissler2004}
{Geissler}, P.~E., {McEwen}, A., {Phillips}, C., {Kesthelyi}, L., and
  {Spencer}, J. (2004).
\newblock {Surface Changes on Io during the Galileo Mission}.
\newblock {\em {Icarus}}, 169:29--64.

\bibitem[{Geissler} and {McMillan}, 2008]{geissler2008}
{Geissler}, P.~E. and {McMillan}, M.~T. (2008).
\newblock {Galileo observations of volcanic plumes on Io}.
\newblock {\em Icarus}, 197:505--518.

\bibitem[{Graps}, 2001]{graps2001a}
{Graps}, A.~L. (2001).
\newblock {\em Io revealed in the Jovian dust streams}.
\newblock PhD thesis, Ruprecht-Karls-Universit{\"a}t Heidelberg.

\bibitem[{Graps}, 2006]{graps2006}
{Graps}, A.~L. (2006).
\newblock {Characterization of Jovian plasma-embedded dust particles}.
\newblock {\em Planetary and Space Science}, 54:911--918.

\bibitem[{Graps} et~al., 2001]{graps2001b}
{Graps}, A.~L., {Gr{\"u}n}, E., {Kr\"uger}, H., {Hor\'anyi}, M., and {Svedhem},
  H. (2001).
\newblock Io revealed in the jovian dust streams.
\newblock In {\em Proceedings of the Meteoroids 2001 Conference, ESA SP-495},
  pages 601--608.

\bibitem[{Graps} et~al., 2000]{graps2000a}
{Graps}, A.~L., {Gr\"un}, E., {Svedhem}, H., {Kr\"uger}, H., {Hor\'anyi}, M.,
  {Heck}, A., and {Lammers}, S. (2000).
\newblock {Io as a source of the Jovian dust streams}.
\newblock {\em Nature}, 405:48--50.

\bibitem[{Gr{\"u}n} et~al., 1995a]{gruen1995b}
{Gr{\"u}n}, E., {Baguhl}, M., {Divine}, N., {Fechtig}, H., {Hamilton}, D.~P.,
  {Hanner}, M.~S., {Kissel}, J., {Lindblad}, B.~A., {Linkert}, D., {Linkert},
  G., {Mann}, I., {McDonnell}, J. A.~M., {Morfill}, G.~E., {Polanskey}, C.,
  {Riemann}, R., {Schwehm}, G.~H., {Siddique}, N., {Staubach}, P., and {Zook},
  H.~A. (1995a).
\newblock {Three years of Galileo dust data}.
\newblock {\em Planetary and Space Science}, 43:953--969.
\newblock Paper~II.

\bibitem[{Gr{\"u}n} et~al., 1995b]{gruen1995c}
{Gr{\"u}n}, E., {Baguhl}, M., {Divine}, N., {Fechtig}, H., {Hamilton}, D.~P.,
  {Hanner}, M.~S., {Kissel}, J., {Lindblad}, B.~A., {Linkert}, D., {Linkert},
  G., {Mann}, I., {McDonnell}, J. A.~M., {Morfill}, G.~E., {Polanskey}, C.,
  {Riemann}, R., {Schwehm}, G.~H., {Siddique}, N., {Staubach}, P., and {Zook},
  H.~A. (1995b).
\newblock {Two years of Ulysses dust data}.
\newblock {\em Planetary and Space Science}, 43:971--999.
\newblock Paper~III.

\bibitem[{Gr{\"u}n} et~al., 1995c]{gruen1995a}
{Gr{\"u}n}, E., {Baguhl}, M., {Hamilton}, D.~P., {Kissel}, J., {Linkert}, D.,
  {Linkert}, G., and {Riemann}, R. (1995c).
\newblock {Reduction of Galileo and Ulysses dust data}.
\newblock {\em Planetary and Space Science}, 43:941--951.
\newblock Paper~I.

\bibitem[{Gr{\"u}n} et~al., 1996a]{gruen1996b}
{Gr{\"u}n}, E., {Baguhl}, M., {Hamilton}, D.~P., {Riemann}, R., {Zook}, H.~A.,
  {Dermott}, S.~F., {Fechtig}, H., {Gustafson}, B.~A., {Hanner}, M.~S.,
  {Hor\'anyi}, M., {Khurana}, K.~K., {Kissel}, J., {Kivelson}, M., {Lindblad},
  B.~A., {Linkert}, D., {Linkert}, G., {Mann}, I., {McDonnell}, J. A.~M.,
  {Morfill}, G.~E., {Polanskey}, C., {Schwehm}, G.~H., and {Srama}, R. (1996a).
\newblock {Constraints from Galileo observations on the origin of Jovian dust
  streams}.
\newblock {\em Nature}, 381:395--398.

\bibitem[{Gr{\"u}n} et~al., 1992a]{gruen1992a}
{Gr{\"u}n}, E., {Fechtig}, H., {Hanner}, M.~S., {Kissel}, J., {Lindblad},
  B.~A., {Linkert}, D., {Maas}, D., {Morfill}, G.~E., and {Zook}, H.~A.
  (1992a).
\newblock {The Galileo dust detector}.
\newblock {\em Space Science Reviews}, 60:317--340.

\bibitem[{Gr{\"u}n} et~al., 1992b]{gruen1992b}
{Gr{\"u}n}, E., {Fechtig}, H., {Kissel}, J., {Linkert}, D., {Maas}, D.,
  {McDonnell}, J. A.~M., {Morfill}, G.~E., {Schwehm}, G.~H., {Zook}, H.~A., and
  {Giese}, R.~H. (1992b).
\newblock {The Ulysses dust experiment}.
\newblock {\em Astronomy and Astrophysics, Supplement}, 92:411--423.

\bibitem[{Gr{\"u}n} et~al., 1996b]{gruen1996c}
{Gr{\"u}n}, E., {Hamilton}, D.~P., {Riemann}, R., {Dermott}, S.~F., {Fechtig},
  H., {Gustafson}, B.~A., {Hanner}, M.~S., {Heck}, A., {Hor\'anyi}, M.,
  {Kissel}, J., {Kivelson}, M., {Kr{\"u}ger}, H., {Lindblad}, B.~A., {Linkert},
  D., {Linkert}, G., {Mann}, I., {McDonnell}, J. A.~M., {Morfill}, G.~E.,
  {Polanskey}, C., {Schwehm}, G.~H., {Srama}, R., and {Zook}, H.~A. (1996b).
\newblock {Dust measurements during Galileo's approach to Jupiter and Io
  encounter}.
\newblock {\em Science}, 274:399--401.

\bibitem[{Gr{\"u}n} et~al., 1998]{gruen1998}
{Gr{\"u}n}, E., {Kr{\"u}ger}, H., {Graps}, A., {Hamilton}, D.~P., {Heck}, A.,
  {Linkert}, G., {Zook}, H., {Dermott}, S.~F., {Fechtig}, H., {Gustafson}, B.,
  {Hanner}, M., {Hor\'anyi}, M., {Kissel}, J., {Lindblad}, B., {Linkert}, G.,
  {Mann}, I., {McDonnell}, J. A.~M., {Morfill}, G.~E., {Polanskey}, C.,
  {Schwehm}, G.~H., and {Srama}, R. (1998).
\newblock {Galileo observes electromagnetically coupled dust in the Jovian
  magnetosphere}.
\newblock {\em Journal of Geophysical Research}, 103:20011--20022.

\bibitem[{Gr{\"u}n} et~al., 1997a]{gruen1997b}
{Gr{\"u}n}, E., {Kr{\"u}ger}, P., {Dermott}, S.~F., {Fechtig}, H., {Graps},
  A.~L., {Gustafson}, B.~A., {Hamilton}, D.~P., {Heck}, A., {Hor\'anyi}, M.,
  {Kissel}, J., {Lindblad}, B.~A., {Linkert}, D., {Linkert}, G., {Mann}, I.,
  {McDonnell}, J. A.~M., {Morfill}, G.~E., {Polanskey}, C., {Schwehm}, G.~H.,
  {Srama}, R., and {Zook}, H.~A. (1997a).
\newblock {Dust measurements in the Jovian magnetosphere}.
\newblock {\em Geophysical Research Letters}, 24:2171--2174.

\bibitem[{Gr{\"u}n} et~al., 1997b]{gruen1997a}
{Gr{\"u}n}, E., {Staubach}, P., {Baguhl}, M., {Hamilton}, D.~P., {Zook}, H.~A.,
  {Dermott}, S.~F., {Gustafson}, B.~A., {Fechtig}, H., {Kissel}, J., {Linkert},
  D., {Linkert}, G., {Srama}, R., {Hanner}, M.~S., {Polanskey}, C.,
  {Hor\'anyi}, M., {Lindblad}, B.~A., {Mann}, I., {McDonnell}, J. A.~M.,
  {Morfill}, G.~E., and {Schwehm}, G.~H. (1997b).
\newblock {South-North and Radial Traverses through the Interplanetary Dust
  Cloud}.
\newblock {\em Icarus}, 129:270--288.

\bibitem[{Gr{\"u}n} et~al., 1993]{gruen1993a}
{Gr{\"u}n}, E., {Zook}, H.~A., {Baguhl}, M., {Balogh}, A., {Bame}, S.~J.,
  {Fechtig}, H., {Forsyth}, R., {Hanner}, M.~S., {Hor\'anyi}, M., {Kissel}, J.,
  {Lindblad}, B.~A., {Linkert}, D., {Linkert}, G., {Mann}, I., {McDonnell}, J.
  A.~M., {Morfill}, G.~E., {Phillips}, J.~L., {Polanskey}, C., {Schwehm},
  G.~H., {Siddique}, N., {Staubach}, P., {Svestka}, J., and {Taylor}, A.
  (1993).
\newblock {Discovery of Jovian dust streams and interstellar grains by the
  Ulysses spacecraft}.
\newblock {\em Nature}, 362:428--430.

\bibitem[{Hamilton} and {Burns}, 1993]{hamilton1993a}
{Hamilton}, D.~P. and {Burns}, J.~A. (1993).
\newblock {Ejection of dust from Jupiter's gossamer ring}.
\newblock {\em Nature}, 364:695--699.

\bibitem[{Hamilton} and {Kr\"uger}, 2008]{hamilton2008}
{Hamilton}, D.~P. and {Kr\"uger}, H. (2008).
\newblock {Jupiter's shadow sculpts its gossamer rings}.
\newblock {\em Nature}, 453:72--75.

\bibitem[Heck, 1998]{heck1998}
Heck, A. (1998).
\newblock {\em Modellierung und Analyse der von der Raumsonde Galileo im
  Jupitersystem vorgefundenen Mikrometeoroiden-Populationen}.
\newblock PhD thesis, Ruprecht-Karls-Universit{\"a}t Heidelberg.

\bibitem[{Hor\'anyi} et~al., 1997]{horanyi1997}
{Hor\'anyi}, M., {Gr{\"u}n}, E., and {Heck}, A. (1997).
\newblock {Modeling the Galileo dust measurements at Jupiter}.
\newblock {\em Geophysical Research Letters}, 24:2175--2178.

\bibitem[{Hor\'anyi} et~al., 1993]{horanyi1993a}
{Hor\'anyi}, M., {Morfill}, G.~E., and {Gr{\"u}n}, E. (1993).
\newblock {Mechanism for the acceleration and ejection of dust grains from
  Jupiter's magnetosphere}.
\newblock {\em Nature}, 363:144--146.

\bibitem[{Johnson} et~al., 1992]{johnson1992}
{Johnson}, T.~V., {Yeates}, C., and {Young}, R. (1992).
\newblock {Galileo Mission Overview}.
\newblock {\em Space Science Reviews}, 60:3--21.

\bibitem[{Kempf} et~al., 2006]{kempf2006a}
{Kempf}, S., {Beckmann}, U., {Srama}, R., {Horanyi}, M., {Auer}, S., and
  {Gr{\"u}n}, E. (2006).
\newblock {The electrostatic potential of E ring particles}.
\newblock {\em Planetary and Space Science}, 54:999--1006.

\bibitem[{Kempf} et~al., 2004]{kempf2004}
{Kempf}, S., {Srama}, R., {Altobelli}, N., {Auer}, S., {Tschernjawski}, V.,
  {Bradley}, J., {Burton}, M.~E., {Helfert}, S., {Johnson}, T.~V., {Kr{\"
  u}ger}, H., {Moragas-Klostermeyer}, G., and {Gr{\" u}n}, E. (2004).
\newblock {Cassini between Earth and asteroid belt: first in-situ charge
  measurements of interplanetary grains}.
\newblock {\em Icarus}, 171:317--335.

\bibitem[{Krivov} et~al., 2002a]{krivov2002a}
{Krivov}, A.~V., {Kr{\"u}ger}, H., {Gr{\"u}n}, E., {Thiessenhusen}, K.-U., and
  {Hamilton}, D.~P. (2002a).
\newblock {A tenuous dust ring of Jupiter formed by escaping ejecta from the
  Galilean satellites}.
\newblock {\em Journal of Geophysical Research}, 107:E1, 10.1029/2000JE001434.

\bibitem[{Krivov} et~al., 2003]{krivov2003}
{Krivov}, A.~V., {Srem\v{c}evi{\'c}}, M., {Spahn}, F., {Dikarev}, V.~V., and
  {Kholshevnikov}, K.~V. (2003).
\newblock {Impact-generated dust clouds around planetary satellites:
  Spherically-symmetric case}.
\newblock {\em Planetary and Space Science}, 51:251--269.

\bibitem[{Krivov} et~al., 2002b]{krivov2002b}
{Krivov}, A.~V., {Wardinski}, I., {Spahn}, F., {Kr{\"u}ger}, H., and
  {Gr{\"u}n}, E. (2002b).
\newblock {Dust on the outskirts of the Jovian system}.
\newblock {\em Icarus}, 157:436--455.

\bibitem[{Kr{\"u}ger} et~al., 2009]{krueger2009b}
{Kr{\"u}ger}, H.~{Hamilton}, D.~P., {Moissl}, R., and {Gr{\"u}n}, E. (2009).
\newblock {Galileo in-situ dust measurements in Jupiter's gossamer rings}.
\newblock {\em Icarus}.
\newblock in press.

\bibitem[{Kr{\"u}ger}, 2003]{krueger2003c}
{Kr{\"u}ger}, H. (2003).
\newblock {\em {Jupiter's Dust Disc, An Astrophysical Laboratory}}.
\newblock Shaker Verlag Aachen, ISBN 3-8322-2224-3.
\newblock {Habilitation Thesis Ruprecht-Karls-Universit\"at Heidelberg}.

\bibitem[{Kr{\"u}ger} et~al., 2006a]{krueger2006b}
{Kr{\"u}ger}, H., {Altobelli}, N., {Anweiler}, B., {Dermott}, S.~F., {Dikarev},
  V., {Graps}, A.~L., {Gr{\"u}n}, E., {Gustafson}, B.~A., {Hamilton}, D.~P.,
  {Hanner}, M.~S., {Hor\'anyi}, M., {Kissel}, J., {Landgraf}, M., {Lindblad},
  B., {Linkert}, D., {Linkert}, G., {Mann}, I., {McDonnell}, J. A.~M.,
  {Morfill}, G.~E., {Polanskey}, C., {Schwehm}, G.~H., {Srama}, R., and {Zook},
  H.~A. (2006a).
\newblock {Five years of Ulysses dust data: 2000 to 2004}.
\newblock {\em Planetary and Space Science}, 54:932--956.
\newblock Paper~IX.

\bibitem[{Kr{\"u}ger} et~al., 2006b]{krueger2006a}
{Kr{\"u}ger}, H., {Bindschadler}, D., {Dermott}, S.~F., {Graps}, A.~L.,
  {Gr{\"u}n}, E., {Gustafson}, B.~A., {Hamilton}, D.~P., {Hanner}, M.~S.,
  {Hor\'anyi}, M., {Kissel}, J., {Lindblad}, B., {Linkert}, D., {Linkert}, G.,
  {Mann}, I., {McDonnell}, J. A.~M., {Moissl}, R., {Morfill}, G.~E.,
  {Polanskey}, C., {Schwehm}, G.~H., {Srama}, R., and {Zook}, H.~A. (2006b).
\newblock {Galileo dust data from the jovian system: 1997 to 1999}.
\newblock {\em Planetary and Space Science}, 54:879--910.
\newblock Paper~VIII.

\bibitem[{Kr{\"u}ger} et~al., 2010]{krueger2010b}
{Kr{\"u}ger}, H., {Dikarev}, V., {Anweiler}, B., {Dermott}, S.~F., {Graps},
  A.~L., {Gr{\"u}n}, E., {Gustafson}, B.~A., {Hamilton}, D.~P., {Hanner}, M.
  M.~S., {Hor\'anyi}, M., {Kissel}, J., {Linkert}, D., {Linkert}, G., {Mann},
  I., {McDonnell}, J. A.~M., {Morfill}, G.~E., {Polanskey}, C., {Schwehm},
  G.~H., and {Srama}, R. (2010).
\newblock {Three years of Ulysses dust data: 2005 to 2007}.
\newblock {\em Planetary and Space Science}.
\newblock Paper~XI, submitted.

\bibitem[{Kr{\"u}ger} et~al., 2003a]{krueger2003d}
{Kr{\"u}ger}, H., {Geissler}, P., {Hor{\'a}nyi}, M., {Graps}, A.~L., {Kempf},
  S., {Srama}, R., {Moragas-Klostermeyer}, G., {Moissl}, R., {Johnson}, T.~V.,
  and {Gr{\"u}n}, E. (2003a).
\newblock {Jovian dust streams: A Monitor of Io's volcanic plume activity}.
\newblock {\em Geophysical Research Letters}, 30:2101--2105.

\bibitem[{Kr{\"u}ger} et~al., 2006c]{krueger2006c}
{Kr{\"u}ger}, H., {Graps}, A.~L., {Hamilton}, D.~P., {Flandes}, A., {Forsyth},
  R.~J., {Hor\'anyi}, M., and {Gr{\"u}n}, E. (2006c).
\newblock {Ulysses jovian latitude scan of high-velocity dust streams
  originating from the jovian system}.
\newblock {\em Planetary and Space Science}, 54:919--931.

\bibitem[{Kr{\"u}ger} et~al., 2001a]{krueger2001a}
{Kr{\"u}ger}, H., {Gr{\"u}n}, E., {Graps}, A.~L., {Bindschadler}, D.~L.,
  {Dermott}, S.~F., {Fechtig}, H., {Gustafson}, B.~A., {Hamilton}, D.~P.,
  {Hanner}, M.~S., {Hor\'anyi}, M., {Kissel}, J., {Lindblad}, B., {Linkert},
  D., {Linkert}, G., {Mann}, I., {McDonnell}, J. A.~M., {Morfill}, G.~E.,
  {Polanskey}, C., {Schwehm}, G.~H., {Srama}, R., and {Zook}, H.~A. (2001a).
\newblock {One year of Galileo dust data from the jovian system: 1996}.
\newblock {\em Planetary and Space Science}, 49:1285--1301.
\newblock Paper~VI.

\bibitem[{Kr{\"u}ger} et~al., 1999a]{krueger1999a}
{Kr{\"u}ger}, H., {Gr{\"u}n}, E., {Hamilton}, D.~P., {Baguhl}, M., {Dermott},
  S.~F., {Fechtig}, H., {Gustafson}, B.~A., {Hanner}, M.~S., {Hor\'anyi}, M.,
  {Kissel}, J., {Lindblad}, B.~A., {Linkert}, D., {Linkert}, G., {Mann}, I.,
  {McDonnell}, J. A.~M., {Morfill}, G.~E., {Polanskey}, C., {Riemann}, R.,
  {Schwehm}, G.~H., {Srama}, R., and {Zook}, H.~A. (1999a).
\newblock {Three years of Galileo dust data: II. 1993 to 1995}.
\newblock {\em Planetary and Space Science}, 47:85--106.
\newblock Paper~IV.

\bibitem[{Kr{\"u}ger} et~al., 1999b]{krueger1999c}
{Kr{\"u}ger}, H., {Gr{\"u}n}, E., {Heck}, A., and {Lammers}, S. (1999b).
\newblock {Analysis of the sensor characteristics of the Galileo dust detector
  with collimated Jovian dust stream particles}.
\newblock {\em Planetary and Space Science}, 47:1015--1028.

\bibitem[{Kr{\"u}ger} et~al., 1999c]{krueger1999b}
{Kr{\"u}ger}, H., {Gr{\"u}n}, E., {Landgraf}, M., {Baguhl}, M., {Dermott},
  S.~F., {Fechtig}, H., {Gustafson}, B.~A., {Hamilton}, D.~P., {Hanner}, M.~S.,
  {Hor\'anyi}, M., {Kissel}, J., {Lindblad}, B., {Linkert}, D., {Linkert}, G.,
  {Mann}, I., {McDonnell}, J. A.~M., {Morfill}, G.~E., {Polanskey}, C.,
  {Schwehm}, G.~H., {Srama}, R., and {Zook}, H.~A. (1999c).
\newblock {Three years of Ulysses dust data: 1993 to 1995}.
\newblock {\em Planetary and Space Science}, 47:363--383.
\newblock Paper~V.

\bibitem[{Kr{\"u}ger} et~al., 2001b]{krueger2001b}
{Kr{\"u}ger}, H., {Gr{\"u}n}, E., {Landgraf}, M., {Dermott}, S.~F., {Fechtig},
  H., {Gustafson}, B.~A., {Hamilton}, D.~P., {Hanner}, M.~S., {Hor\'anyi}, M.,
  {Kissel}, J., {Lindblad}, B., {Linkert}, D., {Linkert}, G., {Mann}, I.,
  {McDonnell}, J. A.~M., {Morfill}, G.~E., {Polanskey}, C., {Schwehm}, G.~H.,
  {Srama}, R., and {Zook}, H.~A. (2001b).
\newblock {Four years of Ulysses dust data: 1996 to 1999}.
\newblock {\em Planetary and Space Science}, 49:1303--1324.
\newblock Paper~VII.

\bibitem[{Kr{\"u}ger} et~al., 2005]{krueger2005a}
{Kr{\"u}ger}, H., {Gr{\"u}n}, E., {Linkert}, D., {Linkert}, G., and {Moissl},
  R. (2005).
\newblock {Galileo long-term dust monitoring in the jovian magnetosphere}.
\newblock {\em Planetary and Space Science}, 53:1109--1120.

\bibitem[{Kr{\"u}ger} et~al., 2003b]{krueger2003a}
{Kr{\"u}ger}, H., {Hor{\'a}nyi}, M., and {Gr{\"u}n}, E. (2003b).
\newblock {Jovian dust streams: Probes of the Io plasma torus}.
\newblock {\em Geophysical Research Letters}, 30:1058--1061.

\bibitem[{Kr{\"u}ger} et~al., 2004]{krueger2004a}
{Kr{\"u}ger}, H., {Hor{\'a}nyi}, M., {Krivov}, A.~V., and {Graps}, A.~L.
  (2004).
\newblock {\em {Jovian dust: streams, clouds and rings}}, pages 219--240.
\newblock Jupiter.~The planet, satellites and magnetosphere.~ Edited by Fran
  Bagenal, Timothy E.~Dowling, William B.~McKinnon.~Cambridge planetary
  science, Vol.~1, Cambridge, UK: Cambridge University Press, ISBN
  0-521-81808-7, 2004.

\bibitem[{Kr{\"u}ger} et~al., 2000]{krueger2000a}
{Kr{\"u}ger}, H., {Krivov}, A.~V., and {Gr{\"u}n}, E. (2000).
\newblock {A dust cloud of Ganymede maintained by hypervelocity impacts of
  interplanetary micrometeoroids}.
\newblock {\em Planetary and Space Science}, 48:1457--1471.

\bibitem[{Kr{\"u}ger} et~al., 1999d]{krueger1999d}
{Kr{\"u}ger}, H., {Krivov}, A.~V., {Hamilton}, D.~P., and {Gr{\"u}n}, E.
  (1999d).
\newblock {Detection of an impact-generated dust cloud around Ganymede}.
\newblock {\em Nature}, 399:558--560.

\bibitem[{Kr{\"u}ger} et~al., 2003c]{krueger2003b}
{Kr{\"u}ger}, H., {Krivov}, A.~V., {Srem\v{c}evi\'c}, M., and {Gr{\"u}n}, E.
  (2003c).
\newblock {Galileo measurements of impact-generated dust clouds surrounding the
  Galilean satellites}.
\newblock {\em Icarus}, 164:170--187.

\bibitem[{McEwen} et~al., 2004]{mcewen2004}
{McEwen}, A., {Keszthelyi}, L., {Lopes}, R., {Schenk}, P., and {Spencer}, J.
  (2004).
\newblock {The Lithosphere and Surface of Io}.
\newblock In Bagenal, F., McKinnon, B., and Dowling, T., editors, {\em
  {Jupiter: Planet, Satellites \& Magnetosphere}}, pages 307--328. Cambridge
  University Press.

\bibitem[{McEwen} et~al., 1998]{mcewen1998}
{McEwen}, A.~S., {Keszthelyi}, L., {Geissler}, P., {Simonelli}, D.~P., {Carr},
  M.~H., {Johnson}, T.~V., {Klaasen}, K.~P., {Breneman}, H.~H., {Jones}, T.~J.,
  {Kaufman}, J.~M., {Magee}, K.~P., {Senske}, D.~A., {Belton}, M. J.~S., and
  {Schubert}, G. (1998).
\newblock {Active volcanism on Io as seen by Galileo SSI}.
\newblock {\em Icarus}, 135:181--219.

\bibitem[{Moissl}, 2005]{moissl2005}
{Moissl}, R. (2005).
\newblock {\em Galileos Staubmessungen in Jupiters Gossamer-Ringen}.
\newblock Ruprecht-Karls-Universit{\"a}t Heidelberg.
\newblock Diplom thesis.

\bibitem[{Porco} et~al., 2003]{porco2003}
{Porco}, C.~C., {West}, R.~A., {McEwen}, A., {Del Genio}, A.~D., {Ingersoll},
  A.~P., {Thomas}, P., {Squyres}, S., {Dones}, L., {Murray}, C.~D., {Johnson},
  T.~V., {Burns}, J.~A., {Brahic}, A., {Neukum}, G., {Veverka}, J., {Barbara},
  J.~M., {Denk}, T., {Evans}, M., {Ferrier}, J.~J., {Geissler}, P.,
  {Helfenstein}, P., {Roatsch}, T., {Throop}, H., {Tiscareno}, M., and
  {Vasavada}, A.~R. (2003).
\newblock {Cassini Imaging of Jupiter's Atmosphere, Satellites, and Rings}.
\newblock {\em Science}, 299:1541--1547.

\bibitem[{Postberg} et~al., 2006]{postberg2006}
{Postberg}, F., {Kempf}, S., {Green}, S.~F., {Hillier}, J.~K., {McBride}, N.,
  and {Gr{\"u}n}, E. (2006).
\newblock {Composition of jovian dust stream particles}.
\newblock {\em Icarus}, 183:122--134.

\bibitem[{Showalter} et~al., 2008]{showalter2008}
{Showalter}, M.~R., {de Pater}, I., {Verbanac}, G., {Hamilton}, D.~P., and
  {Burns}, J.~A. (2008).
\newblock {Properties and Dynamics of Jupiter's Gossamer Rings from Galileo,
  Voyager, Hubble and Keck Images}.
\newblock {\em Icarus}, 195:361--377.

\bibitem[{Spencer} et~al., 2007]{spencer2007}
{Spencer}, J.~R., {Stern}, S.~A., {Cheng}, A.~F., {Weaver}, H.~A., {Reuter},
  D.~C., {Retherford}, K., {Lunsford}, A., {Moore}, J.~M., {Abramov}, O.,
  {Lopes}, R.~M.~C., {Perry}, J.~E., {Kamp}, L., {Showalter}, M., {Jessup},
  K.~L., {Marchis}, F., {Schenk}, P.~M., and {Dumas}, C. (2007).
\newblock {Io Volcanism Seen by New Horizons: A Major Eruption of the Tvashtar
  Volcano}.
\newblock {\em Science}, 318:240--243.

\bibitem[{Srem\v{c}evi\'c} et~al., 2005]{sremcevic2005}
{Srem\v{c}evi\'c}, M., {Krivov}, A.~V., {Kr\"uger}, H., and {Spahn}, F. (2005).
\newblock {Impact-generated dust clouds around planetary satellites: model
  versus Galileo data}.
\newblock {\em Planetary and Space Science}, 53:625--641.

\bibitem[{Srem\v{c}evi\'c} et~al., 2003]{sremcevic2003}
{Srem\v{c}evi\'c}, M., {Krivov}, A.~V., and {Spahn}, F. (2003).
\newblock {Impact-generated dust clouds around planetary satellites: Asymmetry
  effects}.
\newblock {\em Planetary and Space Science}, 51:455--471.

\bibitem[{Svestka} et~al., 1996]{svestka1996}
{Svestka}, J., {Auer}, S., {Baguhl}, M., and {Gr{\"u}n}, E. (1996).
\newblock {Measurements of dust electric charges by the Ulysses and Galileo
  dust detectors}.
\newblock In Gustafson, B.~A. and Hanner, M.~S., editors, {\em Physics,
  Chemistry and Dynamics of Interplanetary Dust, ASP Conference Series}, volume
  104, pages 481--484.

\bibitem[{Thiessenhusen} et~al., 2000]{thiessenhusen2000}
{Thiessenhusen}, K.-U., {Kr{\"u}ger}, H., {Spahn}, F., and {Gr{\"u}n}, E.
  (2000).
\newblock {Dust grains around Jupiter -- The observations of the Galileo Dust
  Detector}.
\newblock {\em Icarus}, 144:89--98.

\bibitem[{Willis} et~al., 2005]{willis2005}
{Willis}, M.~J., {Burchell}, M., {Ahrens}, T.~J., {Kr\"uger}, H., and {Gr\"un},
  E. (2005).
\newblock {Decreased values of cosmic dust number density estimates in the
  solar system}.
\newblock {\em Icarus}, 176:440--452.

\bibitem[{Willis} et~al., 2004]{willis2004}
{Willis}, M.~J., {Burchell}, M., {Cole}, M., and {McDonnell}, J. A.~M. (2004).
\newblock {Influence of impact ionization detection methods on determination of
  dust particle flux in space}.
\newblock {\em Planetary and Space Science}, 52:711--725.

\bibitem[{Zook} et~al., 1996]{zook1996}
{Zook}, H.~A., {Gr\"un}, E., {Baguhl}, M., {Hamilton}, D.~P., {Linkert}, G.,
  {Linkert}, D., {Liou}, J.-C., {Forsyth}, R., and {Phillips}, J.~L. (1996).
\newblock {Solar wind magnetic field bending of Jovian dust trajectories}.
\newblock {\em Science}, 274:1501--1503.

\end{thebibliography}

\clearpage

\begin{table}[h]
\caption{\label{papers}
Summary of Galileo dust data papers and significant mission events.
}
\small
\begin{tabular}{lll}
&&\\
\hline
\hline
Time Interval & Significant Mission Events                        &      Paper Number \\
\hline
1989 -- 1992  & Galileo launch (18 Oct 1989)                      & II \citep{gruen1995b}\\[1.5ex]
1993 -- 1995  & Jupiter orbit insertion (7 Dec 1995)              & IV \citep{krueger1999a}    \\[1.5ex]
1996          & Galileo orbits G1 -- E4                           & VI \citep{krueger2001a}\\[1.5ex]
1997 -- 1999  & Galileo orbits J5 -- I25                          & VIII \citep{krueger2006a}\\[1.5ex]
2000 -- 2003  & Galileo orbits E26 -- J35,                         & X (this paper) \\
              & First gossamer ring passage (5 Nov 2002),         &  \\
              & second gossamer ring passage and                  & \\
              & Galileo Jupiter impact (21 Sep 2003)              & \\
\hline \hline
\end{tabular}
\end{table}

\begin{table}%[htb]
\caption{\label{event_table_1} Galileo mission and dust detector (DDS) 
configuration, tests and other events (2000-2003). See text for details. 
}
{\footnotesize
  \begin{tabular*}{15cm}{lccl}
   \hline
   \hline \\[-2.0ex]
%Titelzeile
Yr-day& 
Date&
Time& 
Event \\[0.7ex]
\hline \\[-2.0ex]
89-291& 18 Oct 1989& 16:52& Galileo launch \\
95-341& 07 Dec 1995& 21:54& Galileo Jupiter closest approach, distance: $\rm 4.0\,R_J$  \\
%98-080& 21 Mar 1998& 20:00& DDS configuration: HV=2, EVD=I, SSEN=0,1,1,1 \\
99-345& 11 Dec 1999& 02:07& DDS configuration: HV=4, EVD=I, SSEN=0,1,1,1 \\
00-001& 01 Jan 2000& 00:00& DDS begin RTS data \\
00-003& 03 Jan 2000& 17:28& DDS end RTS data, begin record data \\
00-003& 03 Jan 2000& 18:00& {\bf Galileo Europa 26 (E26) closest approach}, altitude 351\,km \\ 
00-003& 03 Jan 2000& 18:30& DDS end record data, begin RTS data \\
00-004& 04 Jan 2000& 03:33& Galileo Jupiter closest approach, distance $\rm 5.8 R_J$ \\
00-006& 06 Jan 2000& 02:00& Galileo turn $5^{\circ}$, duration 3\,h, return to nominal attitude \\
00-007& 07 Jan 2000& 19:00& Galileo OTM-82, size of turn $3^{\circ}$, duration 5\,h, return to nominal attitude \\
00-009& 09 Jan 2000& 11:49& DDS end RTS data \\
00-029& 29 Jan 2000& 04:00& Galileo OTM-83, size of turn $3^{\circ}$, duration 3\,h, return to nominal attitude \\
00-033& 02 Feb 2000& 18:00& Galileo turn $4^{\circ}$, new nominal attitude \\
00-038& 07 Feb 2000& 23:00& Galileo turn $7^{\circ}$, duration 3\,h, return to nominal attitude \\
00-049& 18 Feb 2000& 16:36& Galileo OTM-84, no attitude change \\
00-050& 19 Feb 2000& 12:00& DDS begin RTS data \\
00-053& 22 Feb 2000& 12:30& Galileo Jupiter closest approach, distance $\rm 5.9 R_J$ \\
00-053& 22 Feb 2000& 13:02& DDS end RTS data, begin record data \\
00-053& 22 Feb 2000& 13:47& {\bf Galileo Io 27 (I27) closest approach}, altitude 198\,km \\ 
00-053& 22 Feb 2000& 14:25& DDS end record data, begin RTS data \\
00-054& 23 Feb 2000& 19:20& DDS last RTS data before spacecraft anomaly\\
00-055& 24 Feb 2000& 04:00& Galileo turn $15^{\circ}$, duration 30\,h, return to nominal attitude \\
00-055& 24 Feb 2000& 21:00& Galileo spacecraft anomaly \\
00-056& 25 Feb 2000& 04:16& DDS begin RTS data after spacecraft anomaly \\
00-057& 25 Feb 2000& 14:00& Galileo OTM-85, size of turn $19^{\circ}$, duration 11\,h, return to nominal attitude \\
00-059& 28 Feb 2000& 23:56& DDS end RTS data \\
00-070& 10 Mar 2000& 10:00& Galileo turn $8^{\circ}$, new nominal attitude \\
00-088& 28 Mar 2000& 00:00& Galileo turn $3^{\circ}$, duration 3\,h, return to nominal attitude \\ 
00-098& 07 Apr 2000& 12:00& Galileo OTM-86, no attitude change \\
00-116& 25 Apr 2000& 20:00& Galileo turn $8^{\circ}$, new nominal attitude \\
00-117& 26 Apr 2000& 01:11& DDS last MRO before solar conjunction \\
00-118& 27 Apr 2000&      & Start solar conjunction period \\
00-138& 17 May 2000&      & End solar conjunction period \\
00-138& 17 May 2000& 09:30& DDS begin RTS data \\
00-139& 18 May 2000& 10:00& Galileo OTM-87, no attitude change \\
00-141& 20 May 2000& 09:39& DDS end RTS data, begin record data \\
00-141& 20 May 2000& 10:10& {\bf Galileo Ganymede 28 (G28) closest approach}, altitude 808\,km \\ 
\hline
\end{tabular*}\\[1.5ex]
}
Abbreviations used: MRO: DDS
memory readout; HV: channeltron high voltage step; EVD: event definition,
ion- (I), channeltron- (C), or electron-channel (E); SSEN: detection thresholds,
ICP, CCP, ECP and PCP; OTM: orbit trim maneuver; RTS: Realtime science. 
\end{table}

\setcounter{table}{1}

\begin{table}%[htb]
\caption{\label{event_table_2} 
Continued.
}
{\footnotesize
  \begin{tabular*}{15cm}{lccl}
   \hline
   \hline \\[-2.0ex]
%Titelzeile
Yr-day& 
Date&
Time& 
Event \\[0.7ex]
\hline \\[-2.0ex]
00-141& 20 May 2000& 10:40& DDS end record data, begin RTS data \\
00-142& 21 May 2000& 04:52& Galileo Jupiter closest approach, distance $\rm 6.7 R_J$ \\
00-143& 22 May 2000& 08:00& Galileo turn $2^{\circ}$, duration 1\,h, return to nominal attitude \\
00-146& 25 May 2000& 08:00& Galileo OTM-88, no attitude change \\
00-152& 31 May 2000& 07:00& Galileo turn $8^{\circ}$, new nominal attitude \\
00-170& 18 Jun 2000& 23:57& DDS end RTS data \\
00-176& 23 Jun 2000& 03:00& Galileo turn $7^{\circ}$, duration 2\,h, return to nominal attitude \\ 
00-189& 07 Jul 2000& 19:00& Galileo turn $9^{\circ}$, new nominal attitude \\
00-209& 27 Jul 2000& 11:11& DDS configuration: HV = 5  \\
00-216& 03 Aug 2000& 18:31& DDS begin RTS data \\
00-223& 10 Aug 2000& 02:00& Galileo turn $9^{\circ}$, new nominal attitude \\
00-244& 31 Aug 2000& 16:00& Galileo turn $2^{\circ}$, duration 2\,h, return to nominal attitude \\
00-252& 08 Sep 2000& 18:00& Galileo OTM-89, size of turn $2^{\circ}$, duration 3\,h, return to nominal attitude \\
00-253& 09 Sep 2000& 08:11& DDS end RTS data \\
00-300& 26 Oct 2000& 10:01& DDS begin RTS data \\
00-302& 28 Oct 2000& 01:00& Galileo OTM-90, no attitude change \\
00-342& 07 Dec 2000& 23:00& Galileo turn $2^{\circ}$, duration 1\,h, return to nominal attitude \\
00-353& 18 Dec 2000& 05:00& Galileo turn $20^{\circ}$, new nominal attitude \\
00-356& 21 Dec 2000& 20:15& Galileo OTM-91, no attitude change \\
00-363& 28 Dec 2000& 08:25& {\bf Galileo Ganymede 29 (G29) closest approach}, altitude 2,337\,km \\ 
00-364& 29 Dec 2000& 03:27& Galileo Jupiter closest approach, distance $\rm 7.5 R_J$ \\
01-002& 02 Jan 2001& 20:00& Galileo OTM-92, no attitude change \\
01-004& 04 Jan 2001& 03:30& Galileo turn $10^{\circ}$, return to nominal attitude \\
01-086& 27 Mar 2001& 23:00& DDS end RTS data \\
01-094& 04 Apr 2001& 04:00& Galileo turn $8^{\circ}$, new nominal attitude \\
01-114& 24 Apr 2001& 03:00& Galileo turn $8^{\circ}$, duration 9\,h, return to nominal attitude \\
01-130& 10 May 2001& 16:00& Galileo OTM-94, no attitude change \\
01-133& 13 May 2001& 01:00& Galileo turn $4^{\circ}$, new nominal attitude \\
01-142& 22 May 2001& 23:00& Galileo turn $6^{\circ}$, return to nominal attitude \\
01-143& 23 May 2001& 12:01& DDS begin RTS data \\
01-143& 23 May 2001& 17:33& Galileo Jupiter closest approach, distance $\rm 7.3 R_J$ \\
01-145& 25 May 2001& 11:24& {\bf Galileo Callisto 30 (C30) closest approach}, altitude 138\,km \\ 
01-146& 26 May 2001& 07:15& DDS end RTS data \\
01-146& 26 May 2001& 09:00& Galileo turn $5^{\circ}$, duration 2\,h, return to nominal attitude \\
01-146& 26 May 2001& 22:00& Galileo turn $12^{\circ}$, duration 13\,h, return to nominal attitude \\
01-148& 28 May 2001& 19:00& Galileo OTM-95, size of turn $1^{\circ}$ \\
01-152& 01 Jun 2001& 11:00& Galileo turn $5^{\circ}$, new nominal attitude \\
01-154& 03 Jun 2001& 16:58& DDS last MRO before solar conjunction \\
01-155& 04 Jun 2001&      & Start solar conjunction period \\
01-174& 23 Jun 2001&      & End solar conjunction period \\
\hline
\end{tabular*}\\[1.5ex]
}
%Abbreviations used: MRO: DDS
%memory readout; HV: channeltron high voltage step; EVD: event definition,
%ion- (I), channeltron- (C), or electron-channel (E); SSEN: detection thresholds,
%ICP, CCP, ECP and PCP; OTM: orbit trim maneuver; RTS: Realtime science. 
\end{table}

\setcounter{table}{1}

\begin{table}%[htb]
\caption{\label{event_table_3} 
Continued.
}
{\footnotesize
  \begin{tabular*}{15cm}{lccl}
   \hline
   \hline \\[-2.0ex]
%Titelzeile
Yr-day& 
Date&
Time& 
Event \\[0.7ex]
\hline \\[-2.0ex]
01-177& 26 Jun 2001& 22:58& DDS fist MRO after solar conjunction \\
01-183& 02 Jul 2001& 21:00& Galileo turn $4^{\circ}$, new nominal attitude \\
01-186& 05 Jul 2001& 05:00& Galileo turn $5^{\circ}$, duration 5\,h, return to nominal attitude \\
01-194& 13 Jul 2001& 06:00& Galileo OTM-97, size of turn $3^{\circ}$, duration 4\,h, return to nominal attitude \\ 
01-201& 20 Jul 2001& 07:00& Galileo turn $4^{\circ}$, new nominal attitude \\
01-215& 03 Aug 2001& 23:00& Galileo OTM-98, no attitude change \\
01-217& 05 Aug 2001& 05:12& DDS begin RTS data \\
01-218& 06 Aug 2001& 04:52& Galileo Jupiter closest approach, distance $\rm 5.9 R_J$ \\
01-218& 06 Aug 2001& 04:59& {\bf Galileo Io 31 (I31) closest approach}, altitude 193\,km \\ 
01-219& 07 Aug 2001& 16:08& DDS end RTS data \\
01-220& 08 Aug 2001& 03:00& Galileo turn $7^{\circ}$, duration 2\,h, return to nominal attitude \\
01-222& 10 Aug 2001& 19:30& Galileo OTM-99, no attitude change \\
01-224& 12 Aug 2001& 08:00& Galileo turn $3^{\circ}$, new nominal attitude \\
01-236& 24 Aug 2001& 07:00& Galileo turn $3^{\circ}$, duration 2\,h, return to nominal attitude \\
01-245& 02 Sep 2001& 03:00& Galileo OTM-100, no attitude change \\
01-246& 03 Sep 2001& 05:00& Galileo turn $4^{\circ}$, new nominal attitude \\
01-261& 18 Sep 2001& 12:00& Galileo OTM-101, no attitude change \\
01-270& 27 Sep 2001& 04:00& Galileo turn $3^{\circ}$, new nominal attitude \\
01-286& 13 Oct 2001& 18:00& Galileo OTM-102, no attitude change \\
01-287& 14 Oct 2001& 02:04& DDS begin RTS data \\
01-288& 15 Oct 2001& 23:56& Galileo Jupiter closest approach, distance $\rm 5.8 R_J$ \\
01-289& 16 Oct 2001& 01:23& {\bf Galileo Io 32 (I32) closest approach}, altitude 184\,km \\ 
01-290& 17 Oct 2001& 12:00& Galileo turn $3^{\circ}$, duration 2\,h, return to nominal attitude \\
01-293& 20 Oct 2001& 04:00& Galileo OTM-103, size of turn $1^{\circ}$, duration 2\,h, return to nominal attitude \\
01-301& 28 Oct 2001& 23:30& DDS end RTS data \\
01-324& 20 Nov 2001& 00:00& Galileo turn $4^{\circ}$, duration 2\,h, return to nominal attitude \\
01-335& 01 Dec 2001& 00:00& Galileo OTM-104, no attitude change \\
01-340& 06 Dec 2001& 15:00& Galileo turn $2^{\circ}$, new nominal attitude \\
01-352& 18 Dec 2001& 00:00& DDS configuration: HV = 6 \\
01-357& 23 Dec 2001& 16:00& Galileo turn $2^{\circ}$, new nominal attitude \\
02-004& 04 Jan 2002& 12:25& DDS begin RTS data \\
02-010& 10 Jan 2002& 00:00& Galileo turn $5^{\circ}$, duration 2\,h, new nominal attitude \\
02-017& 17 Jan 2002& 09:15& DDS end RTS data \\
02-017& 17 Jan 2002& 13:40& Galileo spacecraft anomaly \\
02-017& 17 Jan 2002& 14:08& {\bf Galileo Io 33 (I33) closest approach}, altitude 102 \,km \\
02-017& 17 Jan 2002& 16:23& Galileo Jupiter closest approach, distance $\rm 5.5\,R_J$ \\
02-017& 17 Jan 2002& 23:50& DDS begin RTS data after spacecraft anomaly \\
02-021& 21 Jan 2002& 12:00& Galileo OTM-106, no attitude change \\
02-029& 29 Jan 2002& 23:00& Galileo turn $2^{\circ}$, new nominal attitude \\
02-032& 01 Feb 2002& 01:00& Galileo turn $44^{\circ}$, duration 90\,h, return to nominal attitude \\
02-047& 16 Feb 2002& 19:09& Galileo spacecraft anomaly \\
02-051& 20 Feb 2002& 17:54& DDS begin RTS data after spacecraft anomaly \\
02-094& 04 Apr 2002& 00:00& Galileo turn $3^{\circ}$, duration 3\,h, return to nominal attitude \\
\hline
\end{tabular*}\\[1.5ex]
}
%Abbreviations used: MRO: DDS
%memory readout; HV: channeltron high voltage step; EVD: event definition,
%ion- (I), channeltron- (C), or electron-channel (E); SSEN: detection thresholds,
%ICP, CCP, ECP and PCP; OTM: orbit trim maneuver; RTS: Realtime science.
\end{table}

\setcounter{table}{1}

\begin{table}%[htb]
\caption{\label{event_table_4} 
Continued.
}
{\footnotesize
  \begin{tabular*}{15cm}{lccl}
   \hline
   \hline \\[-2.0ex]
%Titelzeile
Yr-day& 
Date&
Time& 
Event \\[0.7ex]
\hline \\[-2.0ex]
02-102& 12 Apr 2002& 05:00& Galileo turn $3^{\circ}$, new nominal attitude \\
02-124& 04 May 2002& 23:00& Galileo turn $4^{\circ}$, new nominal attitude \\
02-131& 11 May 2002& 12:00& Galileo turn $1^{\circ}$, duration 2\,h, return to nominal attitude \\
02-146& 26 May 2002& 02:00& Galileo turn $4^{\circ}$, new nominal attitude \\
02-157& 06 Jun 2002& 02:00& Galileo turn $6^{\circ}$, duration 3\,h, return to nominal attitude \\
02-165& 14 Jun 2002& 16:00& Galileo OTM-107, size of turn $3^{\circ}$, duration 3\,h, return to nominal attitude \\
02-182& 01 Jul 2002& 11:00& Galileo turn $12^{\circ}$, new nominal attitude \\
02-190& 09 Jul 2002&      & Begin solar conjunction \\
02-209& 28 Jul 2002&      & End solar conjunction \\
02-214& 02 Aug 2002& 08:00& Galileo turn $5^{\circ}$, new nominal attitude \\
02-232& 20 Aug 2002& 15:00& Galileo turn $4^{\circ}$, new nominal attitude \\
02-250& 07 Sep 2002& 12:00& Galileo turn $4^{\circ}$, new nominal attitude \\
02-274& 01 Oct 2002& 19:30& Galileo spacecraft anomaly \\
02-275& 02 Oct 2002& 23:54& DDS begin RTS after spacecraft anomaly \\
02-281& 08 Oct 2002& 02:00& Galileo turn $6^{\circ}$, duration 3\,h, return to nominal attitude \\
02-285& 12 Oct 2002& 15:00& Galileo turn $9^{\circ}$, new nominal attitude \\
02-298& 25 Oct 2002& 03:00& Galileo turn $2^{\circ}$, duration 2\,h, return to nominal attitude \\
02-309& 05 Nov 2002& 02:44& DDS end RTS data \\
02-309& 05 Nov 2002& 02:44& DDS begin record data \\
02-309& 05 Nov 2002& 06:19& {\bf Galileo Amalthea 34 (A34) closest approach}, 244 \,km distance \\ 
      &            &      &   from moon's centre\\
02-309& 05 Nov 2002& 06:35& Galileo spacecraft anomaly, end record data \\
02-309& 05 Nov 2002& 07:25& Galileo Jupiter closest approach, distance $\rm 2.0\,R_J$ \\
02-318& 14 Nov 2002& 08:00& Galileo turn $9^{\circ}$, duration 3\,h, return to nominal attitude \\
02-322& 18 Nov 2002& 14:29& DDS first MRO after spacecraft anomaly \\
02-363& 29 Dec 2002& 07:53& DDS final MRO \\
03-003& 03 Jan 2003& 22:00& Galileo turn $20^{\circ}$, duration 6\,h, return to nominal attitude \\
03-004& 04 Jan 2003& 21:00& Galileo turn $20^{\circ}$, duration 5\,h, return to nominal attitude\\
03-008& 08 Jan 2003& 08:00& Galileo turn $17^{\circ}$, duration 2\,h, return to nominal attitude \\
03-010& 10 Jan 2003& 22:00& Galileo turn $22^{\circ}$, duration 82\,h, return to nominal attitude\\
03-015& 15 Jan 2003& 10:00& Galileo turn $22^{\circ}$, new nominal attitude \\          
03-063& 04 Mar 2003& 15:00& DDS begin RTS data \\
03-070& 11 Mar 2003& 21:50& DDS end RTS data \\
03-255& 12 Sep 2003& 21:43& DDS begin RTS data \\
03-256& 13 Sep 2003& 18:57& DDS end RTS data \\
03-263& 20 Sep 2003& 13:44& DDS begin RTS data \\
03-263& 20 Sep 2003& 14:26& DDS end RTS data \\
03-264& 21 Sep 2003& 12:10& DDS begin RTS data \\
03-264& 21 Sep 2003& 17:59& DDS end RTS data \\
03-264& 21 Sep 2003& 18:57& {\bf Galileo Jupiter impact, end of mission} \\          
\hline
\end{tabular*}\\[1.5ex]
}
%Abbreviations used: MRO: DDS
%memory readout; HV: channeltron high voltage step; EVD: event definition,
%ion- (I), channeltron- (C), or electron-channel (E); SSEN: detection thresholds,
%ICP, CCP, ECP and PCP; OTM: orbit trim maneuver; RTS: Realtime science. 
\end{table}

\clearpage

\begin{table}[h]
\caption{\label{tab_fov}
Dust detector sensitive area and field-of-view (FOV) for different dust data sets.
}
\small
\begin{tabular}{lccl}
&&\\
\hline
\hline
Dust data set              &   FOV        & Sensor area         &     Comment \\
                           & ($^{\circ}$) & ($\mathrm{cm^2}$)   &       \\ 
\hline
Stream particles class 2   & 140          &  1000               & Nominal target FOV  \citep{gruen1992a}\\
Stream particles class 3   & 96           &  110                & Reduced target FOV  \citep{krueger1999c}    \\
All other                  & 180          &  1000               & Target plus side wall FOV \citep{altobelli2004a}\\
\hline \hline
\end{tabular}
\end{table}

\begin{table}%[htb]
\begin{center}
\caption[Galileo dust data transmission modes]
{\label{tab_data_modes} Details of Galileo dust data transmission modes
during the Jupiter mission. See text for details.
}
\small
\begin{tabular}{lccccc}
                        &
                        &
                        &
                        &
                        &
                        \\[-1.8ex]
\hline
\hline
                        &
                        &
                        &
                        &
                        &
                        \\[-2.2ex]
                        &
                        & \mc{2}{c}{Realtime Science (RTS)}      & Record    & MROs      \\[0.5ex]
                        & & Low rate          & High rate          &           &           \\       
\hline & & & & & \\[-1.9ex]
Data rate (bits\,$\rm s^{-1}$) & &   1.1          &    3.4          &   24      & $\sim 3\times 10^{-3}\,^{a)}$ \\[1.5ex]
Timing accuracy (min)   & &  21.2                 &     7.1         & $\sim 1$   & 259 \\[1.5ex]
Data frames per readout &       &    \mc{2}{c}{7}                   &   7       & 46          \\[1.5ex]
Mission time coverage (\%)  & & \mc{2}{c}{40}                        & $<0.1$  &  60     \\[1.5ex]
Maximum event rate recordable     & AC21/AC31$^{b)}$ & 3000 & 9000 & 65000 & $\approx 2^{a)}$ \\ 
\mc{1}{c}{by accumulators ($\rm min^{-1}$)}  & All other         & 12   &  36  &  256  & $\approx 0.01^{a)}$         \\[1.5ex]
Maximum event rate for full &   &   &  &  \\
\mc{1}{c}{data set transmission ($\rm min^{-1}$)} &  & $\frac{1}{21.2}$ & $\frac{1}{7.1}$ & $\sim 1$ & $\approx \frac{46}{20 \mathrm{\,days}}$ \\
\hline
\hline 
                        &
                        &
                        &
                        &
                        &
                        \\[-2ex]
\mc{6}{l}{$^{a)}$ One MRO every 20 days assumed.} \\
\mc{6}{l}{$^{b)}$ Since 4 December 1996; the ``All other'' row was valid for all data before this time.} \\
\end{tabular}
\end{center}
\vspace{-3mm}
\end{table}

\begin{table}%[htb]
\caption[Criteria for noise identification in the gossamer ring]
{\label{tab_gossamer_noise}
Criteria for the separation of noise events in classes 1 and 2
from true dust impacts in the region within Io's orbit for Galileo orbits A34
and J35 (gossamer ring passages).
Noise events in the lowest amplitude range (AR1) fulfill 
at least one of the criteria listed, whereas noise events in the 
higher ranges fulfill two criteria  \citep[from][]{moissl2005}.
}
\begin{center}
{\small
  \begin{tabular}{lcccc}
 & & & & \\[-4.5ex]
   \hline
   \hline \\[-2.5ex]

Class, AR              &     
EA - IA                &      
                       &
CA                     &    
EIC                   \\
\hline \\[-2.5ex]
Class~1, AR1             & 
$\leq$ 2 or $\geq 9$     & 
or                       &
$\leq$ 2                 &
--                        \\
Class~1, AR2-6           & 
$\leq$ 2 or $\geq 9$     & 
and                      &
$\leq$ 2                 &
--                        \\
Class~2, AR1             & 
--                       &
                         &
--                       &
= 0                     \\
Class~2, AR2-6           &
$\leq$ 1 or $\geq 7$     &
and                      &
$\leq$ 2                 &
--                        \\[0.3ex]
\hline
\hline
 & & & & \\[-5.5ex]
\end{tabular}
}
\end{center}
\end{table}

\clearpage

%------- Begin Eventstabelle
{\renewcommand{\baselinestretch}{0.6} % Zeilenabstand 0.8 * Defaultwert

\thispagestyle{empty}
\begin{sidewaystable}
\tabcolsep1.5mm
\tiny
\vbox{
%\hspace{-2cm}
\begin{minipage}[t]{14cm}   
\caption{\tiny \label{rate_table} Overview of dust impacts accumulated with Galileo DDS 
between 1 January 2000 and 21 September 2003. 
The jovicentric distance $D_{Jup}$, 
the length of the time interval $\Delta $t (days) from the previous 
table entry, and the corresponding numbers of impacts are given for the 
class 2 and 3 accumulators. The accumulators are arranged with increasing signal 
amplitude ranges (AR), 
e.g.~AC31 means counter for CLN = 3 and AR = 1. The determination of the 
noise contamination $\rm f_{noi}$ in class~2 is described in Paper~VI. 
The $\Delta $t in the first line (day 00-002) 
is the time interval counted from the last entry in Table~2 in 
Paper~VIII. 
The totals of  counted impacts, of impacts with 
complete data, and of all events (noise plus impact events) 
for the entire period are given as well.}
\end{minipage}
}
\bigskip
%\hspace{-2cm}
\begin{tabular}{|r|r|r|r|ccc|ccc|ccc|ccc|}
\hline
&&&& &&&& &&&& &&&  \\
\mc{1}{|c|}{Date}&
\mc{1}{|c|}{Time}&
\mc{1}{|c|}{$D_{Jup}$}&
\mc{1}{|c|}{$\Delta $t}&
%\mc{4}{|c|}{AR-1}&
%\mc{4}{|c|}{AR-2}&
%\mc{4}{|c|}{AR-3}&
%\mc{4}{|c|}{AR-4}&
%\mc{4}{|c|}{AR-5}&
%\mc{4}{|c|}{AR-6}\\
{\scriptsize $\rm f_{noi,AC21}$}&{\scriptsize AC}&{\scriptsize AC}&{\scriptsize $\rm f_{noi,AC22}$}&
{\scriptsize AC}&{\scriptsize AC}&{\scriptsize $\rm f_{noi,AC23}$}&{\scriptsize AC}&
{\scriptsize AC}&{\scriptsize $\rm f_{noi,AC24}$}&{\scriptsize AC}&{\scriptsize AC}\\
%{\scriptsize $\rm f_{noi,AC25}$}&{\scriptsize AC}&{\scriptsize AC}&{\scriptsize $\rm f_{noi,AC26}$}&
%{\scriptsize AC}&{\scriptsize AC}\\
&&[$\rm R_J$]&\mc{1}{c|}{[d]}&
&$21^{\ast}$&$31^{\ast}$&
&22&32&
&23&33&
&24&34\\
%&25&35&
%&26&36\\
&& &&&& &&&& &&&& &\\
\hline
&& &&&& &&&& &&&& &\\
00-002&23:46&      19.28& 21.06& 0.81&8&-&-&-&-&-&-&-&-&-&-\\
00-003&21:55&      7.370& 0.922& 0.27&521&40& 0.00&1&-&-&-&-&-&-&-\\
00-005&10:08&      20.76& 1.508& 0.54&552&9& 0.50&2&1& 0.00&1&2&-&-&-\\
00-052&15:19&      15.96& 47.21& 0.77&47&-& 0.00&1&1&-&-&-& 0.00&1&1\\
00-053&10:26&      6.120& 0.796& 0.56&301&8& 0.00&4&1&-&-&-&-&-&-\\
 && &&&& &&&& &&&& &\\
00-056&22:40&      42.65& 3.510& 0.80&869&-& 0.80&20&2&-&-&-& 0.00&3&2\\
00-074&20:37&      126.1& 17.91& 0.59&17&-& 0.50&2&1& 0.00&1&-&-&-&-\\
00-141&14:34&      12.54& 66.74& 0.96&20&2& 0.25&4&-&-&-&-&-&-&-\\
00-143&23:11&      27.10& 2.359& 0.50&209&-& 0.00&2&-&-&-&-&-&-&-\\
00-149&19:41&      76.24& 5.853& 0.10&89&3&-&-&-&-&-&-&-&-&-\\
 && &&&& &&&& &&&& &\\
00-177&07:37&      193.7& 27.49& 0.23&42&3& 0.50&2&-&-&-&-&-&-&-\\
00-206&19:08&      256.7& 29.47& 0.72&5383&160&-&-&-&-&-&1&-&-&-\\
00-216&23:50&      270.0& 10.19& 0.00&4422&115&-&-&-&-&-&-&-&-&-\\
00-220&17:43&      274.0& 3.745& 0.00&13922&522& 1.00&1&-&-&-&-&-&-&-\\
00-222&23:09&      276.2& 2.226& 0.00&3964&219&-&-&-&-&-&-&-&-&-\\
 && &&&& &&&& &&&& &\\
00-228&05:08&      280.6& 5.249& 0.00&2648&203&-&-&-&-&-&-&-&-&-\\
00-231&18:47&      283.1& 3.568& 0.00&5590&424&-&-&-&-&-&-&-&-&-\\
00-236&10:15&      285.7& 4.644& 0.00&10750&842&-&-&-&-&-&-&-&-&-\\
00-246&03:28&      289.1& 9.717& 0.00&33125&3489& 0.00&2&-&-&-&-&-&-&-\\
00-256&11:57&      289.6& 10.35& 0.01&49616&3149& 0.00&1&-&-&-&-&-&-&-\\
 && &&&& &&&& &&&& &\\
00-266&13:29&      287.1& 10.06& 0.00&47774&2768&-&-&1&-&-&-&-&-&-\\
00-276&00:28&      281.8& 9.457& 0.00&17821&5196&-&-&-&-&-&-&-&-&-\\
00-300&11:26&      254.7& 24.45& 0.00&3200&163&-&-&-&-&-&-&-&-&-\\
00-306&12:10&      244.7& 6.030& 0.01&161&3&-&-&-&-&-&-&-&-&-\\
00-326&08:51&      200.4& 19.86& 0.05&223&9&-&-&-&-&-&-&-&-&-\\
 && &&&& &&&& &&&& &\\
00-363&16:33&      10.84& 37.32& 0.37&68&1& 0.00&1&1&-&-&-&-&-&-\\
00-366&00:28&      27.62& 2.329& 0.18&1014&41& 0.00&3&-&-&-&-&-&-&-\\
00-366&23:49&      37.39& 0.973& 0.04&345&19&-&-&-& 0.00&1&-&-&-&-\\
01-002&14:24&      51.27& 0.607& 0.09&513&47&-&-&-&-&-&-&-&-&-\\
01-006&08:59&      77.49& 3.774& 0.00&661&47&-&-&-&-&-&-&-&-&-\\
 && &&&& &&&& &&&& &\\
01-010&13:29&      100.5& 4.187& 0.05&312&15&-&-&-&-&-&-&-&-&-\\
01-015&12:45&      122.8& 4.969& 0.02&373&17&-&-&-&-&-&-&-&-&-\\
01-026&00:22&      158.6& 10.48& 0.07&365&24&-&-&-&-&-&-&-&-&-\\
01-040&21:05&      191.8& 14.86& 0.09&87&4&-&-&-&-&-&-&-&-&-\\
01-096&14:04&      198.7& 55.70& 0.11&59&3& 0.33&3&-&-&-&1&-&-&-\\
 && &&&& &&&& &&&& &\\
\hline
\end{tabular}
\end{sidewaystable}

\clearpage

\thispagestyle{empty}
\begin{sidewaystable}
\tabcolsep1.5mm
\tiny
\vbox{
%\hspace{-2cm}
\begin{minipage}[t]{14cm}   
Table~\ref{rate_table} continued.
\end{minipage}
}
\bigskip
%\hspace{-2cm}
\begin{tabular}{|r|r|r|r|ccc|ccc|ccc|ccc|}
\hline
&&&& &&&& &&&& && &\\
\mc{1}{|c|}{Date}&
\mc{1}{|c|}{Time}&
\mc{1}{|c|}{$D_{Jup}$}&
\mc{1}{|c|}{$\Delta $t}&
{\scriptsize $\rm f_{noi,AC21}$}&{\scriptsize AC}&{\scriptsize AC}&{\scriptsize $\rm f_{noi,AC22}$}&
{\scriptsize AC}&{\scriptsize AC}&{\scriptsize $\rm f_{noi,AC23}$}&{\scriptsize AC}&
{\scriptsize AC}&{\scriptsize $\rm f_{noi,AC24}$}&{\scriptsize AC}&{\scriptsize AC}\\
%{\scriptsize $\rm f_{noi,AC25}$}&{\scriptsize AC}&{\scriptsize AC}&{\scriptsize $\rm f_{noi,AC26}$}&
%{\scriptsize AC}&{\scriptsize AC}\\
&&[$\rm R_J$]&\mc{1}{c|}{[d]}&
&$21^{\ast}$&$31^{\ast}$&
&22&32&
&23&33&
&24&34\\
%&25&35&
%&26&36\\
&& &&&& &&&& &&&& &\\
\hline
&& &&&& &&&& &&&& &\\
01-143&15:55&      7.390& 47.07& 0.40&3245&45& 0.44&9&2&-&-&-&-&-&-\\
01-146&06:32&      34.34& 2.609& 0.60&2913&129& 0.18&11&1&-&-&-& 0.00&2&-\\
01-154&16:58&      87.12& 8.434& 0.18&16888&756& 0.00&1&-&-&-&-&-&-&-\\
01-177&20:07&      136.1& 23.13& 0.00&38739&1940& 0.00&1&-&-&-&-&-&-&-\\
01-194&13:03&      124.8& 16.70& 0.00&4139&386&-&-&-&-&-&-&-&-&-\\
&& &&&& &&&& &&&& &\\
01-209&11:19&      76.61& 14.92& 0.01&394&31& 0.00&1&1&-&-&-&-&-&-\\
01-218&07:24&      6.280& 8.836& 0.47&351&6& 0.00&5&4& 0.00&1&1& 0.33&3&2\\
01-221&21:01&      44.16& 3.567& 0.45&207&8& 0.33&6&-&-&-&-& 0.00&2&-\\
01-246&07:41&      128.5& 24.44& 0.01&509&9& 0.00&1&-&-&-&1&-&-&-\\
01-288&01:25&      16.92& 41.73& 0.06&243&14&-&-&-&-&-&-&-&-&-\\
&& &&&& &&&& &&&& &\\
01-291&11:32&      34.37& 3.420& 0.73&303&1& 0.00&2&2& 0.00&2&-& 0.50&2&-\\
01-314&06:23&      137.8& 22.78& 0.33&51&-&-&-&-&-&-&-&-&-&-\\
01-348&23:00&      152.6& 34.69& 0.33&19&1& 0.00&2&-&-&-&-&-&-&-\\
02-017&02:52&      12.09& 33.16& 0.57&36&-& 0.00&1&-&-&-&1& 0.00&2&1\\
02-018&00:32&      8.820& 0.902& 0.86&341&-& 0.00&1&2&-&-&-& 0.00&2&2\\
&& &&&& &&&& &&&& &\\
02-032&03:41&      118.4& 14.13& 0.51&80&3& 0.20&5&2& 0.00&1&-&-&-&-\\
02-069&12:20&      247.7& 37.36& 0.30&47&1&-&-&-&-&-&-&-&-&-\\
02-106&03:02&      312.8& 36.61& 0.48&23&-&-&-&-&-&-&-&-&-&-\\
02-151&13:43&      346.5& 45.44& 0.17&29&1& 0.00&1&-&-&-&-&-&-&-\\
02-184&10:32&      343.9& 32.86& 0.27&22&-&-&-&-&-&-&-&-&-&-\\
&& &&&& &&&& &&&& &\\
02-222&06:37&      312.7& 37.83& 0.88&24&1&-&-&-&-&-&-&-&-&-\\
02-265&21:04&      228.9& 43.60& 0.51&19&3&-&-&1&-&-&-&-&-&-\\
02-306&10:19&      42.73& 40.55& 0.06&18&-& 0.00&1&-&-&-&1&-&-&-\\
02-308&21:03&      11.18& 2.447& 0.52&55&3&-&-&2&-&-&1&-&-&-\\
02-309&04:21&      4.540& 0.304& 0.80&17&-& 0.00&2&2& 0.00&3&1&-&-&2\\
&& &&&& &&&& &&&& &\\
02-309&05:32&      3.320& 0.048& 0.83&77&-&-&-&-& 0.00&2&-&-&-&-\\
02-309&06:34&      2.410& 0.043& 0.83&512&5& 0.14&21&1& 0.00&21&1& 0.00&15&5\\
02-323&16:36&      121.7& 14.41& 0.67&9841&80& 0.00&191&2& 0.00&46&-& 0.00&62&2\\
02-357&20:16&      245.2& 34.15& 0.04&92&6&-&-&-&-&-&-&-&-&1\\
03-066&18:08&      357.0& 73.91& 0.86&11&1&-&-&-&-&-&-&-&-&2\\
&& &&&& &&&& &&&& &\\
03-262&11:23&      38.06& 195.7& 0.00&28&1& 0.00&2&5& 0.00&1&-&-&-&1\\
03-264&14:11&      7.190& 2.117& 0.00&36&-& 0.00&14&1&-&-&-& 0.00&1&1\\
03-264&15:43&      5.500& 0.063& 0.53&41&-&-&-&1& 0.00&1&2&-&-&1\\
03-264&17:59&      2.540& 0.094& 0.53&124&-& 0.00&45&8& 0.42&26&10& 0.12&68&36\\
\hline 
\mc{4}{|l|}{}& && &&&& &&&& &\\
\mc{4}{|l|}{Events (counted)}&- &284545&20976&- &372&45&- &107&23&- &163&59\\[1.5ex]
\mc{4}{|l|}{Impacts (complete data)}&-&3291&1865&-&68&38&-&36&16&-&26&28\\[1.5ex]
\mc{4}{|l|}{All events(complete data)}& 0.22&4229&1865& 0.28&95&41& 0.05&38&16& 0.07&28&28\\[1.5ex]
\hline
\end{tabular}
\\  
{\footnotesize $\ast$: AC21 and AC31: Overflows of the 8 bit accumulators were counted with overflow counters so
that no unrecognized overflows occurred in these two channels. \\ $\rm f_{noi}$ has been estimated from 
the data sets transmitted. }
 \\
\end{sidewaystable}

\clearpage

 \thispagestyle{empty}                                                              

 \begin{sidewaystable} 
\tabcolsep1.2mm
\tiny %\hspace{-2cm}
\vbox{
\vspace{3cm}
\begin{minipage}[t]{22.5cm}      
%{\bf Table 5.}
 \caption{\label{dust_impacts}
 DPF data: No., impact time,  TEV (in minutes)
 CLN, AR, SEC, IA, EA, CA, IT, ET, 
 EIT, EIC, ICC, PA, PET, EVD, ICP, ECP, CCP, PCP, HV and 
 evaluated data: LON, LAT, $\rm D_{Jup}$ (in $\mathrm{R_J}$), rotation angle (ROT), instr. pointing ($\rm S_{LON}$, $\rm S_{LAT}$), 
 speed v (in $\mathrm{km\,s^{-1}}$), 
 speed error factor (VEF), mass m (in g) and mass error factor (MEF). Velocity and mass should be 
 considered as lower and upper limits for the true values, respectively, because of the
 strong degradation of the dust instrument electronics (see text for details).
 }
\end{minipage}
}

\bigskip

{\tiny      
\hspace{-2cm}                                                            
\begin{tabular*}{26cm}  {@{\extracolsep{\fill}}                                 
rrrrrrrrrrrrrrrcrrrrcrrrrrrrrrrrrrrrrrrrrr}                                     
 \mc{1}{c}{No.}&               \mc{1}{c}{IMP. DATE}&         \mc{1}{c}{ TEV }&             \mc{1}{c}{C}&                 
 \mc{1}{c}{AR}&                \mc{1}{c}{S}&                 \mc{1}{c}{IA}&                \mc{1}{c}{EA}&                
 \mc{1}{c}{CA}&                \mc{1}{c}{IT}&                \mc{1}{c}{ET}&                \mc{1}{c}{E}&                 
 \mc{1}{c}{E}&                 \mc{1}{c}{I}&                 \mc{1}{c}{PA}&                \mc{1}{c}{P}&                 
 \mc{1}{c}{E}&                 \mc{1}{c}{I}&                 \mc{1}{c}{E}&                 \mc{1}{c}{C}&                 
 \mc{1}{c}{P}&                 \mc{1}{c}{HV}&                \mc{1}{c}{LON}&               \mc{1}{c}{LAT}&               
 \mc{1}{c}{$\rm D_{Jup}$}&     \mc{1}{c}{ROT}&               \mc{1}{c}{$\rm S_{LON}$}&     \mc{1}{c}{$\rm S_{LAT}$}&     
 \mc{1}{c}{V}&                 \mc{1}{c}{VEF}&               \mc{1}{c}{M}&                 \mc{1}{c}{MEF}\\
\hline                                                      
\end{tabular*}} \end{sidewaystable}                                                     
\clearpage       
\thispagestyle{empty}                                                               
\begin{sidewaystable}                                                                   
{\tiny 
\hspace{-2cm} 
\vspace{-3cm}                                                                  
\begin{tabular*}{26cm}  {@{\extracolsep{\fill}}                                 
rrrrrrrrrrrrrrrcrrrrcrrrrrrrrrrrrrrrrrrrrr}                                     
 \mc{1}{c}{No.}&               \mc{1}{c}{IMP. DATE}&         \mc{1}{c}{ TEV }&             \mc{1}{c}{C}&                 
 \mc{1}{c}{AR}&                \mc{1}{c}{S}&                 \mc{1}{c}{IA}&                \mc{1}{c}{EA}&                
 \mc{1}{c}{CA}&                \mc{1}{c}{IT}&                \mc{1}{c}{ET}&                \mc{1}{c}{E}&                 
 \mc{1}{c}{E}&                 \mc{1}{c}{I}&                 \mc{1}{c}{PA}&                \mc{1}{c}{P}&                 
 \mc{1}{c}{E}&                 \mc{1}{c}{I}&                 \mc{1}{c}{E}&                 \mc{1}{c}{C}&                 
 \mc{1}{c}{P}&                 \mc{1}{c}{HV}&                \mc{1}{c}{LON}&               \mc{1}{c}{LAT}&               
 \mc{1}{c}{$\rm D_{Jup}$}&     \mc{1}{c}{ROT}&               \mc{1}{c}{$\rm S_{LON}$}&     \mc{1}{c}{$\rm S_{LAT}$}&     
 \mc{1}{c}{V}&                 \mc{1}{c}{VEF}&               \mc{1}{c}{M}&                 \mc{1}{c}{MEF}\\
\hline                                                      
\end{tabular*}} \end{sidewaystable}                                                     
\clearpage  
\thispagestyle{empty}                                                                    
\begin{sidewaystable}                                                                   
{\tiny        
\hspace{-2cm}  
\vspace{-3cm}                                                          
\begin{tabular*}{26cm}  {@{\extracolsep{\fill}}                                 
rrrrrrrrrrrrrrrcrrrrcrrrrrrrrrrrrrrrrrrrrr}                                     
 \mc{1}{c}{No.}&               \mc{1}{c}{IMP. DATE}&         \mc{1}{c}{ TEV }&             \mc{1}{c}{C}&                 
 \mc{1}{c}{AR}&                \mc{1}{c}{S}&                 \mc{1}{c}{IA}&                \mc{1}{c}{EA}&                
 \mc{1}{c}{CA}&                \mc{1}{c}{IT}&                \mc{1}{c}{ET}&                \mc{1}{c}{E}&                 
 \mc{1}{c}{E}&                 \mc{1}{c}{I}&                 \mc{1}{c}{PA}&                \mc{1}{c}{P}&                 
 \mc{1}{c}{E}&                 \mc{1}{c}{I}&                 \mc{1}{c}{E}&                 \mc{1}{c}{C}&                 
 \mc{1}{c}{P}&                 \mc{1}{c}{HV}&                \mc{1}{c}{LON}&               \mc{1}{c}{LAT}&               
 \mc{1}{c}{$\rm D_{Jup}$}&     \mc{1}{c}{ROT}&               \mc{1}{c}{$\rm S_{LON}$}&     \mc{1}{c}{$\rm S_{LAT}$}&     
 \mc{1}{c}{V}&                 \mc{1}{c}{VEF}&               \mc{1}{c}{M}&                 \mc{1}{c}{MEF}\\
\hline                                                      
\end{tabular*}} \end{sidewaystable}                                                     
\clearpage    
\thispagestyle{empty}                                                                  
\begin{sidewaystable}                                                                   
{\tiny     
\hspace{-2cm}  
\vspace{-3cm}                                                             
\begin{tabular*}{26cm}  {@{\extracolsep{\fill}}                                 
rrrrrrrrrrrrrrrcrrrrcrrrrrrrrrrrrrrrrrrrrr}                                     
 \mc{1}{c}{No.}&               \mc{1}{c}{IMP. DATE}&         \mc{1}{c}{ TEV }&             \mc{1}{c}{C}&                 
 \mc{1}{c}{AR}&                \mc{1}{c}{S}&                 \mc{1}{c}{IA}&                \mc{1}{c}{EA}&                
 \mc{1}{c}{CA}&                \mc{1}{c}{IT}&                \mc{1}{c}{ET}&                \mc{1}{c}{E}&                 
 \mc{1}{c}{E}&                 \mc{1}{c}{I}&                 \mc{1}{c}{PA}&                \mc{1}{c}{P}&                 
 \mc{1}{c}{E}&                 \mc{1}{c}{I}&                 \mc{1}{c}{E}&                 \mc{1}{c}{C}&                 
 \mc{1}{c}{P}&                 \mc{1}{c}{HV}&                \mc{1}{c}{LON}&               \mc{1}{c}{LAT}&               
 \mc{1}{c}{$\rm D_{Jup}$}&     \mc{1}{c}{ROT}&               \mc{1}{c}{$\rm S_{LON}$}&     \mc{1}{c}{$\rm S_{LAT}$}&     
 \mc{1}{c}{V}&                 \mc{1}{c}{VEF}&               \mc{1}{c}{M}&                 \mc{1}{c}{MEF}\\
\hline                                                      
\end{tabular*}} \end{sidewaystable}                                                     
\clearpage 
\thispagestyle{empty}                                                                     
\begin{sidewaystable}                                                                   
{\tiny            
\hspace{-2cm}    
\vspace{-3cm}                                                    
\begin{tabular*}{26cm}  {@{\extracolsep{\fill}}                                 
rrrrrrrrrrrrrrrcrrrrcrrrrrrrrrrrrrrrrrrrrr}                                     
 \mc{1}{c}{No.}&               \mc{1}{c}{IMP. DATE}&         \mc{1}{c}{ TEV }&             \mc{1}{c}{C}&                 
 \mc{1}{c}{AR}&                \mc{1}{c}{S}&                 \mc{1}{c}{IA}&                \mc{1}{c}{EA}&                
 \mc{1}{c}{CA}&                \mc{1}{c}{IT}&                \mc{1}{c}{ET}&                \mc{1}{c}{E}&                 
 \mc{1}{c}{E}&                 \mc{1}{c}{I}&                 \mc{1}{c}{PA}&                \mc{1}{c}{P}&                 
 \mc{1}{c}{E}&                 \mc{1}{c}{I}&                 \mc{1}{c}{E}&                 \mc{1}{c}{C}&                 
 \mc{1}{c}{P}&                 \mc{1}{c}{HV}&                \mc{1}{c}{LON}&               \mc{1}{c}{LAT}&               
 \mc{1}{c}{$\rm D_{Jup}$}&     \mc{1}{c}{ROT}&               \mc{1}{c}{$\rm S_{LON}$}&     \mc{1}{c}{$\rm S_{LAT}$}&     
 \mc{1}{c}{V}&                 \mc{1}{c}{VEF}&               \mc{1}{c}{M}&                 \mc{1}{c}{MEF}\\
\hline                                                      
\end{tabular*}} \end{sidewaystable}                                                     
\vfill 

\clearpage 

%------- Begin Eventstabelle
{\renewcommand{\baselinestretch}{1.1} % Zeilenabstand 0.8 * Defaultwert

\setcounter{table}{5}
\begin{table}%[htb]
\caption
{\label{tab_slopes}
Slopes $\alpha$ of power law fits to the measured dust flux profiles and Io's dust
production.
}
\begin{center}
  \begin{tabular}{ccccc}
 & &  & &\\[-4.5ex]
   \hline
   \hline \\[-2.5ex]
Orbit & Distance  & Time interval & $\alpha$   & Dust Production\\
      &  ($\mathrm{R_J}$)&        &            &  ($\mathrm{kg\,s^{-1}}$)  \\
\hline 
G1    & 48 -- 15  & 96-175 to 96-179 &   --1.9  &  0.1 -- 2  \\
G2    & 128 -- 20 & 96-214 to 96-250 &   --4.3 &  0.2 -- 5   \\
C3    & 113 -- 20 & 96-284 to 96-310 &   --1.9 &  0.2 -- 10   \\
E4    & 49 -- 20  & 96-349 to 96-352 &   --0.7 &  3   -- 20   \\
E6    & 38 -- 15  & 97-048 to 97-051 &   --1.6 &  0.01 -- 0.07   \\
G7    & 59 -- 20  & 97-087 to 97-093 &   --3.5 &  0.1 -- 5   \\
G8    & 48 -- 20  & 97-123 to 97-127 &   --5.4 &  0.2 -- 3   \\
C9    & 50 -- 20  & 97-173 to 97-177 &   --2.0 &  0.07 -- 1   \\
C10   & 55 -- 20  & 97-256 to 97-260 &   --1.7 &  0.1 -- 5   \\
E11   & 57 -- 20  & 97-304 to 97-309 &   --3.7 &  0.1 -- 1   \\
E12   & 50 -- 20  & 97-344 to 97-348 &   --2.1 &  0.05 -- 0.7   \\
E14   & 48 -- 20  & 98-083 to 98-087 &   --6.1 &  0.05 -- 0.3   \\
E16   & 39 -- 20  & 98-198 to 98-200 &   --7.3 &  0.01 -- 0.5   \\
E17   & 44 -- 20  & 98-265 to 98-268 &   --1.4 &  0.05 -- 0.5   \\
E18   & 47 -- 33  & 98-321 to 98-323 &   --5.6 &  0.05 -- 0.5   \\
E19   & 48 -- 20  & 99-027 to 99-030 &   --0.6 &  0.05 -- 0.5   \\
E26   & 15 -- 30  & 00-005 to 00-006 &   --5.3 &  0.01 -- 0.05   \\
G28   & 22 -- 167 & 00-143 to 00-168 &   --1.8 &  0.1 -- 3   \\
G28   & 270 -- 290& 00-227 to 00-252 &    --   &  10 -- 1000 \\
G29   & 290 -- 172& 00-253 to 00-335 &   +10.4 &  0.1 -- 500  \\
G29   & 20 -- 216 & 01-365 to 01-070 &   --2.1 &  0.01 -- 1   \\
I32   & 22 -- 93  & 01-290 to 01-300 &    +0.2 &  0.005 -- 0.5 \\
I33   & 61 -- 348 & 02-023 to 02-164 &   --1.1 &  0.05 -- 5   \\
A34   & 348 -- 22 & 02-164 to 02-308 &   --0.4 &  0.05 -- 1   \\
\hline 
\hline
\end{tabular}
\end{center}
\end{table}

\clearpage

%\section{Figures}

\begin{figure}[ht]
\vspace{-8cm}
\hspace{-2cm}
\includegraphics[scale=0.88]{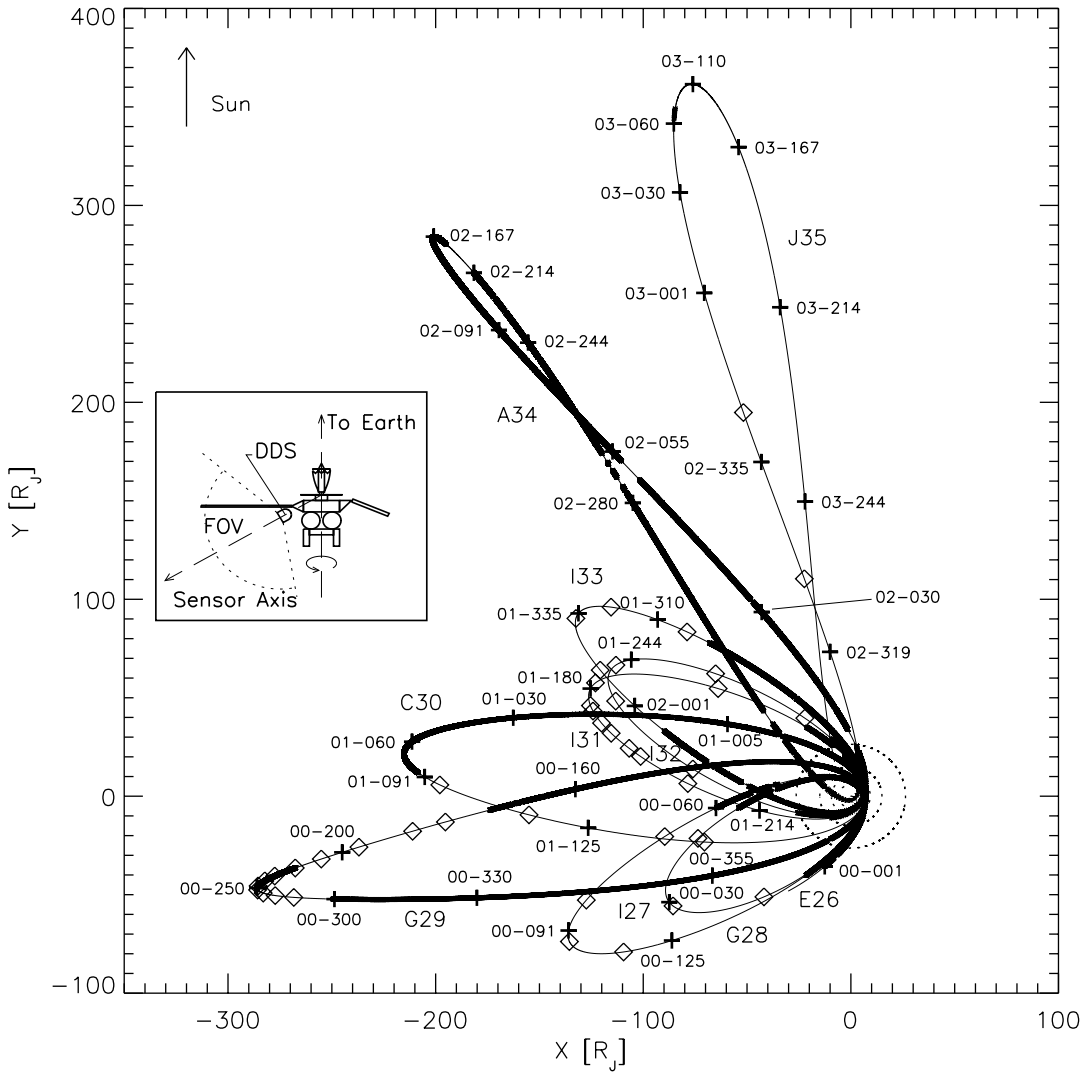}

\vspace{-1.5cm}
        \caption{\label{trajectory}
Galileo's trajectory in the jovian system from 2000 to 2003 in a 
Jupiter-centric coordinate system (thin solid line). 
Crosses mark the spacecraft position at approximately 30 day intervals 
(days of year are indicated).
Periods when RTS data were obtained are shown as thick solid lines,
MROs are marked by diamonds. Galileo's orbits are labelled 'E26', 'I27',
'G28', 'G29', 'C30', 'I31', 'I32', 'I33', 'A34' and 'J35'. 
Sun direction is to the top
and the Sun and Earth directions coincide to within $10^{\circ}$.
The orbits of the Galilean moons are indicated (dotted lines).
The sketch of the Galileo spacecraft shows the dust detector (DDS),
its geometry of dust detection and its field-of-view (FOV). 
The spacecraft antenna usually pointed towards Earth and the 
spacecraft made about 3 revolutions per minute. 
}
\end{figure}

\begin{figure}
\hspace{-1.5cm}
\parbox{0.99\hsize}{
\parbox{0.49\hsize}{
\vspace{-7cm}
\includegraphics[scale=0.45]{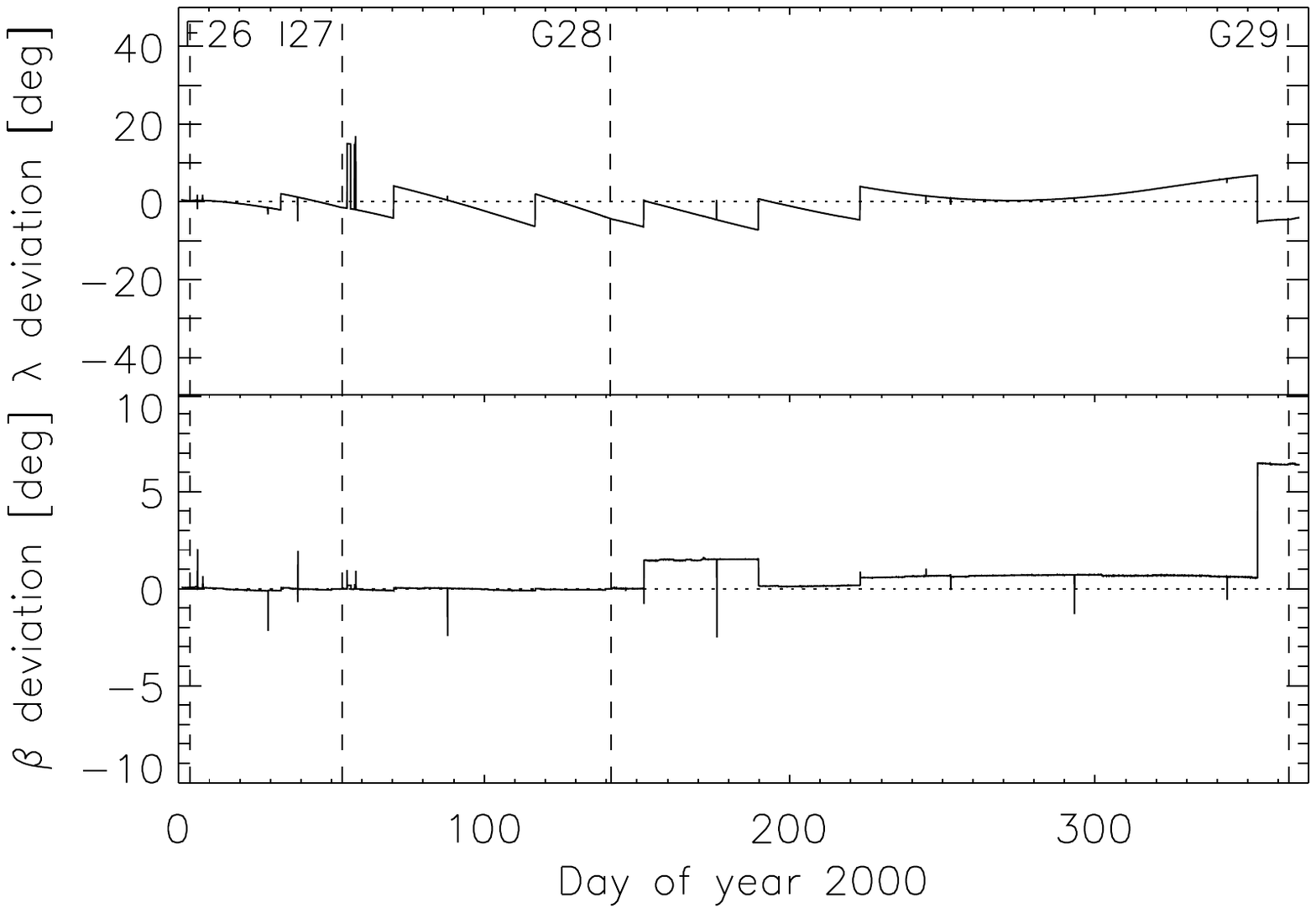}
}
\parbox{0.49\hsize}{
\vspace{-7cm}
\hspace{1cm}
\includegraphics[scale=0.45]{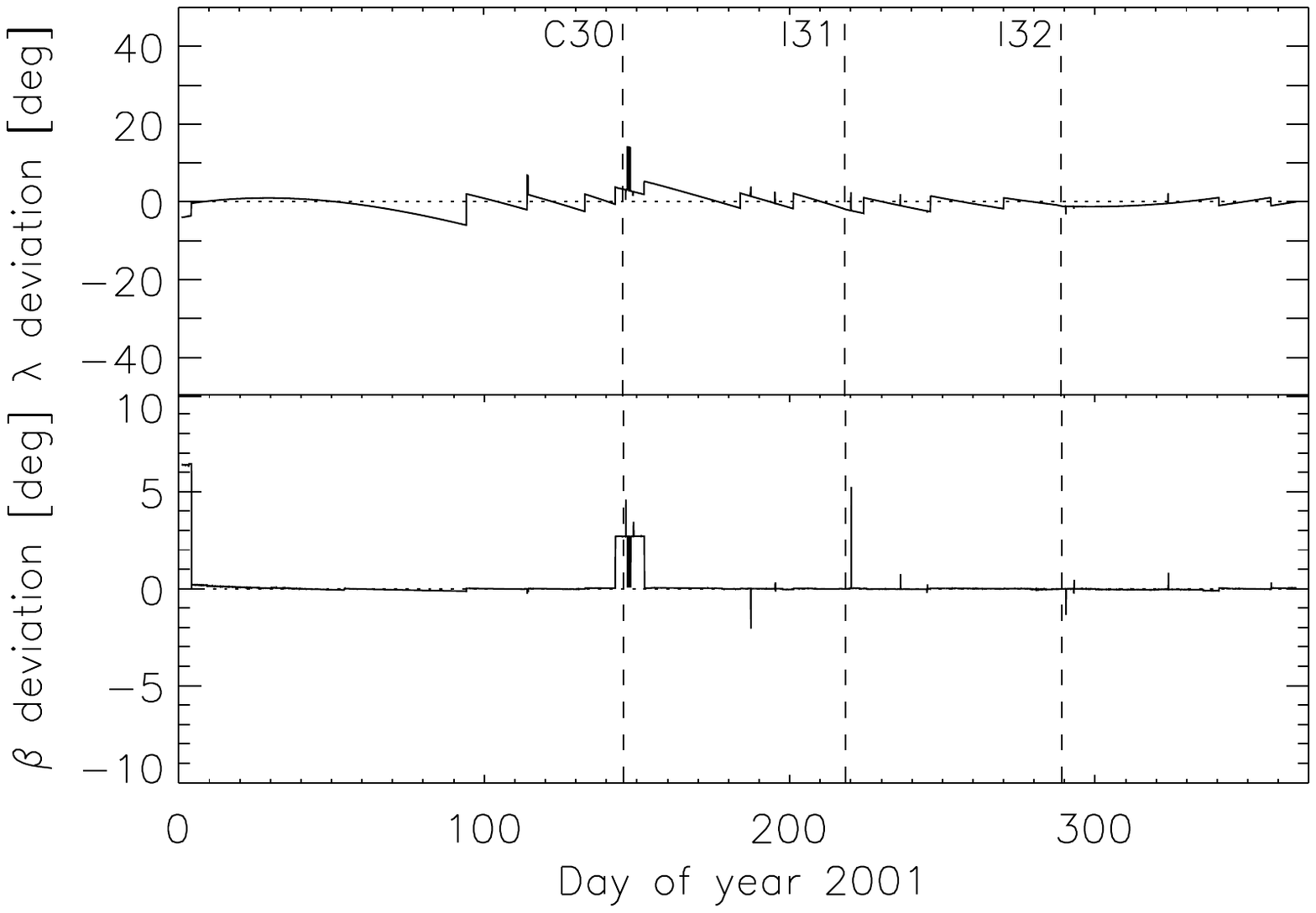}
}
\parbox{0.49\hsize}{
\vspace{-5cm}
\includegraphics[scale=0.45]{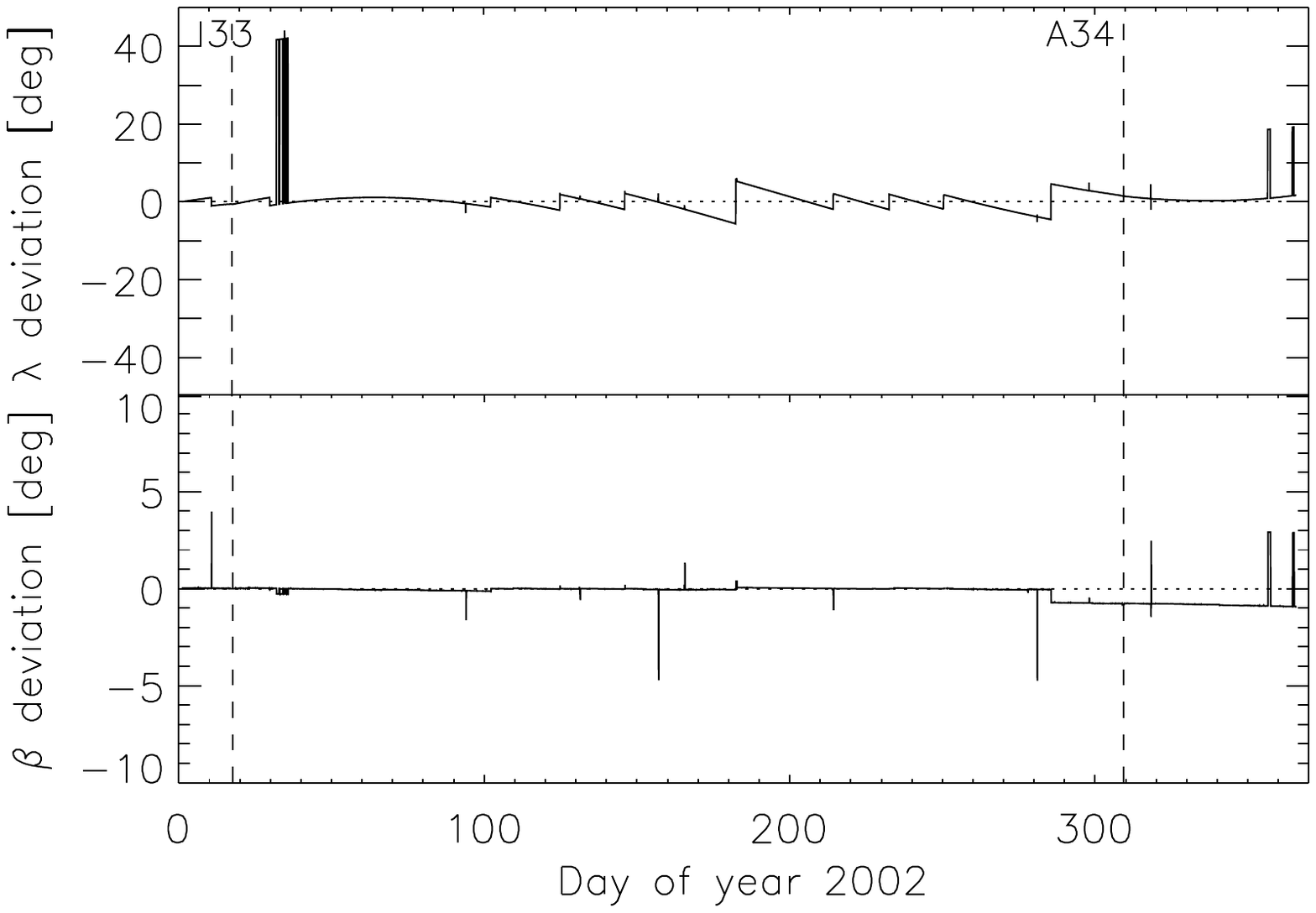}
}
\parbox{0.49\hsize}{
\vspace{-5cm}
\hspace{1.1cm}
\includegraphics[scale=0.45]{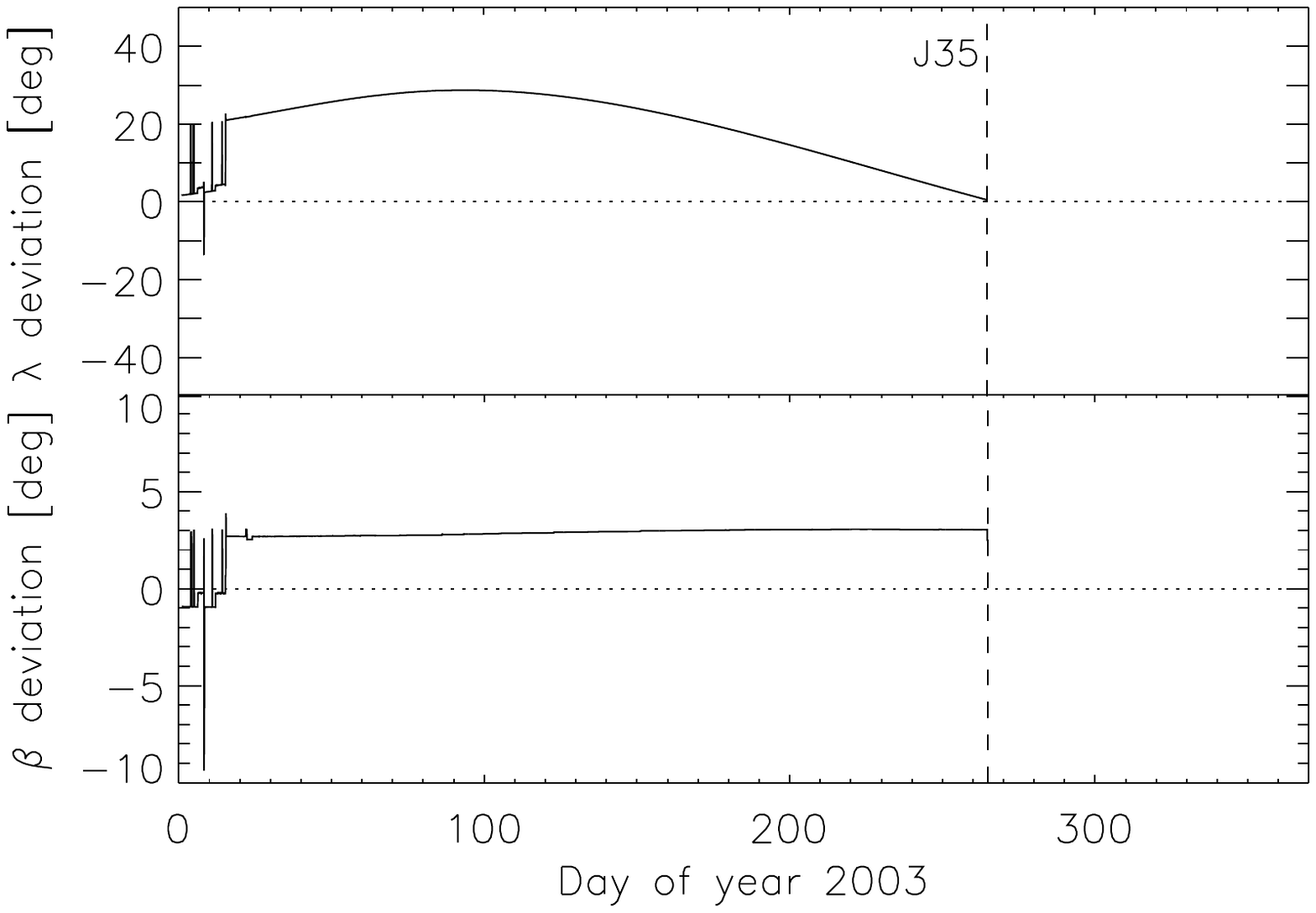}
}
}
        \caption{\label{pointing}
Spacecraft attitude: deviation of the antenna pointing direction 
(i.~e. negative spin axis) from the Earth direction. The angles are 
given in ecliptic longitude ($\lambda$) and latitude ($\beta$, equinox 1950.0).
Dashed vertical lines indicate satellite flybys (E26-A34) or Galileo's Jupiter impact 
(J35).
Sharp spikes are associated with imaging observations with 
Galileo's cameras or orbit trim maneuvers with the spacecraft thrusters.
}
\end{figure}

\begin{figure}
\hspace{-1.5cm}
\parbox{0.99\hsize}{
\parbox{0.49\hsize}{
\vspace{-3cm}
\includegraphics[scale=0.45]{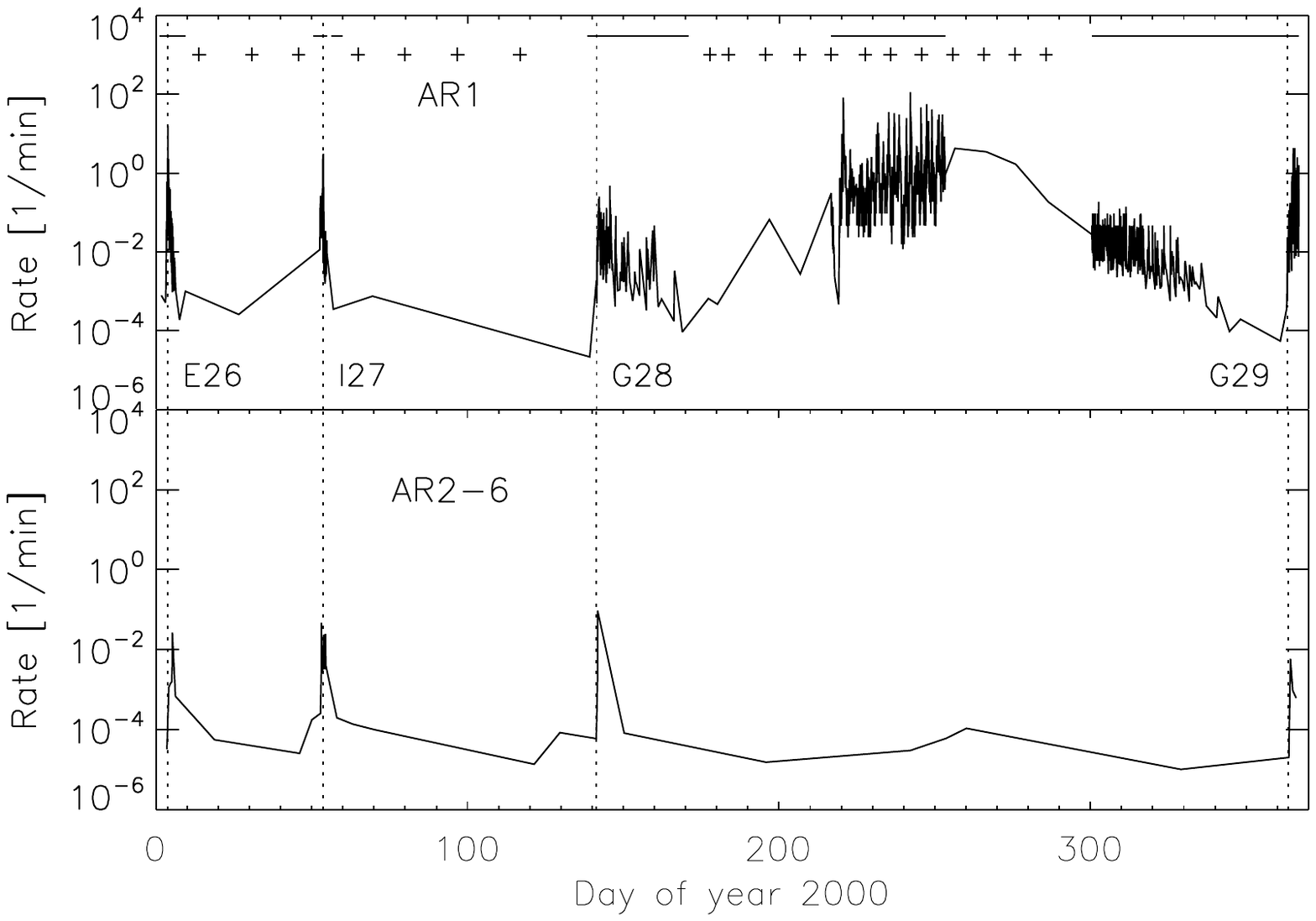}
}
\parbox{0.49\hsize}{
\vspace{-3cm}
\hspace{1cm}
\includegraphics[scale=0.45]{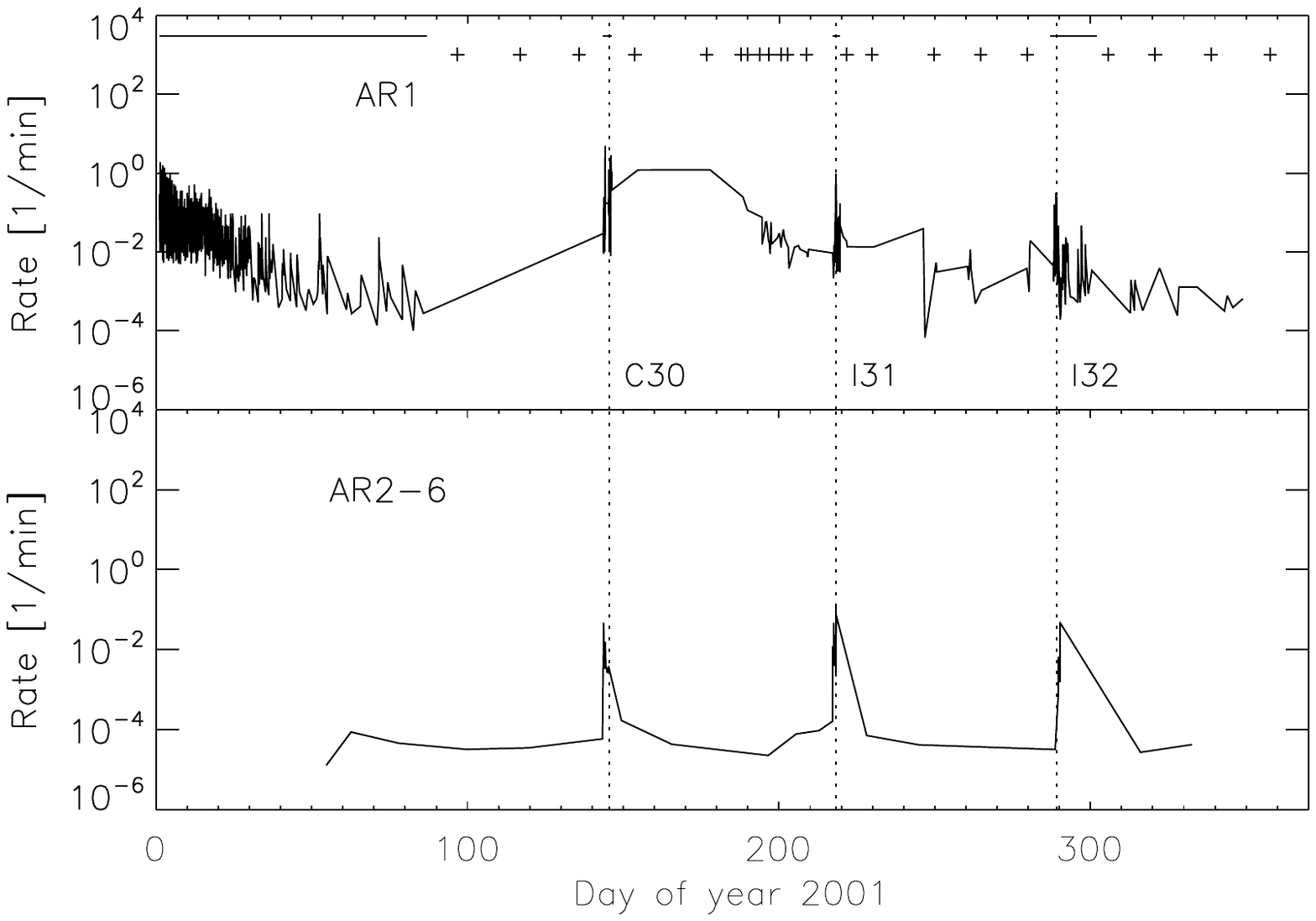}
}
\parbox{0.49\hsize}{
\vspace{-5.5cm}
\includegraphics[scale=0.45]{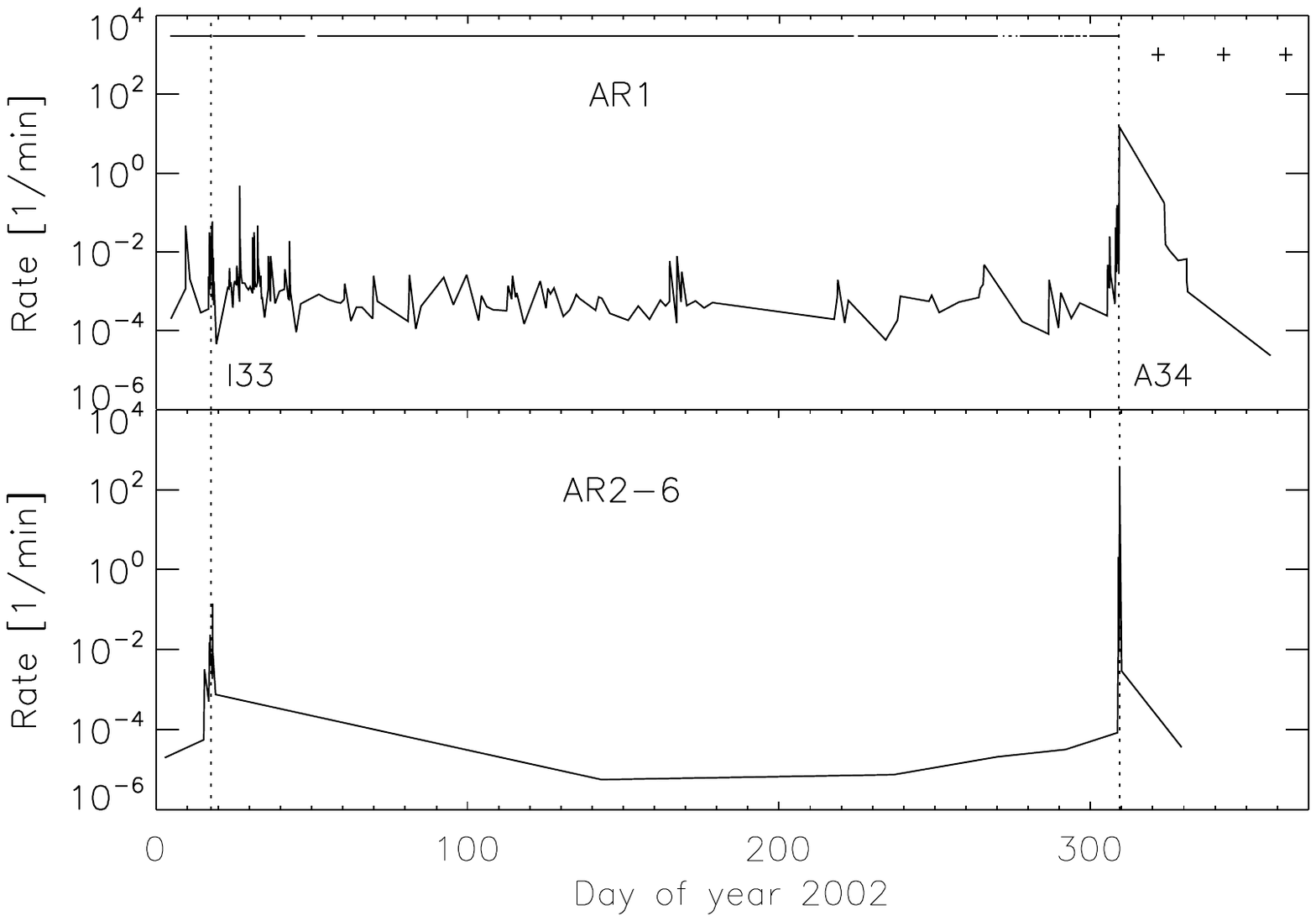}
}
\parbox{0.49\hsize}{
\vspace{-5.5cm}
\hspace{1.1cm}
\includegraphics[scale=0.45]{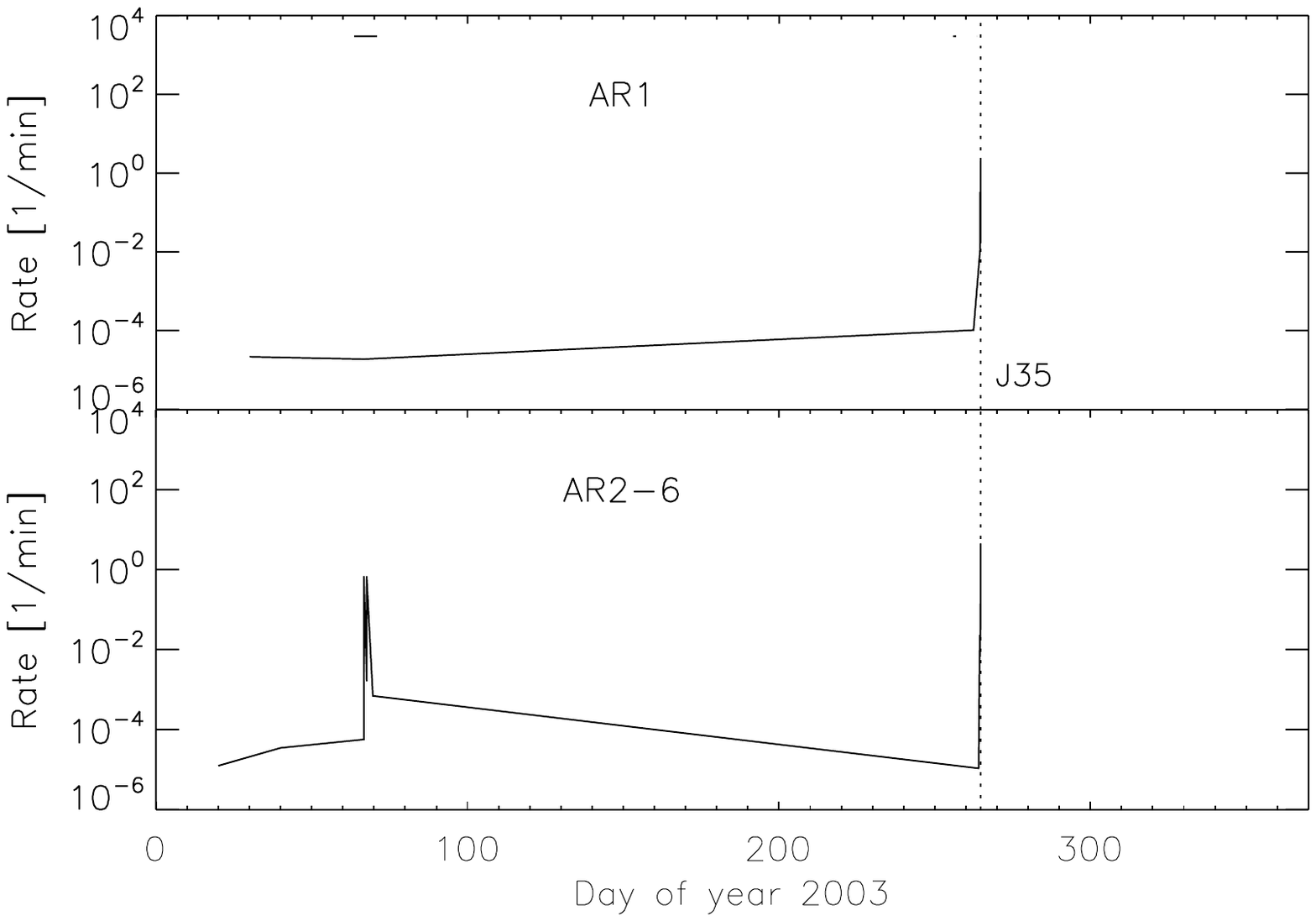}
}
}
\vspace{-5cm}
        \caption{\label{rate}
Dust impact rate detected in 2000-2003. For each year the top panel shows the 
impact rate in AR1 and the bottom panel that  
for the higher amplitude ranges AR2-6. Only data for classes 2 and 3 are shown. 
Dotted lines indicate satellite flybys (E26-A34) or Galileo's Jupiter impact 
(J35).
Perijove passages occurred within two days of the moon closest approaches.
These curves are plotted from the number of impacts with the highest time resolution
which is available only in electronic form. No smoothing was applied to the data. 
In the top panels
(AR1), time intervals with continuous RTS coverage are indicated by horizontal bars,
memory readouts (MROs) are marked by crosses.
}
\end{figure}

\begin{figure}
\hspace{-0.8cm}
\parbox{0.99\hsize}{
\parbox{0.55\hsize}{
\vspace{-9cm}
\includegraphics[scale=0.47]{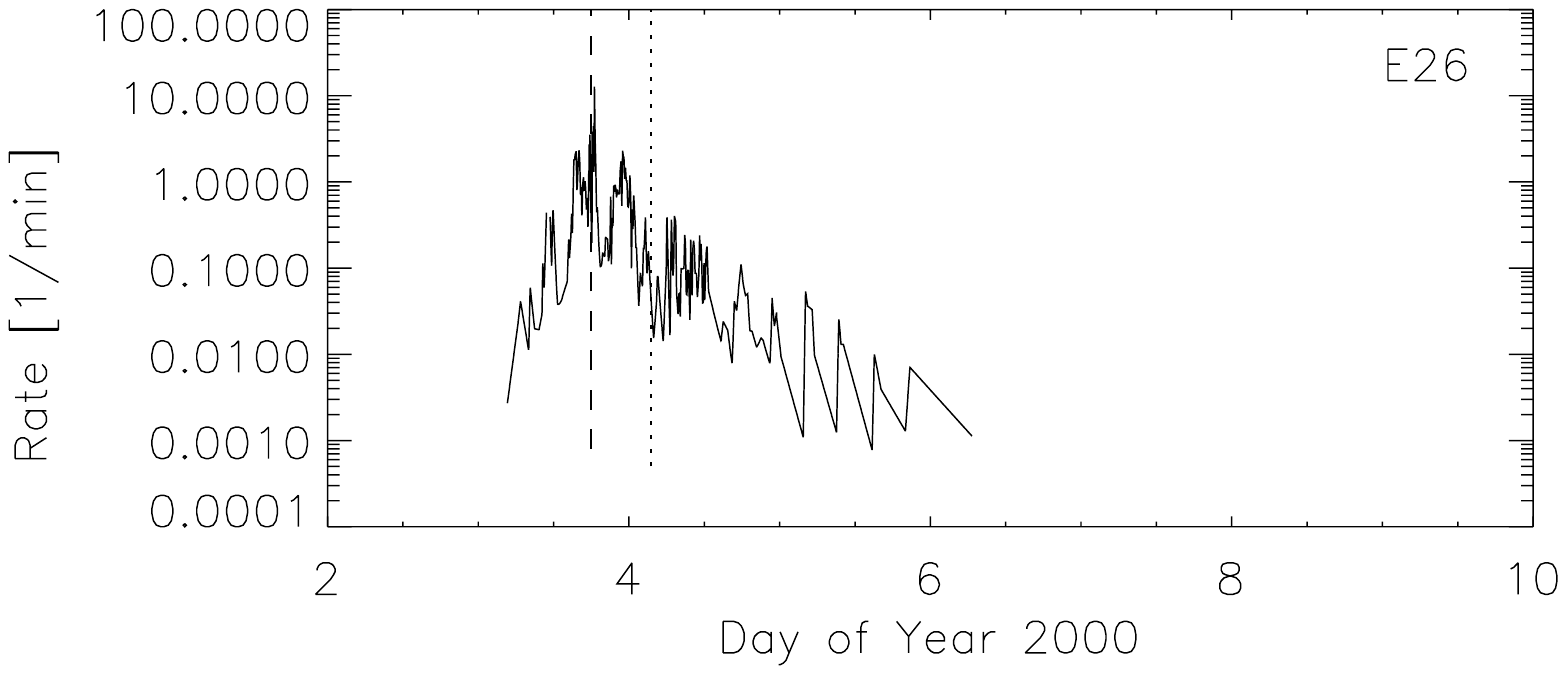}

\vspace{-9.5cm}
\includegraphics[scale=0.47]{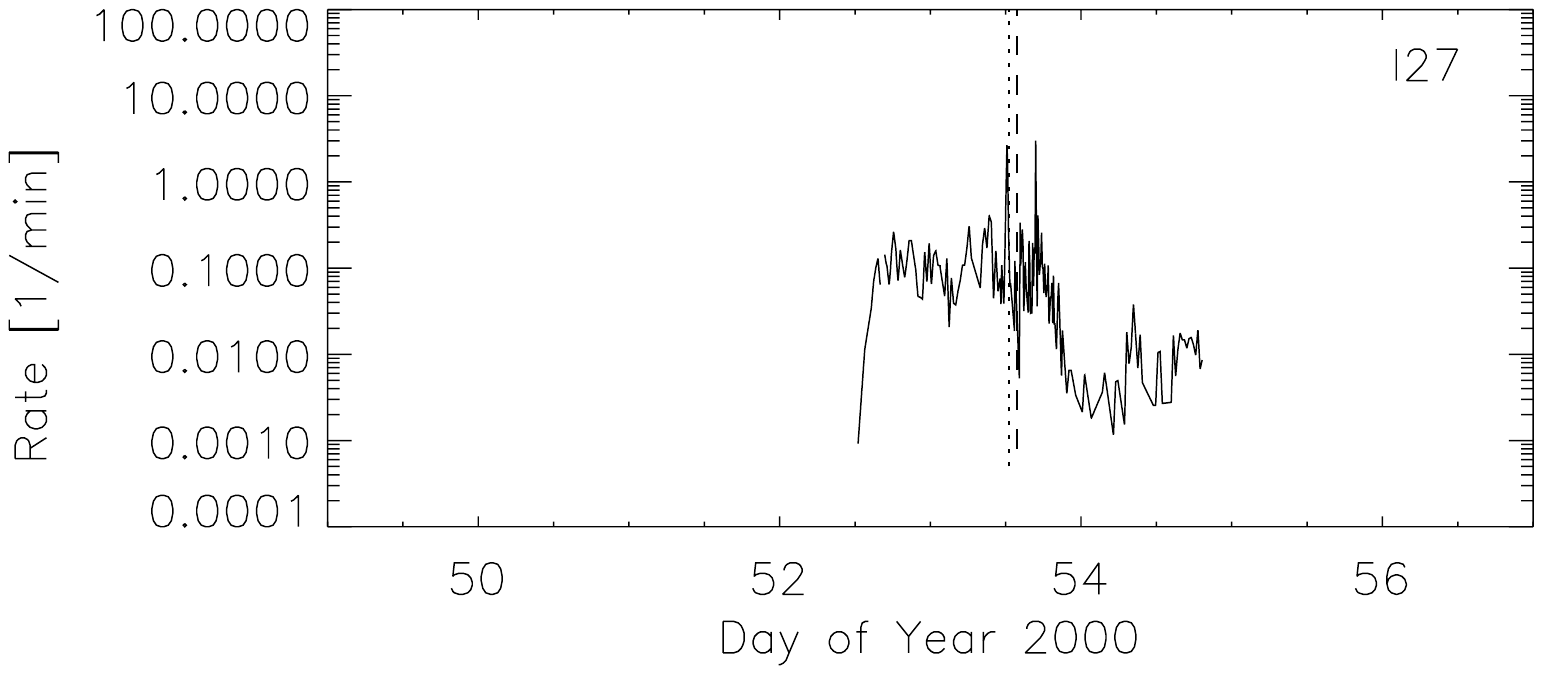}

\vspace{-9.5cm}
\includegraphics[scale=0.47]{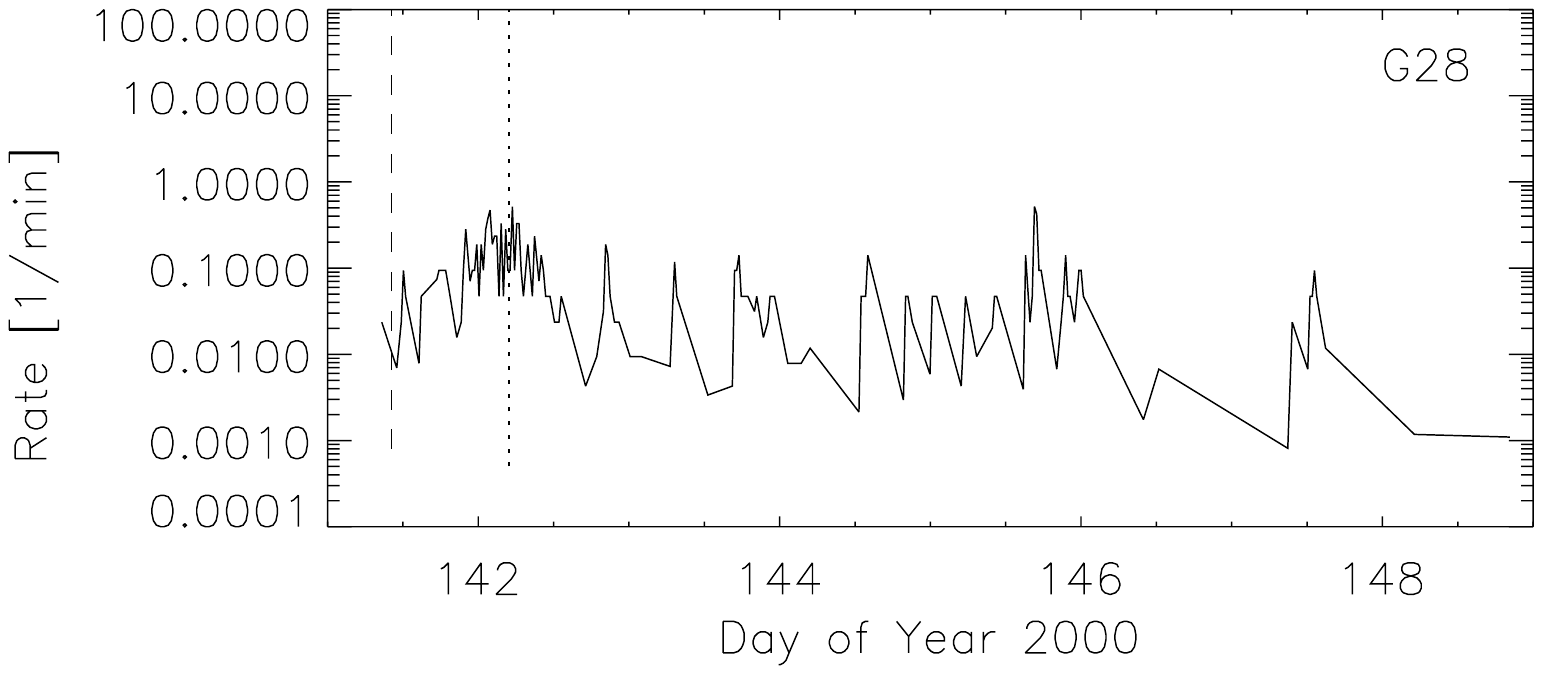}

\vspace{-9.5cm}
\includegraphics[scale=0.47]{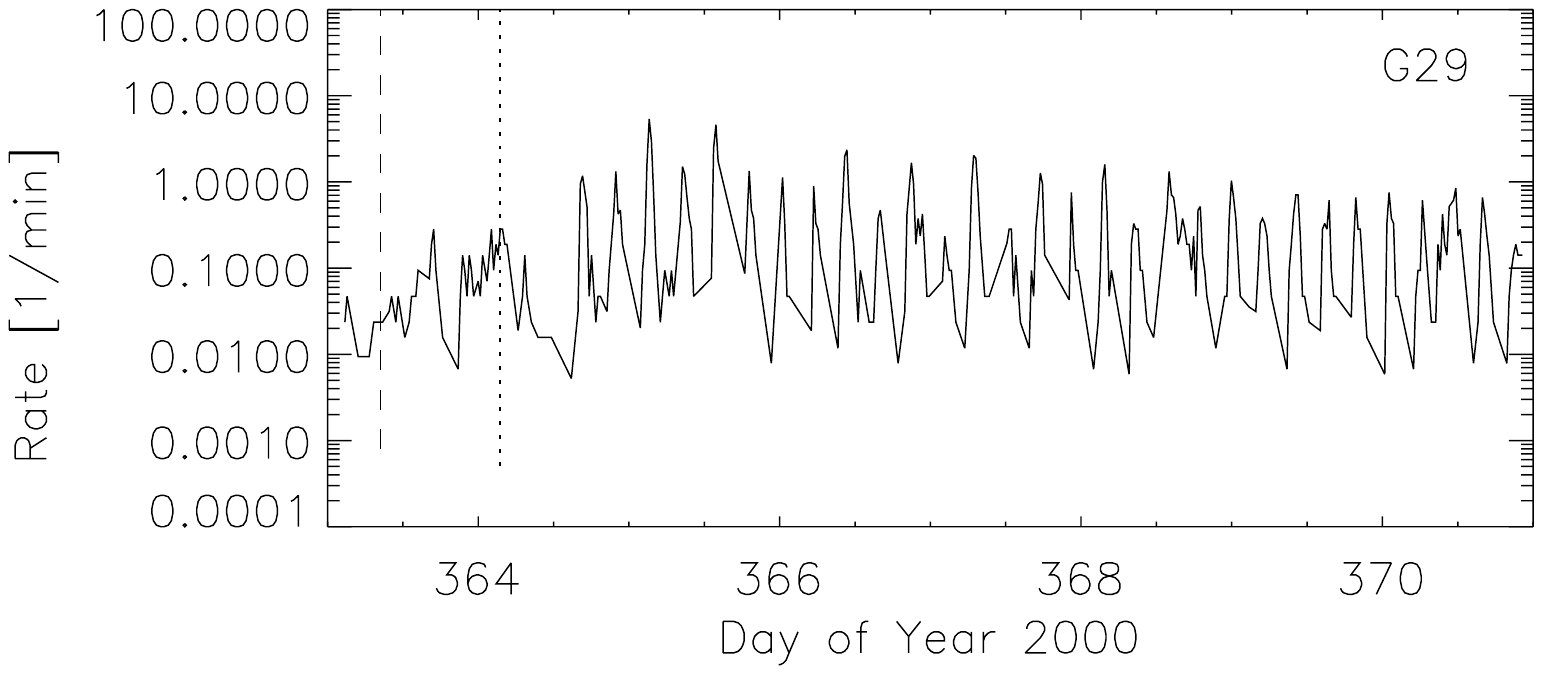}

\vspace{-9.5cm}
\includegraphics[scale=0.47]{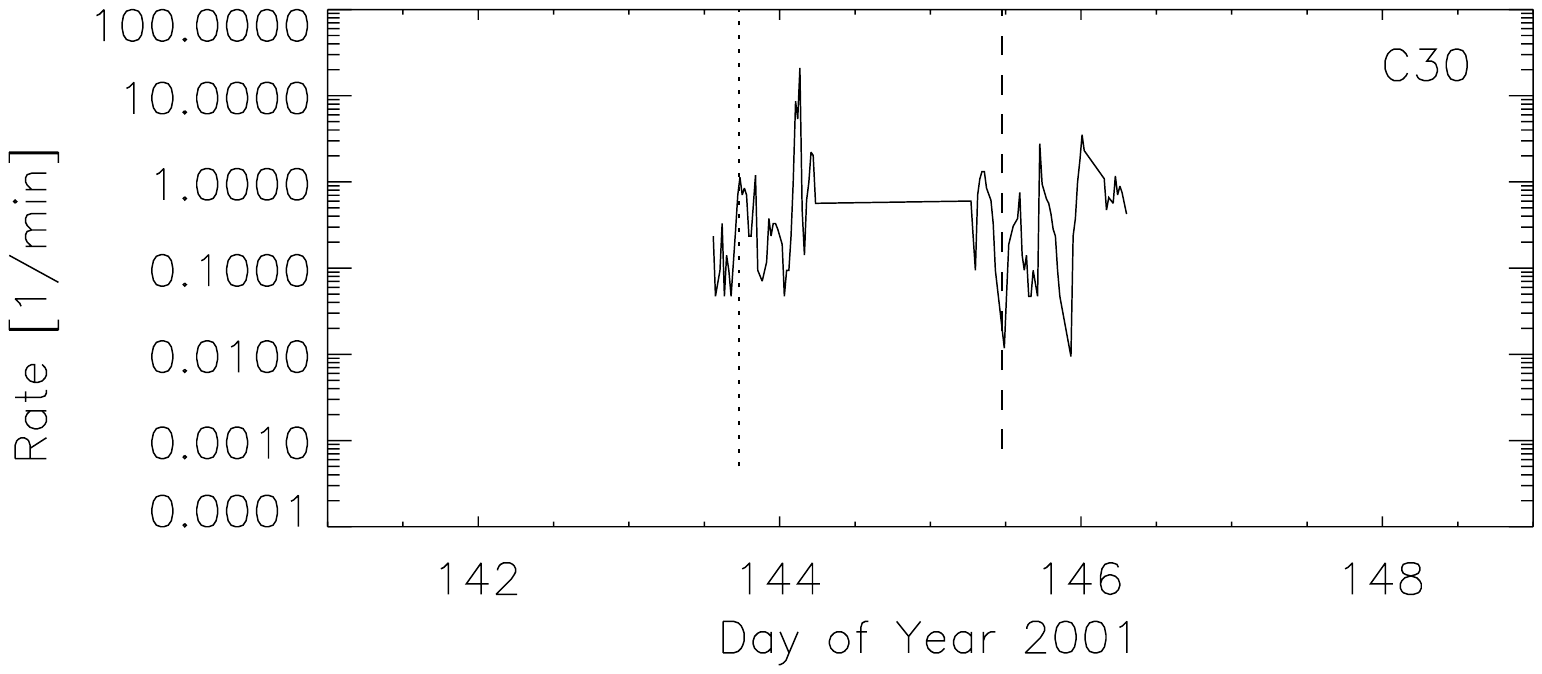}
\vspace{-1.5cm}
}
\parbox{0.55\hsize}{
%\hspace{0.7cm}
\vspace{-9cm}
\includegraphics[scale=0.47]{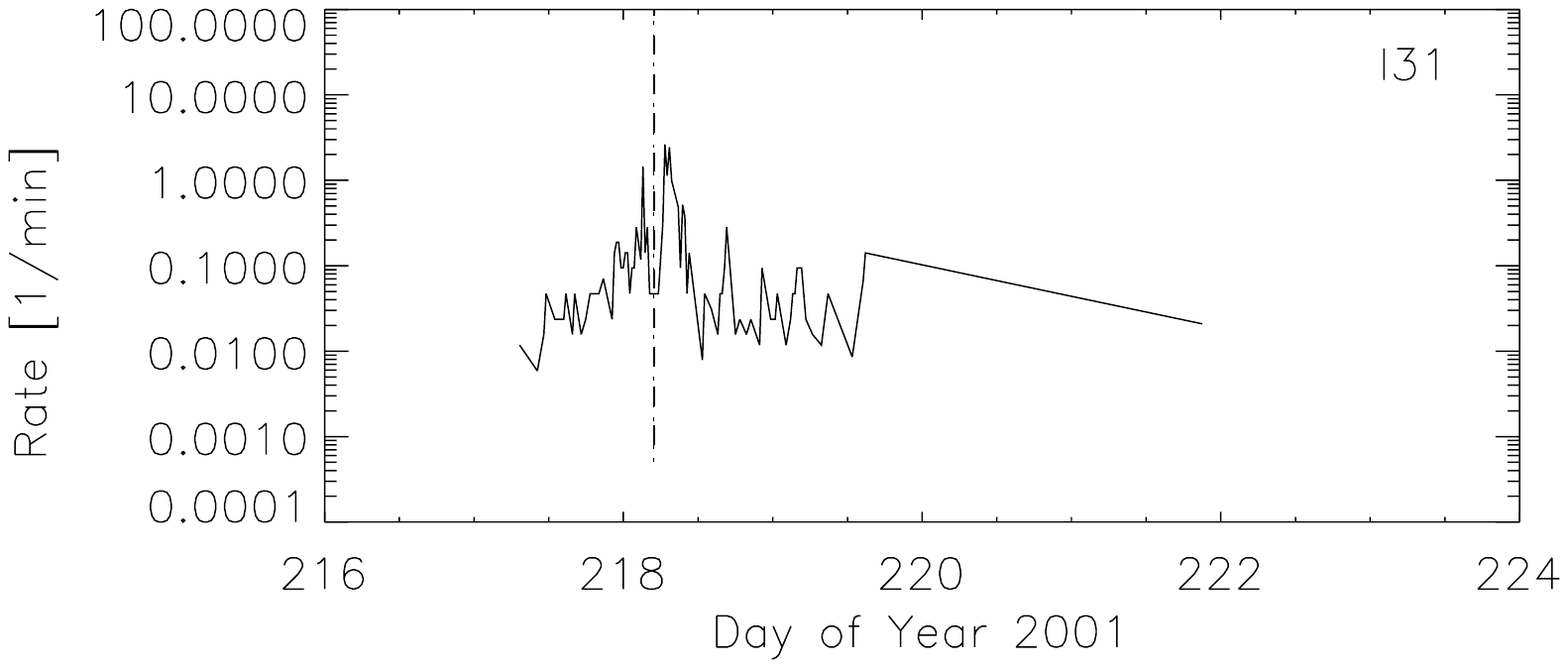}

\vspace{-9.5cm}
\includegraphics[scale=0.47]{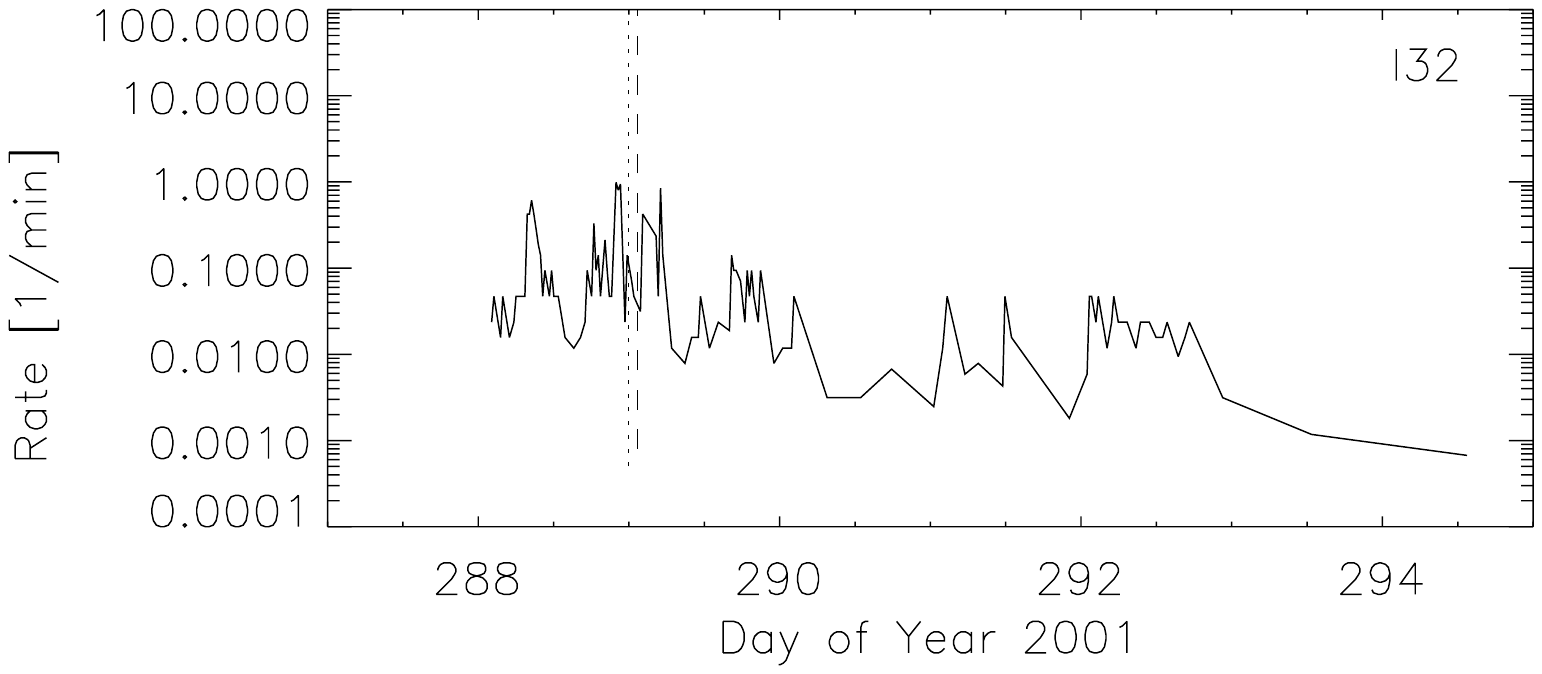}

\vspace{-9.5cm}
\includegraphics[scale=0.47]{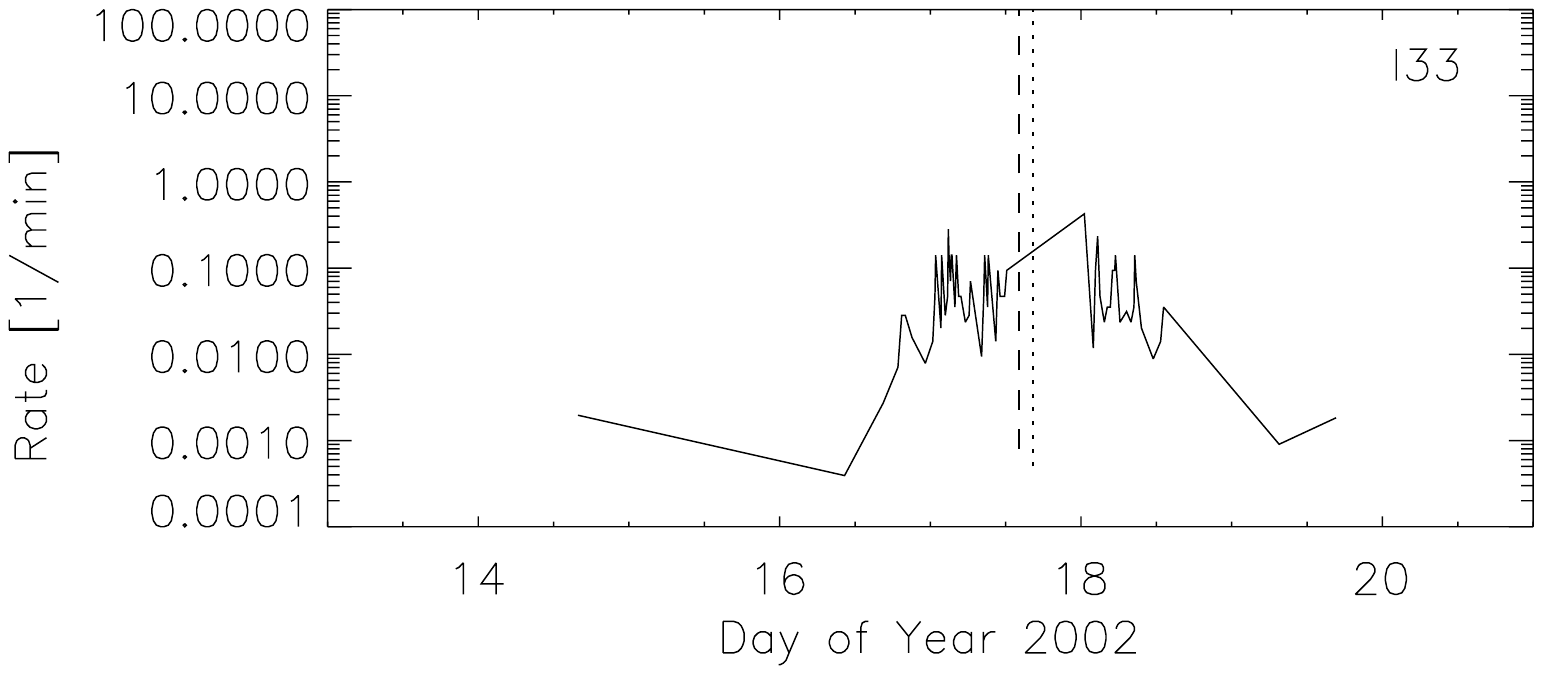}

\vspace{-9.5cm}
\includegraphics[scale=0.47]{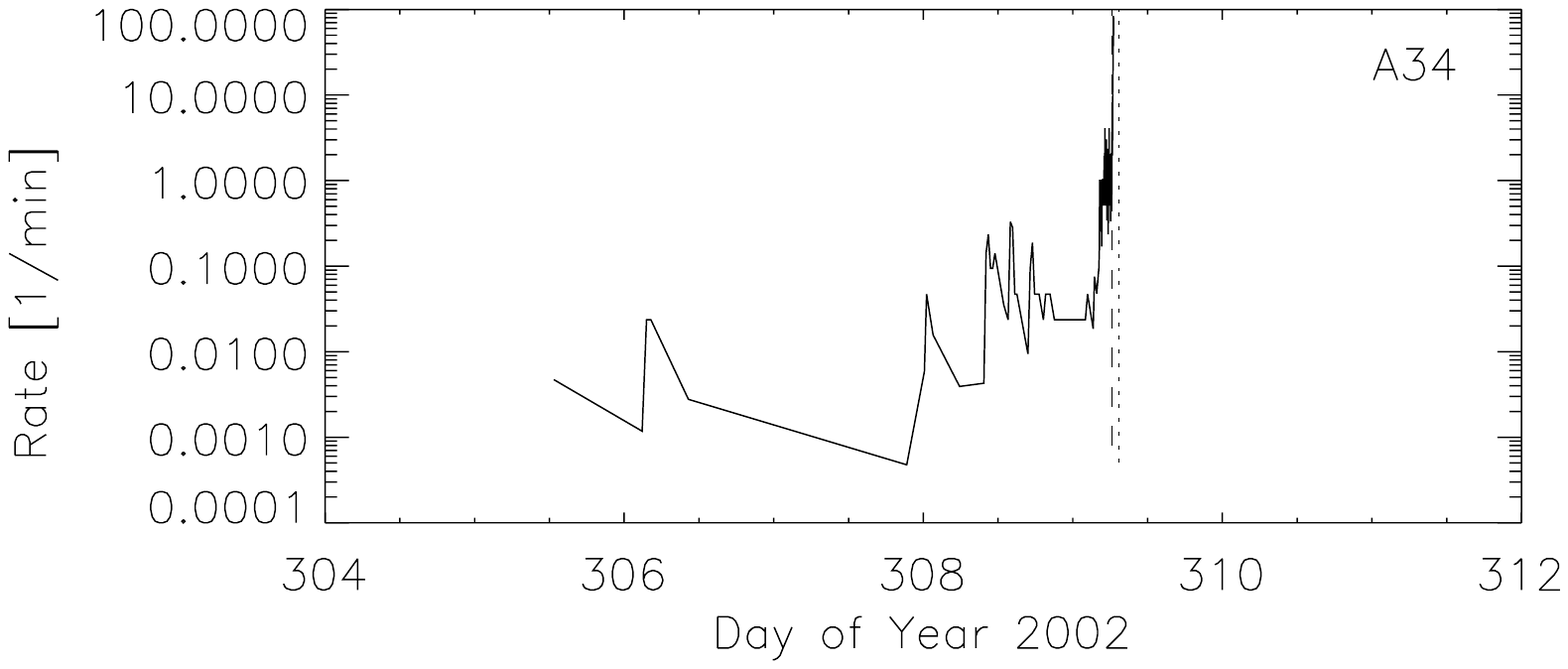}

\vspace{-9.5cm}
\includegraphics[scale=0.47]{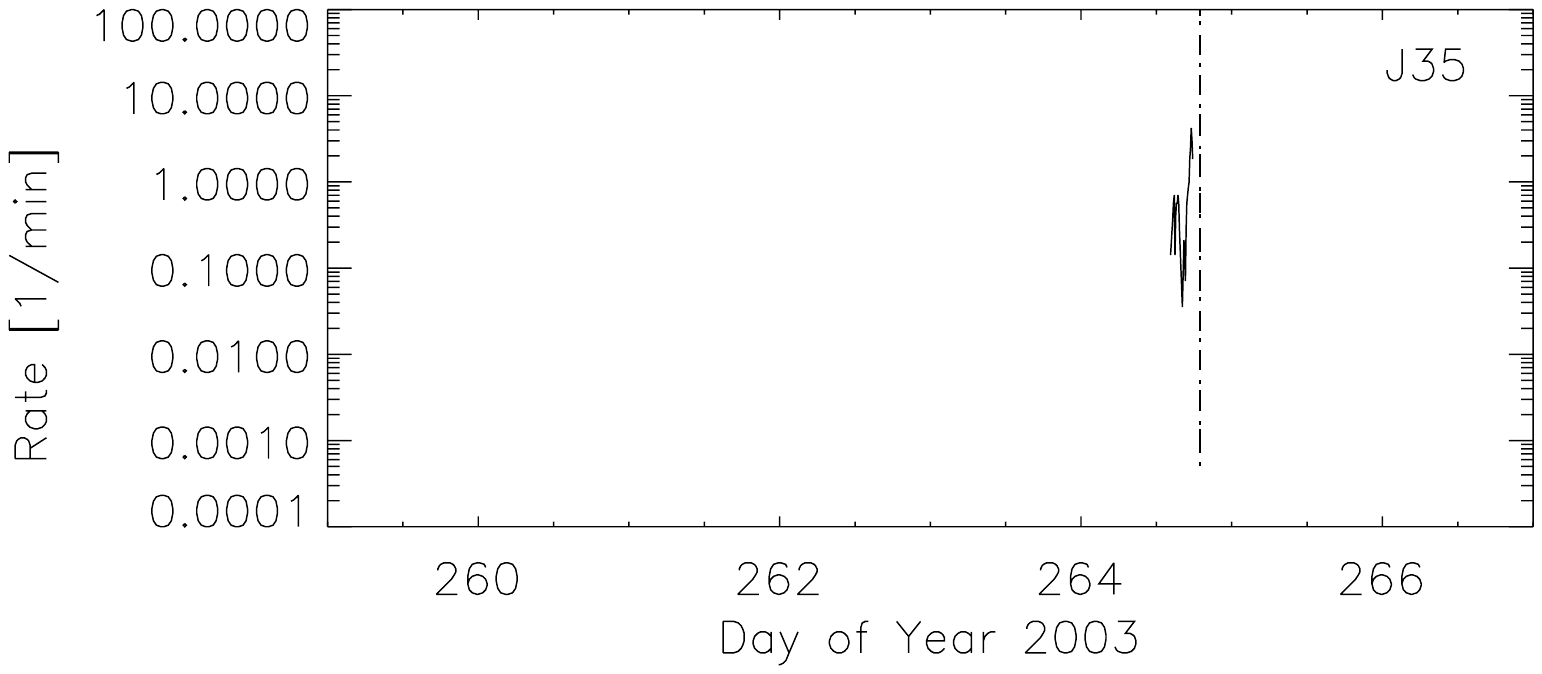}
\vspace{-1.5cm}
}
}
%\vspace{-4cm}
        \caption{\label{rate_highres}
Dust impact rate detected in the inner jovian system in higher time resolution. 
An 8-day interval is shown in each panel. Only data for AR1 (classes~2 and 3) are shown. 
Dotted lines indicate perijove passages of
Galileo, dashed lines satellite closest approaches (E26-A34) or Jupiter impact (J35).
}
\end{figure}

\clearpage

\begin{figure}
\vspace{-3cm}
\parbox{0.99\hsize}{
\includegraphics[scale=0.55]{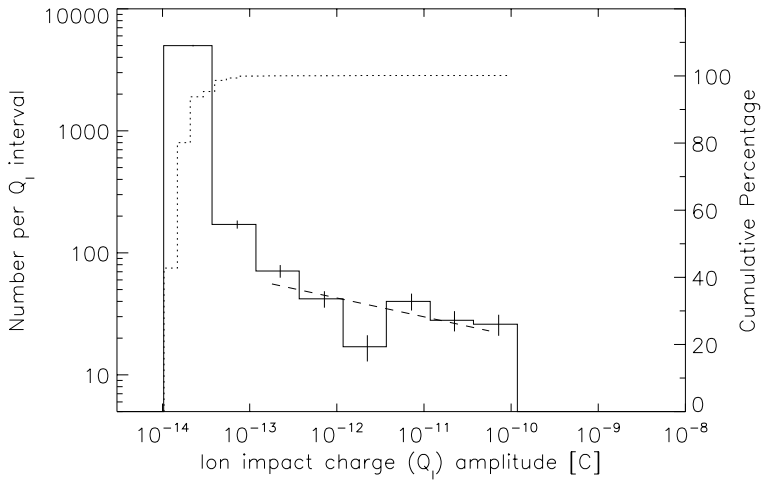}
}
\vspace{-6cm}
        \caption{\label{nqi}
Amplitude distribution of the impact charge $\QI$ for the 5389 dust particles 
detected in 2000-2003. The solid line
indicates the number of impacts per charge interval, whereas the 
dotted line shows the cumulative distribution. Vertical bars
indicate the $\sqrt{n}$ statistical fluctuation. A power law fit
to the data with $\QI > 10^{-13}\,\rm C$ (big particles, AR2-4) 
is shown as a dashed line (Number $N \sim \QI^{-0.15}$). 
}
\end{figure}

\begin{figure}
\vspace{-4cm}
\includegraphics[scale=0.55]{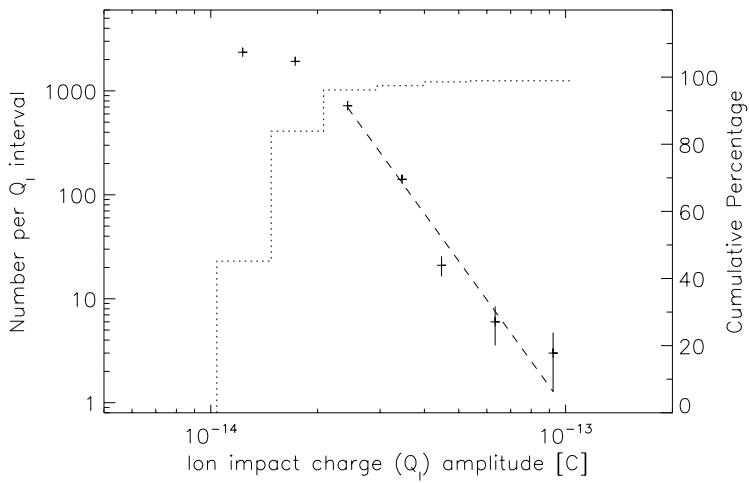}
\vspace{-7cm}
        \caption{\label{nqi2}
Same as Figure~\ref{nqi} but for the 5165 small particles in the lowest 
amplitude range (AR1) only. A power law fit to the data with 
$\rm 2\times 10^{-14}\,C < \QI < 10^{-13}\, C$ is shown as a dashed 
line (Number $N \sim \QI^{-4.72}$).
}
\end{figure}

\begin{figure}
\vspace{-4cm}
\includegraphics[scale=0.65]{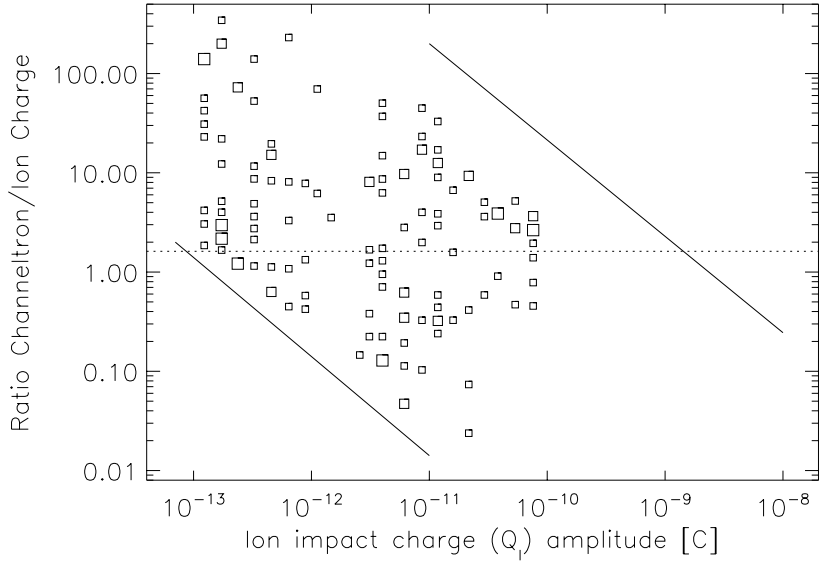}
\vspace{-8cm}
        \caption{\label{qiqc}
Channeltron amplification factor $A = \QC/\QI$
as a function of impact charge $\QI$ for big particles (AR2-6)
detected in 2000-2003. Only impacts measured with a channeltron high 
voltage setting $\mathrm{HV\,=\,6}$ are shown. The solid lines indicate the sensitivity
threshold (lower left) and the saturation limit (upper right) of the channeltron. 
Squares indicate dust particle impacts, and the area of the squares is proportional 
to the number of events (the scaling of the squares is the same as in
Papers~VI and VIII). The dotted horizontal line shows the mean value 
of the channeltron amplification $ A = 1.62$ calculated from 65 impacts 
in the ion impact charge range ${\rm 10^{-12}~C} < \QI < 10^{-10}~\rm{C}$.
}
\end{figure}

\begin{figure}
\vspace{-2cm}
\includegraphics[scale=0.65]{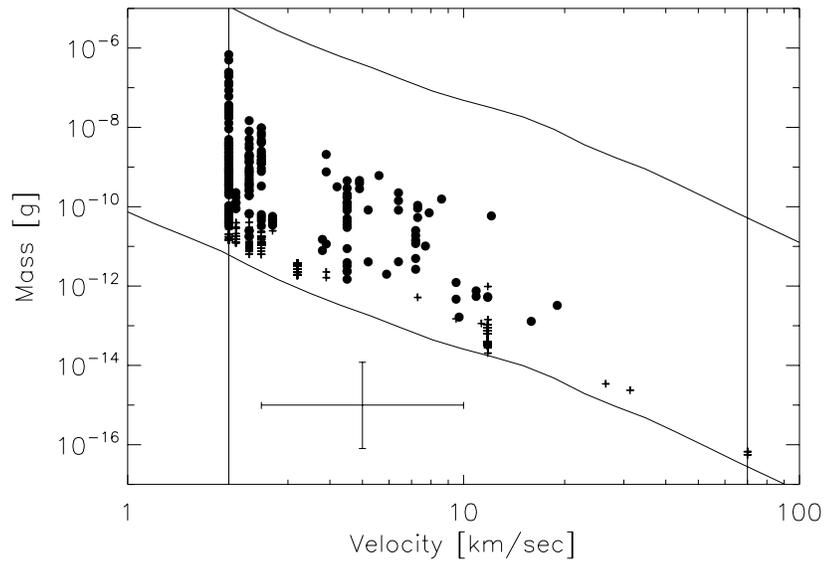}
\vspace{-8cm}
        \caption{\label{mass_speed}
Masses and impact speeds of all 5389 impacts recorded in 2000-2003.
The lower and upper solid lines indicate the threshold and
saturation limits of the detector, respectively, and the vertical lines 
indicate the calibrated velocity range. A sample error bar is shown that indicates
a factor of 2 error for the velocity and a factor of 10 for the mass determination.
Note that all particles are most likely much faster and smaller 
than implied by this diagram (see text for details).
Plus signs show particles in AR1 while filled circles refer to particles 
in AR2-4. No impacts were measured in AR5 or AR6.
}
\end{figure}

\begin{figure}
\hspace{-1.5cm}
\parbox{0.99\hsize}{
\parbox{0.49\hsize}{
\vspace{-3cm}
\includegraphics[scale=0.43]{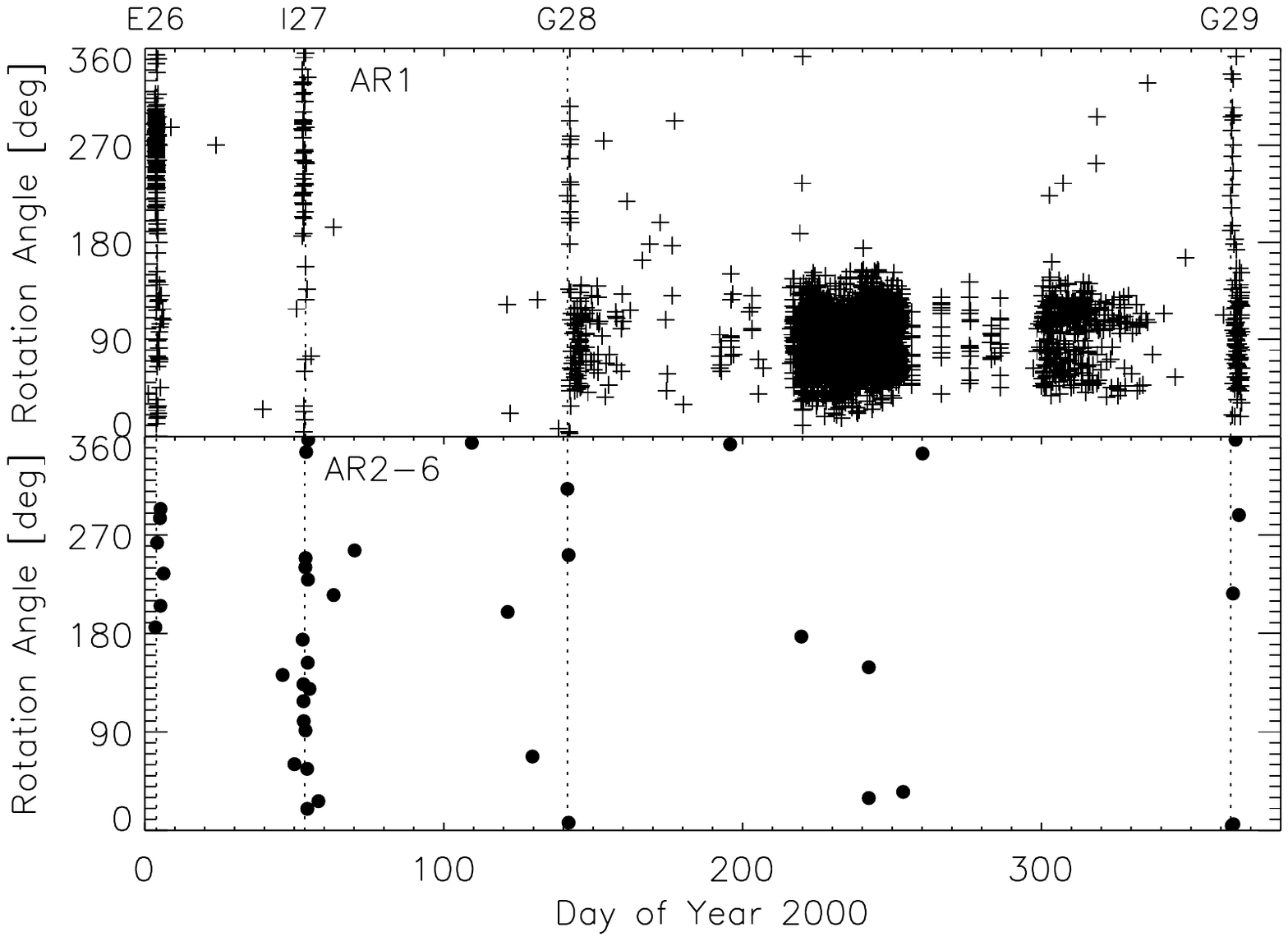}
}
\parbox{0.49\hsize}{
\vspace{-3cm}
\hspace{1cm}
\includegraphics[scale=0.43]{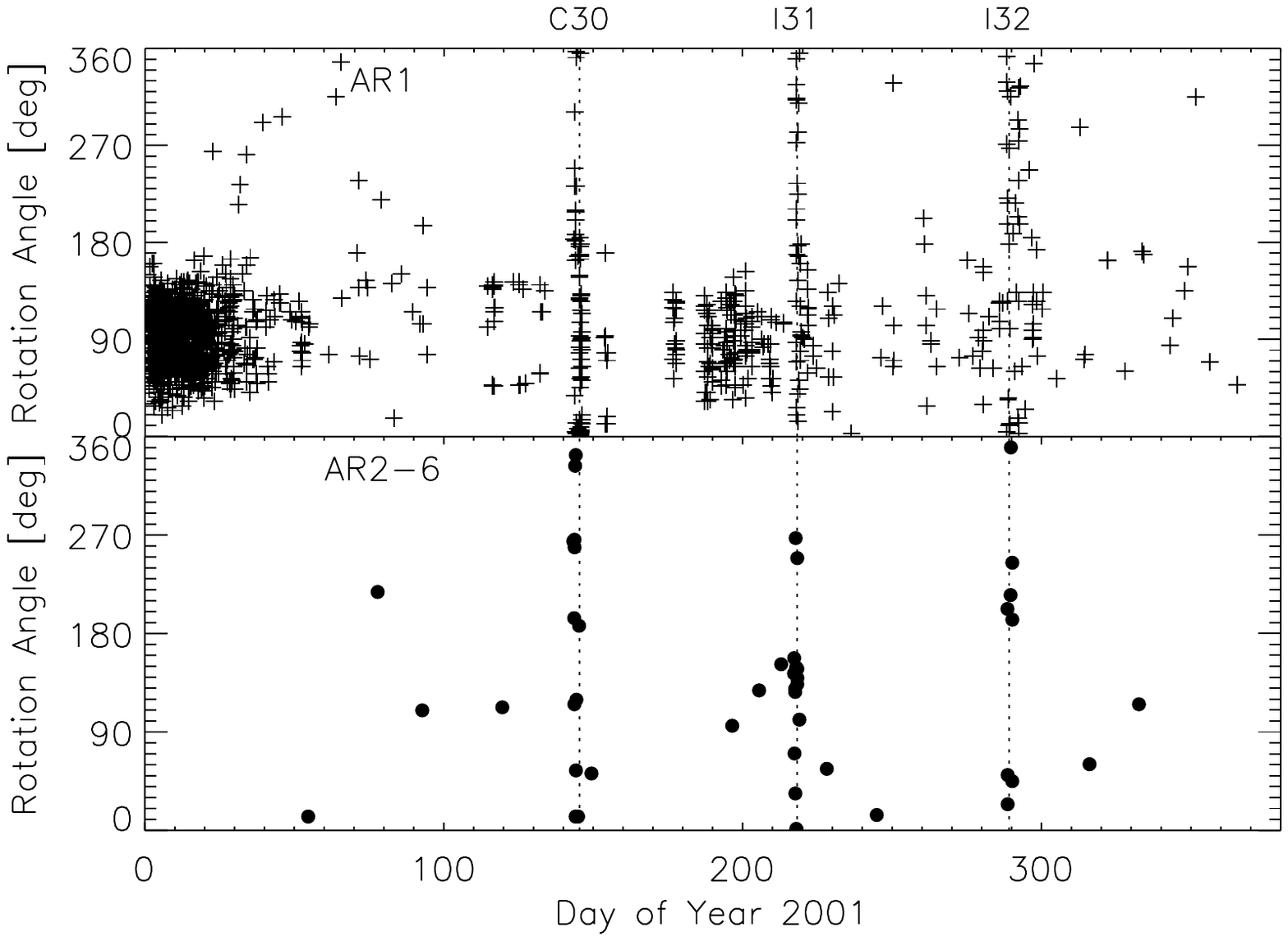}
}
\parbox{0.49\hsize}{
\vspace{-5.5cm}
\includegraphics[scale=0.43]{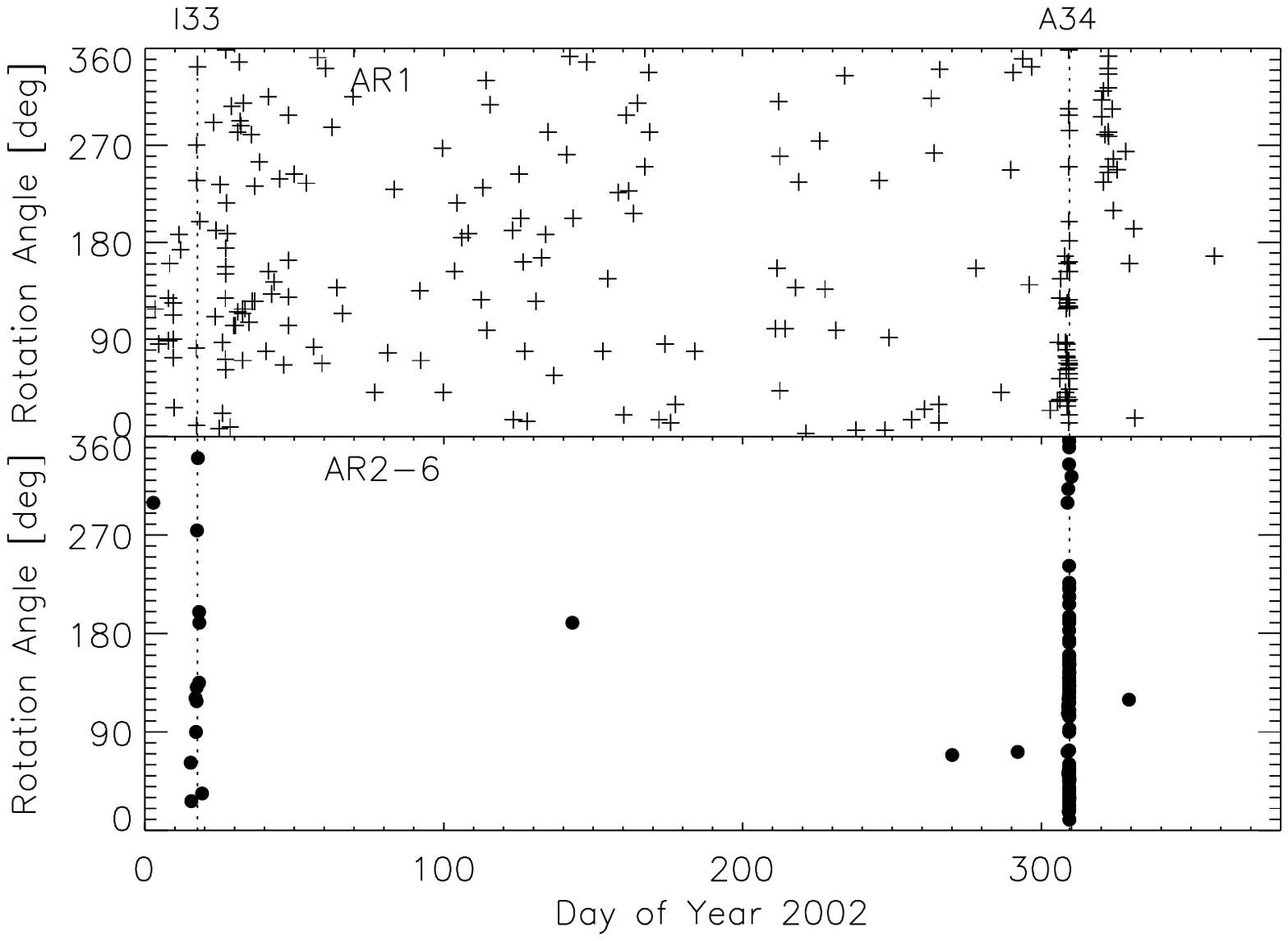}
}
\parbox{0.49\hsize}{
\vspace{-5.5cm}
\hspace{1.1cm}
\includegraphics[scale=0.43]{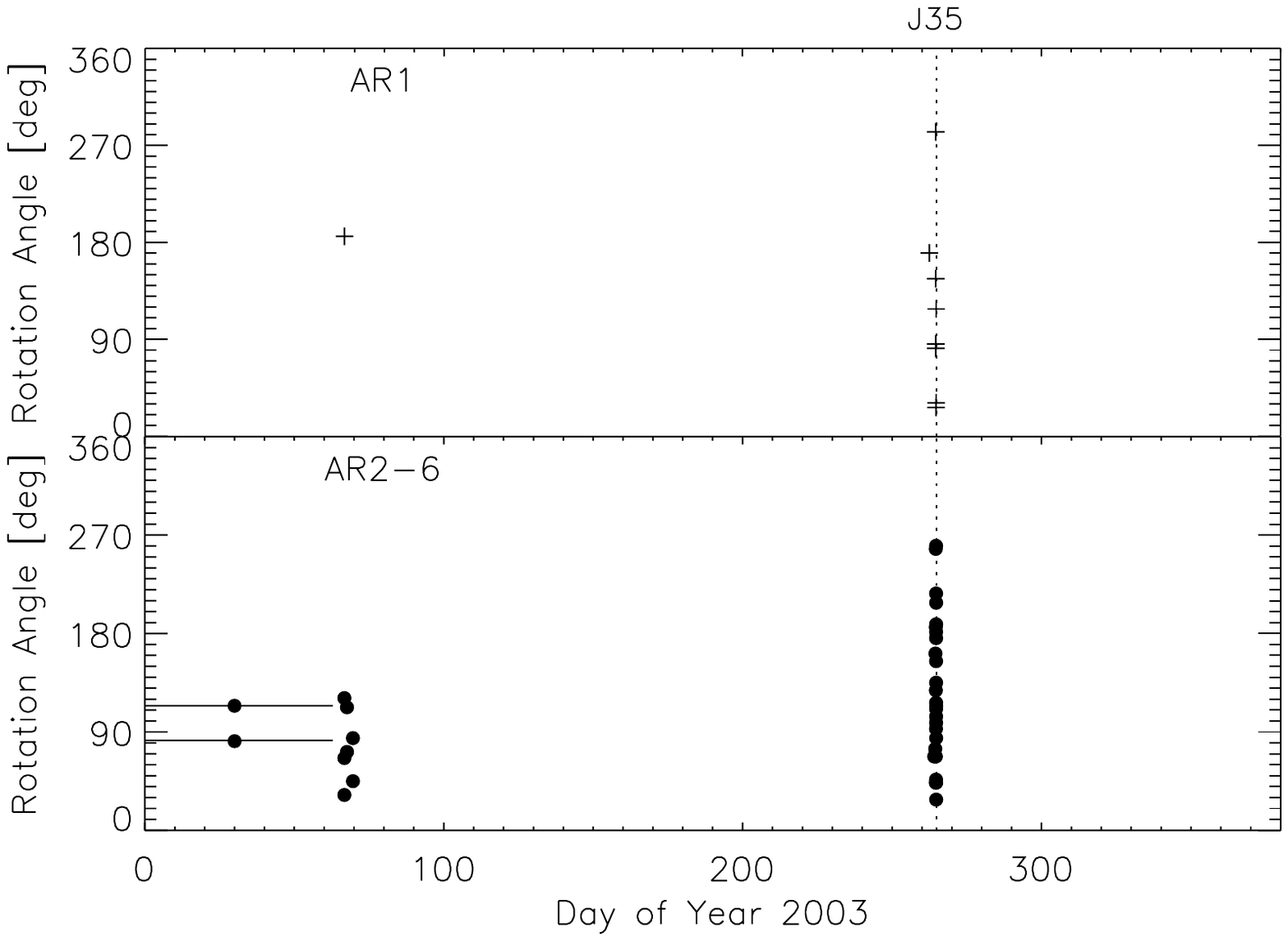}
}
}
\vspace{-5cm}
        \caption{\label{rot_angle} 
Rotation angle vs. time for two different mass ranges.
Upper panel: small particles, AR1; 
%(Io dust stream particles, except during the gossamer ring passages);
lower panel: big particles, AR2-4. See 
Section~\ref{mission} for an explanation of the rotation angle.
Vertical dotted lines indicate Galileo's satellite encounters (E26-A34)
or the spacecraft impact into Jupiter (J35).
No impacts were measured in AR5 or AR6. The uncertainty in the determination 
of the impact time is usually much smaller than the symbol sizes, 
except for two impacts in 2003 which have a very large 
uncertainty (indicated by two horizontal bars).
}
\end{figure}

\begin{figure}
\hspace{-1.3cm}
\parbox{0.99\hsize}{
\parbox{0.51\hsize}{
\vspace{-9cm}
\includegraphics[scale=0.47]{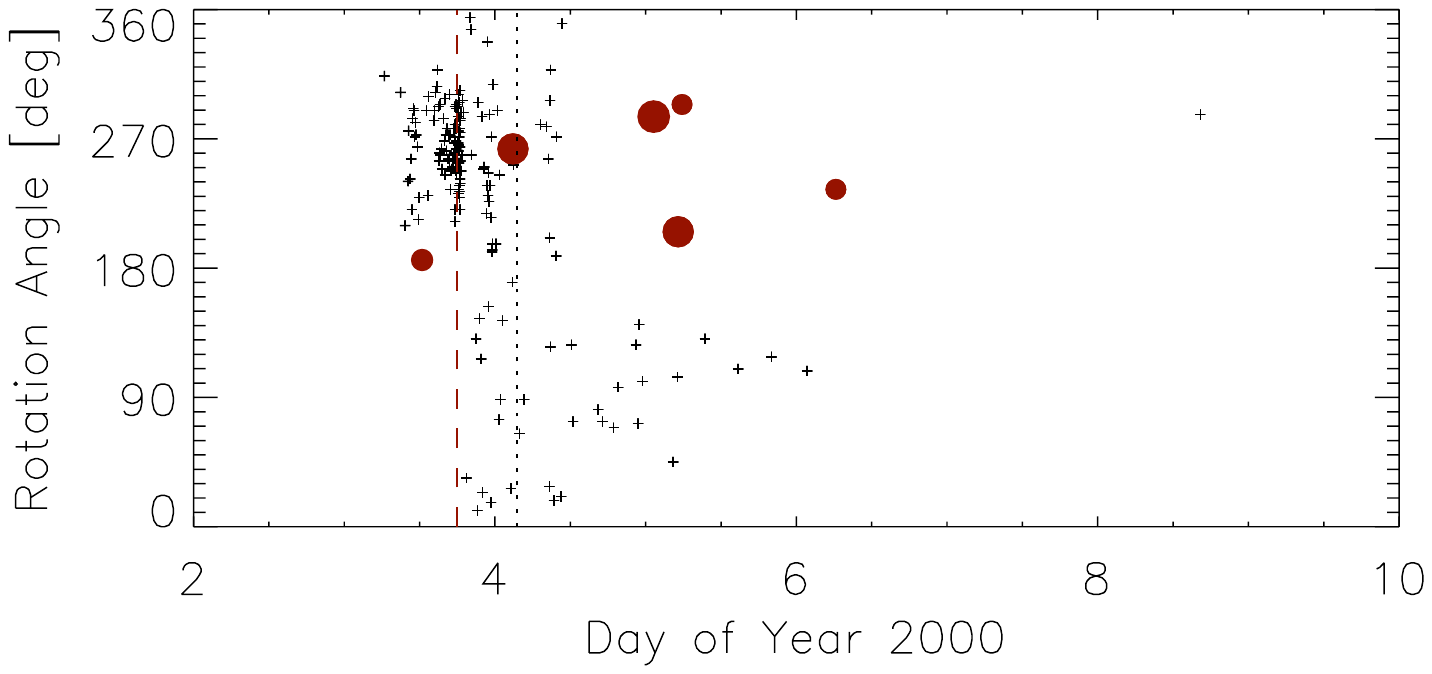}

\vspace{-9.5cm}
\includegraphics[scale=0.47]{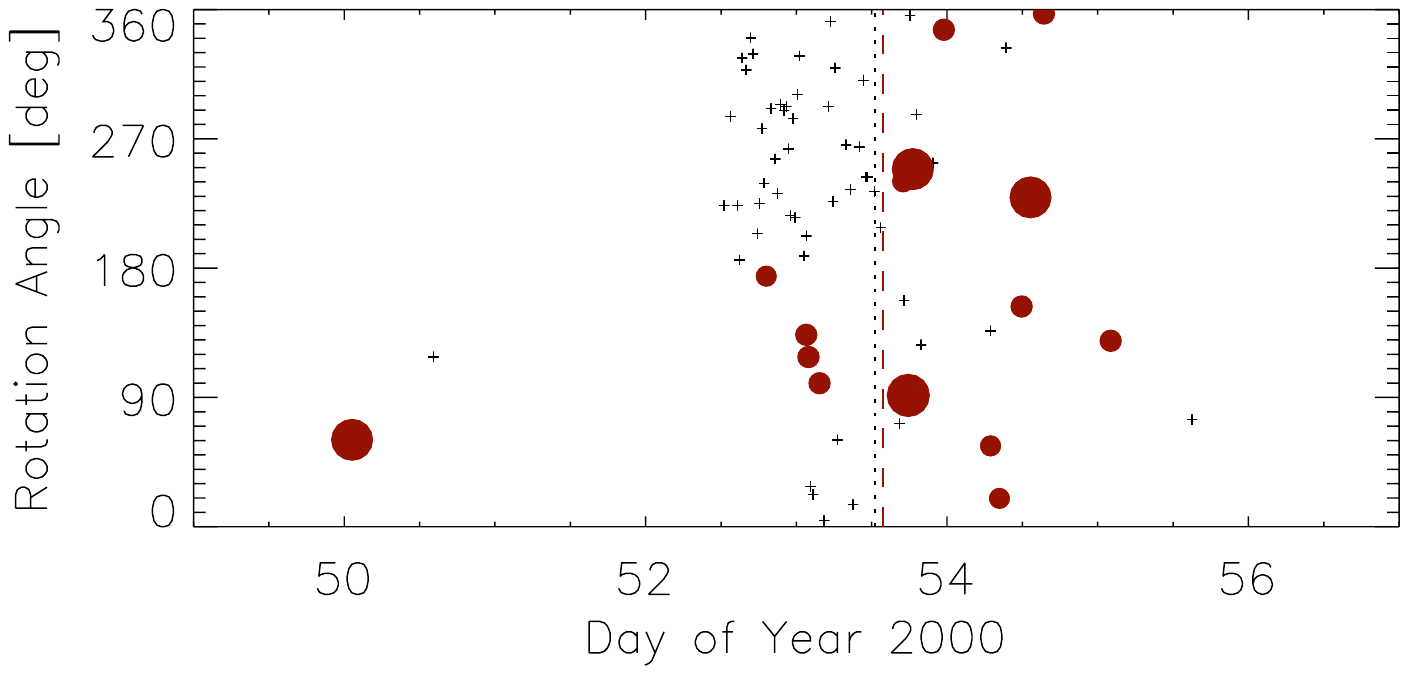}

\vspace{-9.5cm}
\includegraphics[scale=0.47]{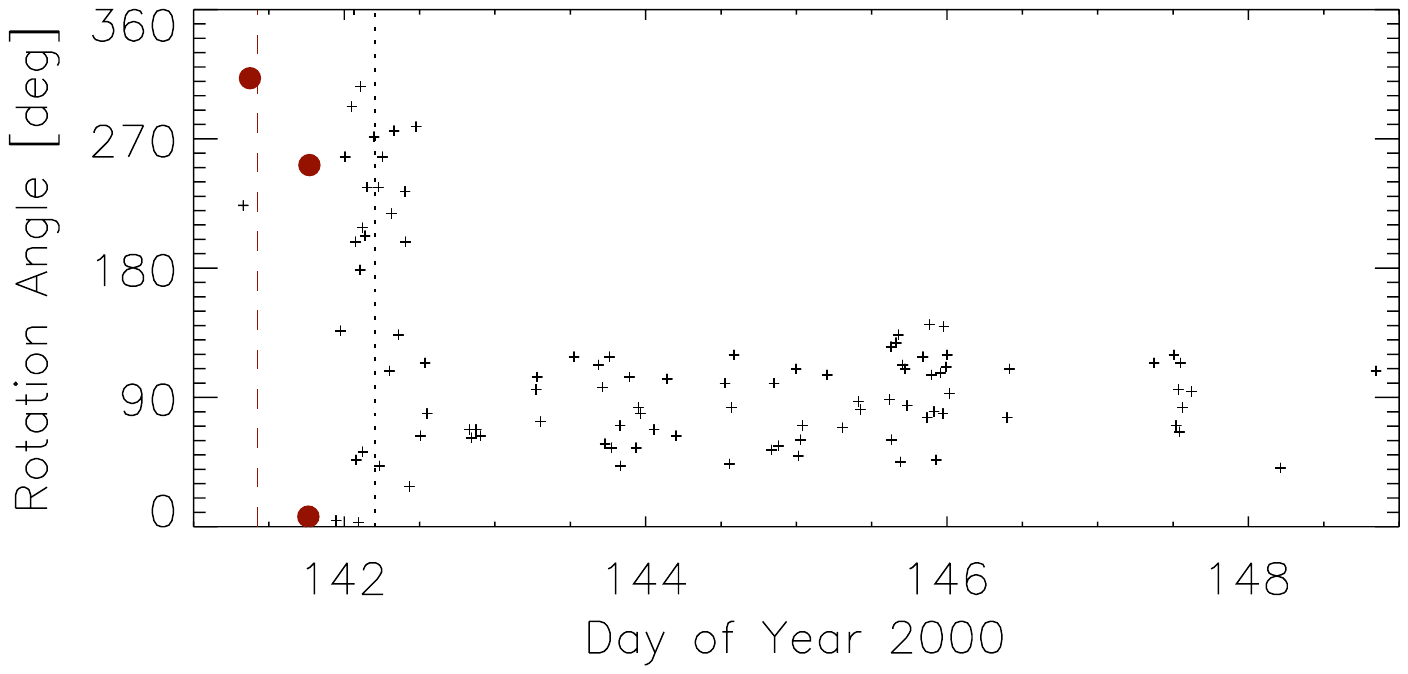}

\vspace{-9.5cm}
\includegraphics[scale=0.47]{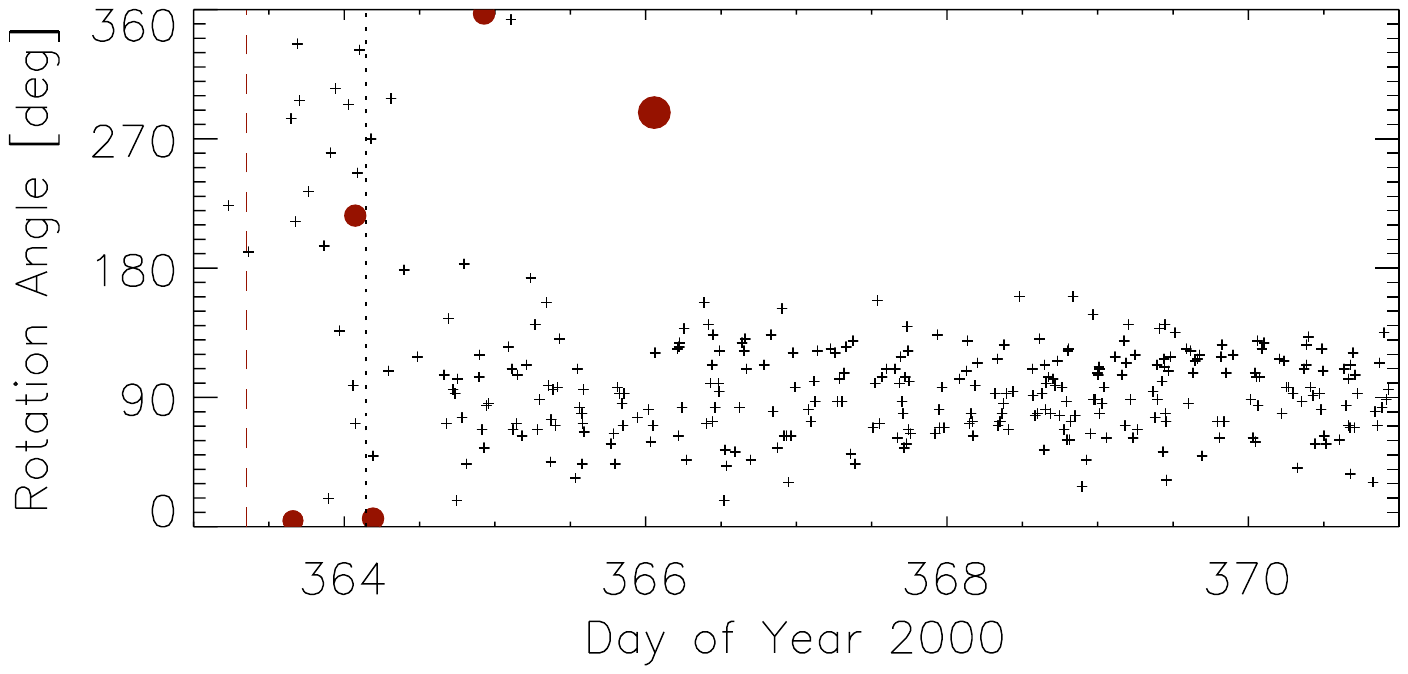}

\vspace{-9.5cm}
\includegraphics[scale=0.47]{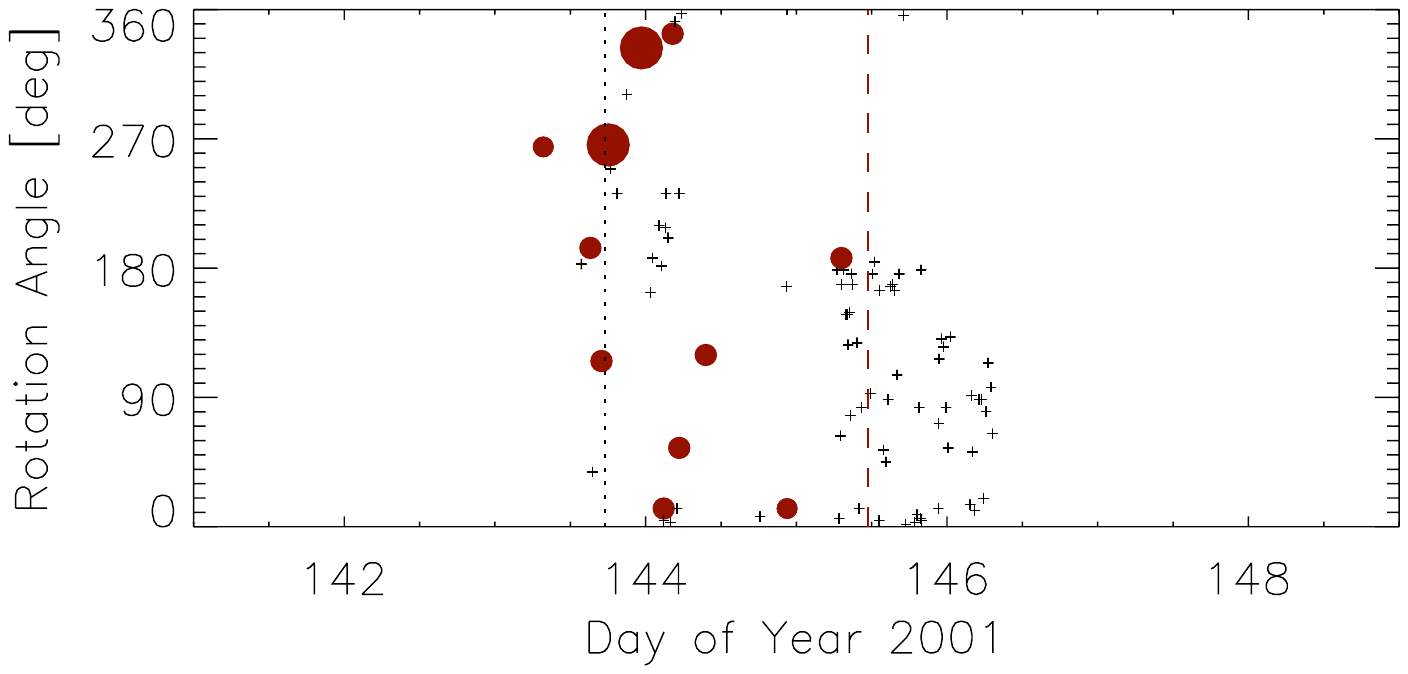}
\vspace{-1.5cm}
}
\parbox{0.51\hsize}{
%\hspace{0.7cm}
\vspace{-9cm}
\includegraphics[scale=0.47]{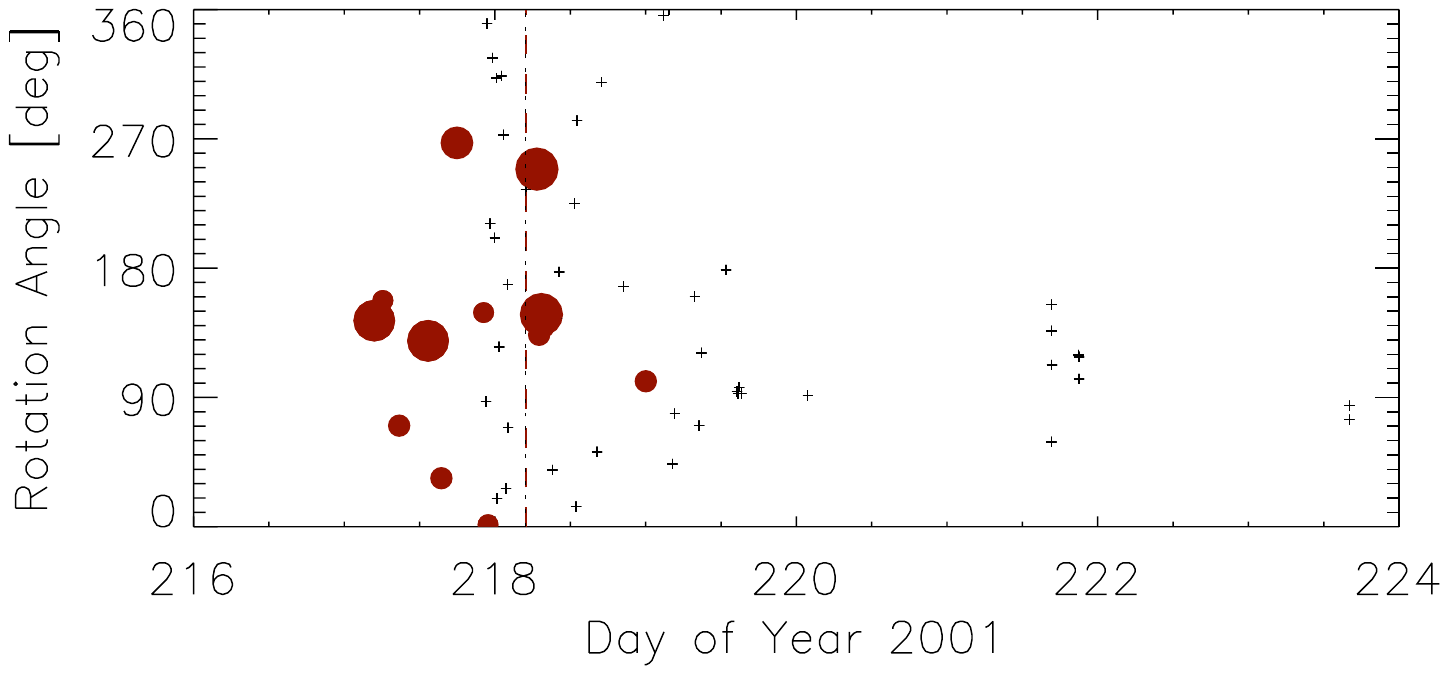}

\vspace{-9.5cm}
\includegraphics[scale=0.47]{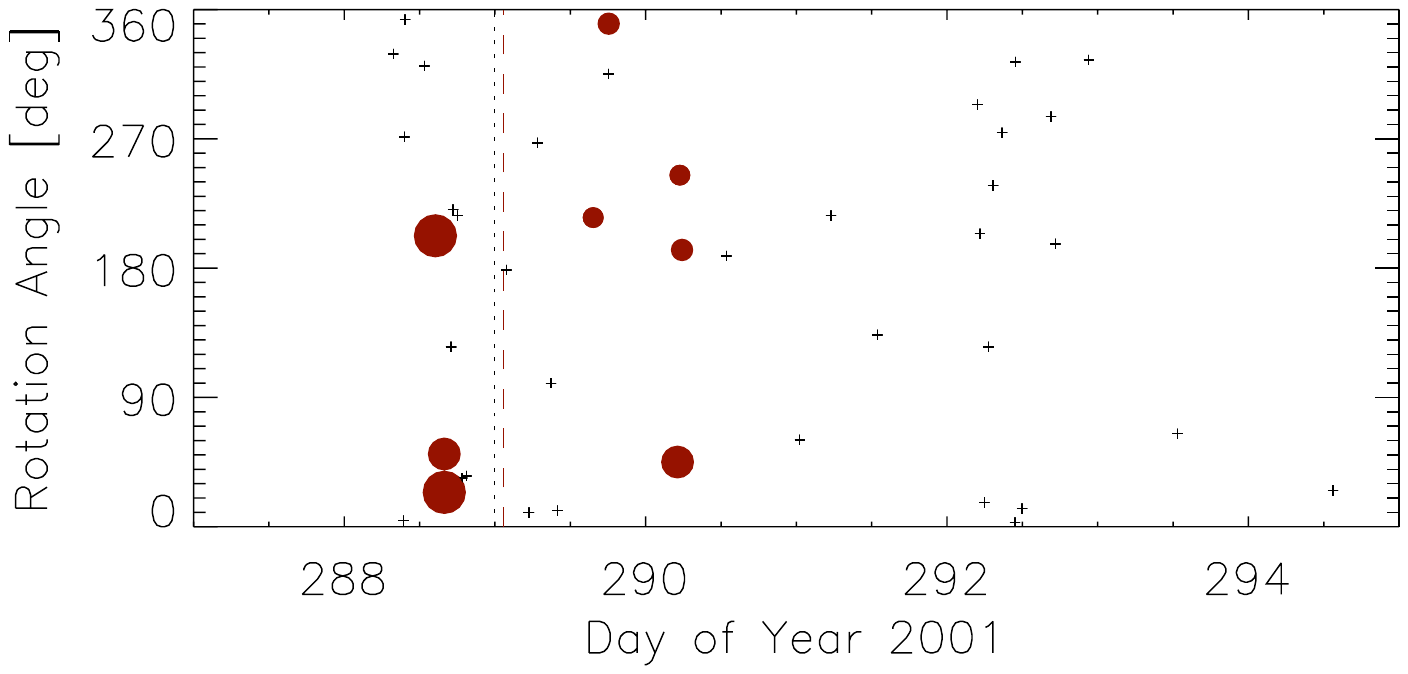}

\vspace{-9.5cm}
\includegraphics[scale=0.47]{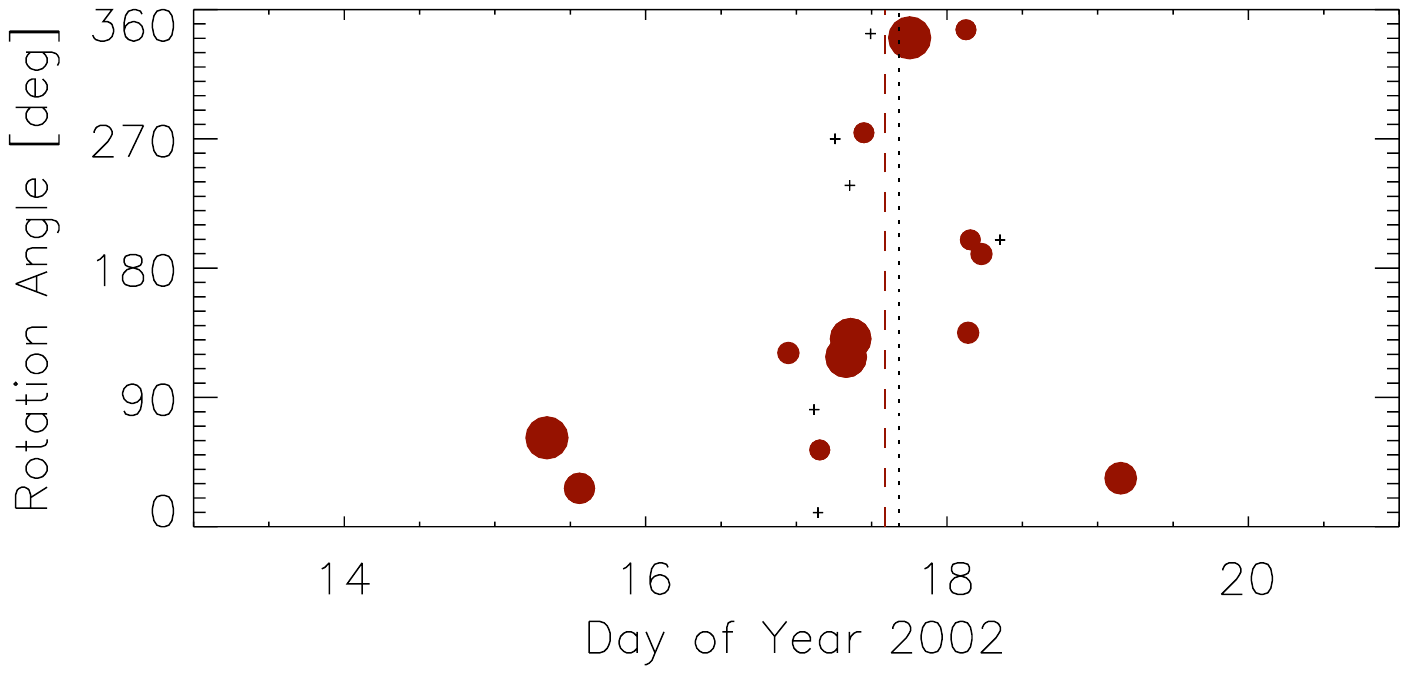}

\vspace{-9.5cm}
\includegraphics[scale=0.47]{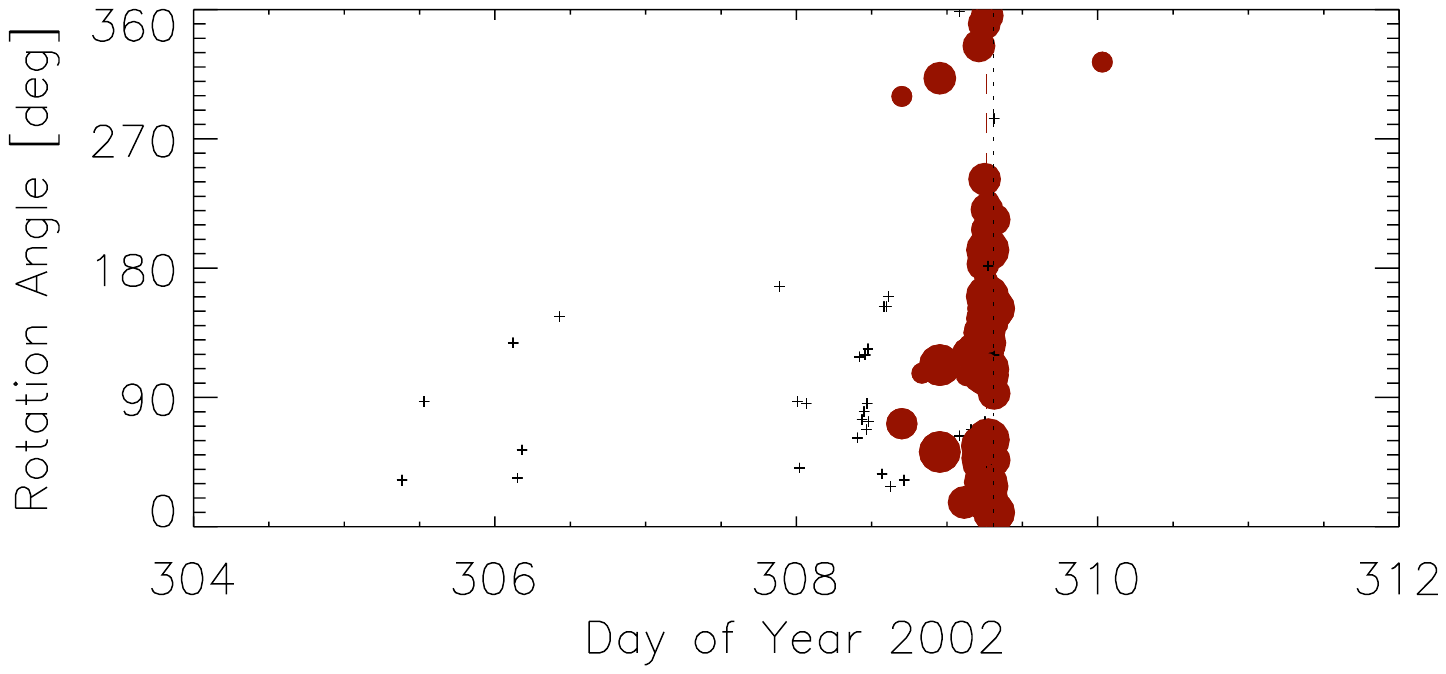}

\vspace{-9.5cm}
\includegraphics[scale=0.47]{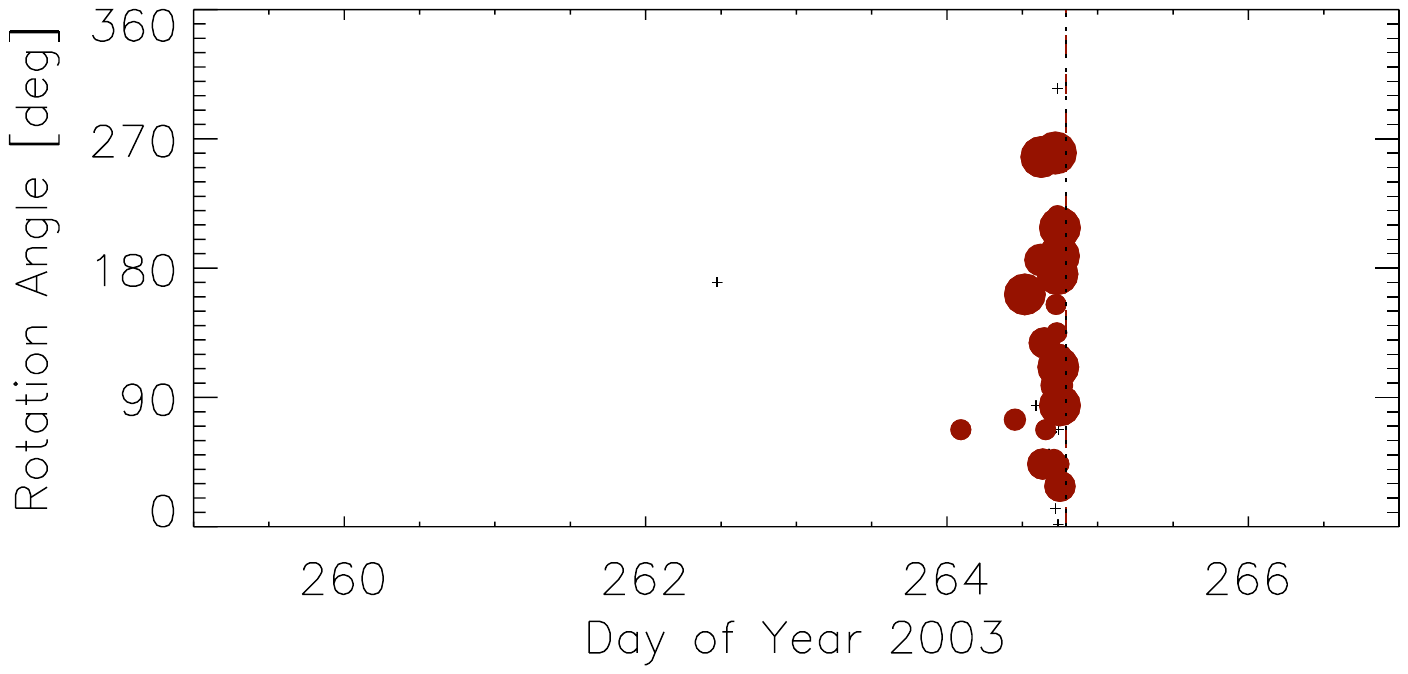}
\vspace{-1.5cm}
}
}
%\vspace{-4cm}
        \caption{\label{rot_angle_highres}
Rotation angle detected by the dust instrument in the inner jovian system in 
higher time resolution. 
Only dust data for classes~2 and 3 are shown. Crosses denote impacts
in AR1, filled circles those in  AR2-4, with the circle size indicating the 
amplitude range. Dotted lines indicate perijove passages of
Galileo, dashed lines satellite closest approaches (E26-A34) or Jupiter impact (J35).
}
\end{figure}

\begin{figure}
\vspace{-15cm}
\hspace{-2.3cm}
%\begin{turn}{90}
\includegraphics[scale=0.88]{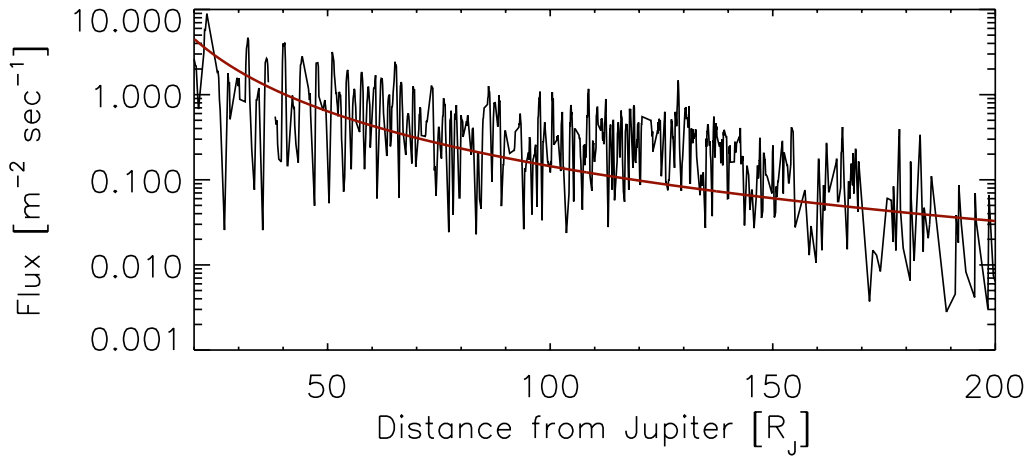}
%\end{turn}
\vspace{-3cm}
        \caption{\label{rate_g29} 
Dust flux measured during Galileo's G29 orbit. The data were smoothed with a 
2-hour boxcar average. A power law fit with slope $-2.14$ is shown.
See text for details.
}
\end{figure}

\begin{figure}
\vspace{-4cm}
\hspace{-2.0cm}
%\begin{turn}{90}
\includegraphics[scale=0.6]{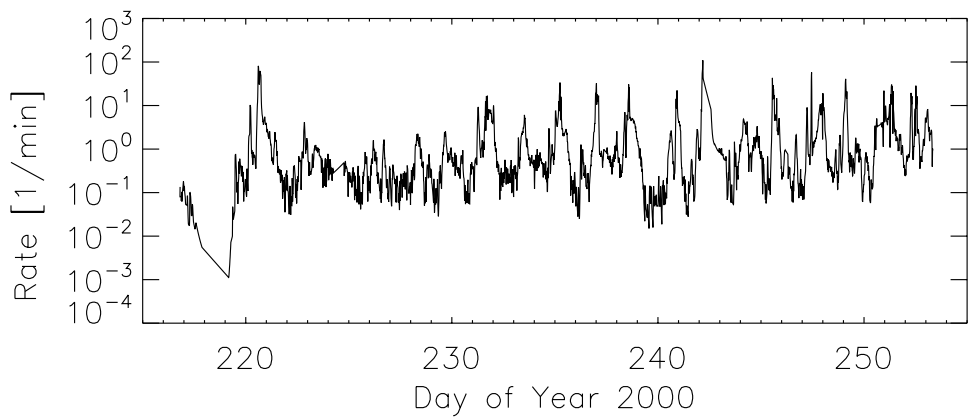}
%\end{turn}
\vspace{-4.5cm}
        \caption{\label{rate_g28} 
Impact rate of jovian dust stream particles (AC21 and AC31) measured during Galileo's G28 orbit
at approximately $\mathrm{280\,R_J}$ from Jupiter (no smoothing applied).
}
\end{figure}

\begin{figure}
\hspace{-1.5cm}
\parbox{0.99\hsize}{
\parbox{0.49\hsize}{
\vspace{-3cm}
\includegraphics[scale=0.45]{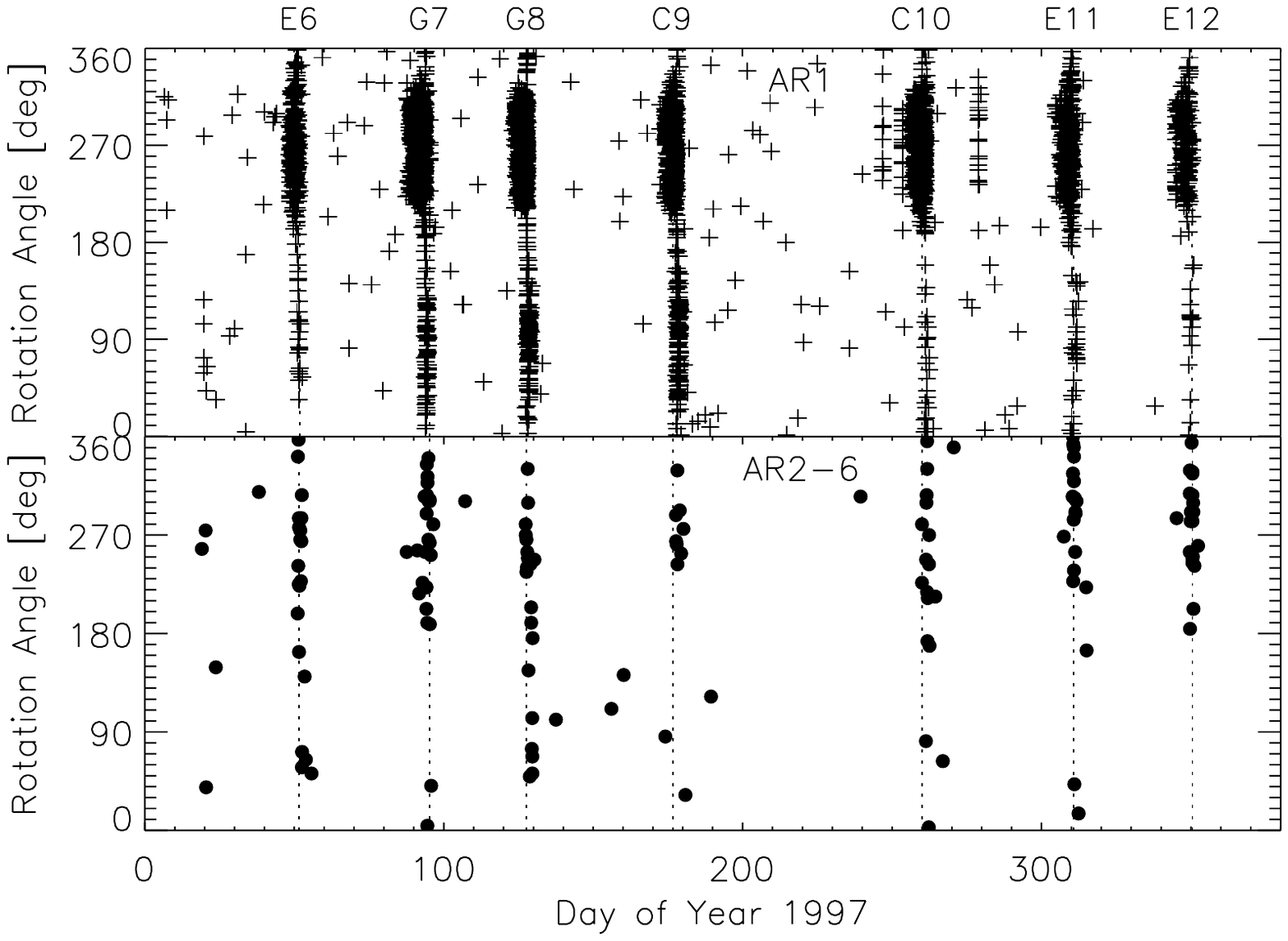}
}
\parbox{0.49\hsize}{
\vspace{-3cm}
\hspace{1cm}
\includegraphics[scale=0.45]{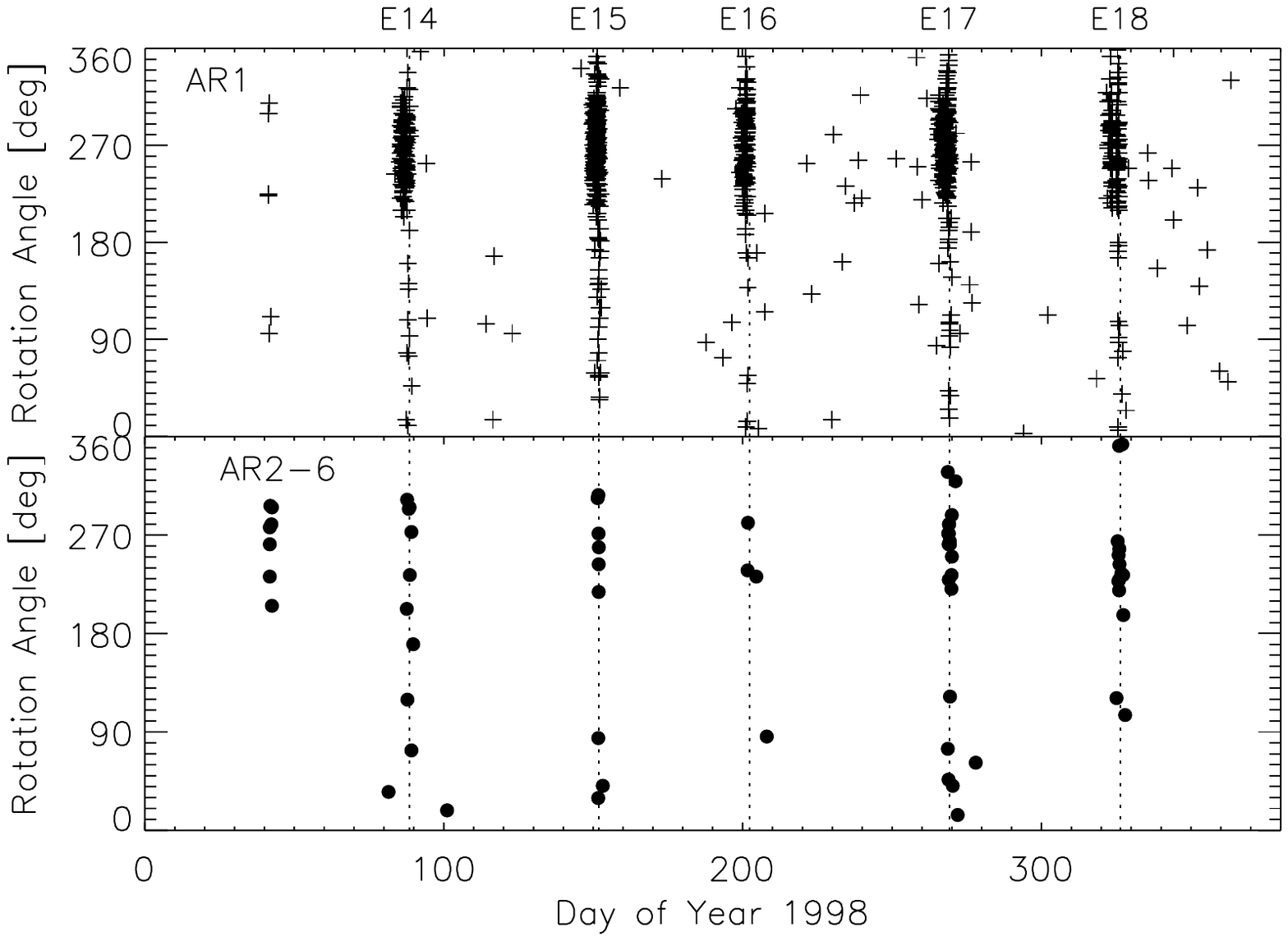}
}
\parbox{0.49\hsize}{
\vspace{-5.5cm}
\includegraphics[scale=0.45]{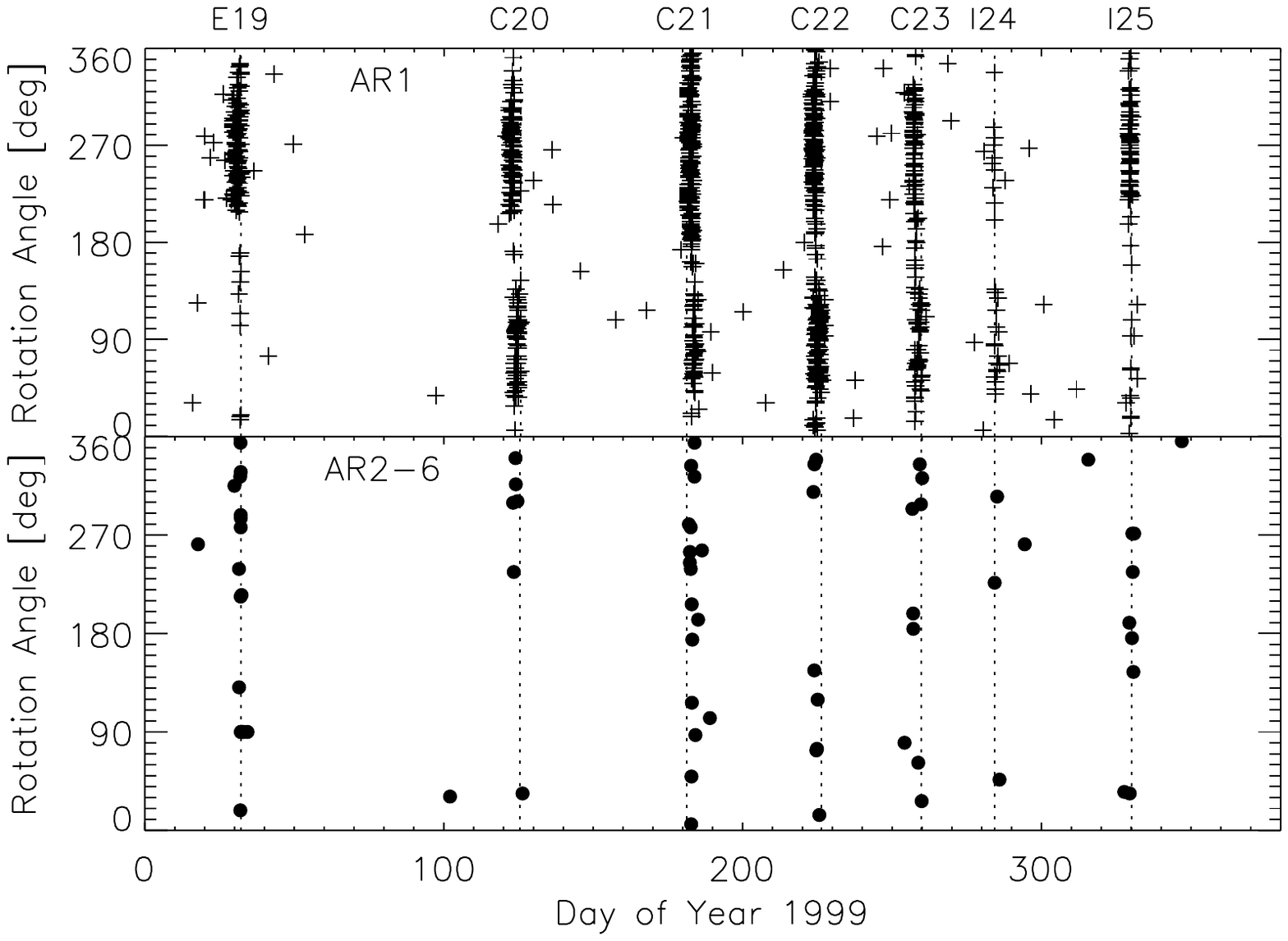}
}
}
\vspace{-5cm}
        \caption{\label{rot_old}
Correction for Paper~VIII: rotation angle vs. time for two different mass ranges
for the time interval 1997-1999.
Upper panel: small particles, AR1 (Io dust stream particles);
lower panel: big particles, AR2-4. 
Vertical dotted lines indicate Galileo's satellite encounters.
}
\end{figure}

\end{document}